\PassOptionsToPackage{warn}{textcomp}
\documentclass[a4paper, 12pt]{article}
\usepackage{iftex}
\ifPDFTeX
  \DeclareUnicodeCharacter{2212}{\ensuremath{-}}
\fi
\usepackage{graphicx}
\newsavebox{\paperTableBox}
\newcommand{\paperfit}[2]{\par\noindent\sbox{\paperTableBox}{#2}\ifdim\wd\paperTableBox>#1\resizebox{#1}{!}{\usebox{\paperTableBox}}\else\usebox{\paperTableBox}\fi\par}
\usepackage{amsmath}
\usepackage{amsfonts}
\usepackage{amssymb}
\usepackage{amsthm}
\usepackage{bm}
\usepackage{accents}
\usepackage{statex}
\usepackage{mathptmx}
\usepackage{marginnote}
\usepackage{enumitem}
\usepackage{rotating}
\usepackage{fancyvrb}
\usepackage{epstopdf}
\usepackage{float}
\usepackage{cases}
\usepackage[doublespacing]{setspace}
\usepackage{natbib}
\usepackage{multirow}
\usepackage{dcolumn}
\usepackage{booktabs}
\usepackage{caption}
\usepackage{threeparttable}
\usepackage{longtable}
\usepackage{tabularx}
\usepackage{makecell}
\usepackage{siunitx}
\usepackage{geometry}
\usepackage{hyperref}
\usepackage{subcaption}
\usepackage{pdfpages}
\usepackage{lscape}
\usepackage{url}
\usepackage{titling}
\usepackage{titlesec}
\usepackage{times}
\usepackage{comment}
\usepackage{datetime}
\usepackage{textcomp}
\usepackage{pdfpages}
\usepackage[symbol]{footmisc}
\usepackage[svgnames]{xcolor}
\usepackage{tikz}
\usetikzlibrary{arrows.meta,positioning,calc,fit,backgrounds}
\hypersetup{colorlinks, citecolor=NavyBlue, filecolor=NavyBlue, linkcolor=NavyBlue, urlcolor=NavyBlue}

\makeatletter
\def\wideubar{\underaccent{{\cc@style\underline{\mskip8mu}}}}
\def\sym#1{\ifmmode^{#1}\else\(^{#1}\)\fi}
\def\@pdfborder{0 0 0}

\numberwithin{corollary}{section}
\numberwithin{lemma}{section}
\numberwithin{proposition}{section}
\numberwithin{theorem}{section}
\numberwithin{remark}{section}

\titlelabel{\thetitle.\quad}
\titleformat{\subsection}{\bf\large\itshape}{\thesubsection.}{1em}{}

\newdateformat{monthyeardate}{%
	\monthname[\THEMONTH] \THEYEAR}

\renewcommand*{\thefootnote}{\fnsymbol{footnote}}
\graphicspath{{./Figures/}}

\begin{document}
	\setlength{\droptitle}{-5em}
	\title{\textbf{Tracking High-Frequency Trading Activity in China and Its Effects on Market Quality}\thanks{
    \footnotesize
    % Li Guo, School of Economics, Fudan University and Shanghai Institute of International Finance and Economics; Frank Weikai Li, City University of Hong Kong; Ding Qi, School of Economics, Fudan University; Lijian Sun, School of Economics, Fudan University. Send correspondence to Ding Qi, dqi24@m.fudan.edu.cn, No. 600 Handan Road, Shanghai, China. We thank Chao Zhang, Shihao Yu and seminar participants at HKUST(Guangzhou) for useful comments. All authors contributed equally to this paper.
%
    Li Guo, School of Economics, Fudan University and Shanghai Institute of International Finance and Economics (guo\_li@fudan.edu.cn); Frank Weikai Li, City University of Hong Kong (weikaili@cityu.edu.hk); Ding Qi, School of Economics, Fudan University (dqi24@m.fudan.edu.cn); Lijian Sun, School of Economics, Fudan University (sunlijian@fudan.edu.cn). We thank Hongzhong Liu, Wendi Wu, Shihao Yu, Chao Zhang, and Guofu Zhou for helpful comments. We also thank seminar participants at HKUST (Guangzhou), Fudan University, Kobe University, and Hitotsubashi University for valuable feedback.
    }}

	\author{ Li Guo
		\hspace{10mm} Frank Weikai Li
		\hspace{10mm} Ding Qi
		\hspace{10mm} Lijian Sun}

	\date{}

	\begin{titlingpage}
		\maketitle

		\vspace{-2cm}

		\begin{center}
			{
				This draft: \monthyeardate\today
			}
		\end{center}
		\thispagestyle{empty}

		\begin{abstract}
        \begin{spacing}{1.0}
We develop a new order-level proxy for high-frequency trading (HFT) activity based on precise order placement and study its effects on market quality. With quotes updating multiple times per second, millisecond-level order-book access makes high-frequency traders more likely to submit active exact-match orders, defined as limit orders priced exactly at the prevailing best opposite-side quote. Using Shenzhen Stock Exchange data, we validate the proxy: active exact-match orders respond faster to quote updates, cancel more rapidly, concentrate when depth imbalance is favorable and spreads are narrow, and predict mid-quote returns from 100 milliseconds through ten minutes. A difference-in-differences design exploiting cross-channel variation in order dissemination latency at the market open shows that lower latency increases active exact-match shares. This variation yields plausibly causal evidence that latency-induced HFT raises short-horizon price variation, wide-spread incidence, and adverse-selection exposure, while realized spreads decline and pricing error falls modestly. Unlike developed markets, China's T+1 constraint, absence of maker-taker rebates, and limits on rapid quote revision tilt speed advantages toward directional liquidity taking, highlighting how market design shapes the channel through which speed affects market quality.
        \medskip

         \medskip

        \end{spacing}
		\end{abstract}

	\end{titlingpage}

	\setcounter{page}{1}
	\renewcommand*{\thefootnote}{\arabic{footnote}}
	\interfootnotelinepenalty=10000 %footnote in one page

	\section{Introduction}

    High-frequency trading (HFT) accounts for a substantial share of trading volume worldwide \citep{sec2010concept,esma2014hft,zaharudinHighfrequencyTradingDefinition2022} and has changed how information enters prices. As \citet{oharaHighFrequencyMarket2015a} observes, ``HFTs also turn speed into information via colocation and other technologies,'' enabling fast traders to observe and react to market updates before slower participants. This growing influence has generated an active debate over HFT's effects on market quality, fairness, and stability. Evidence on market quality is mixed: many studies find improvements in liquidity and price efficiency \citep{hendershottDoesAlgorithmicTrading2011,hasbrouckLowlatencyTrading2013,conradHighfrequencyQuotingTrading2015,boehmerCompetitiveLandscapeHighFrequency2018,malcenieceHighFrequencyTrading2019,boehmerAlgorithmicTradingMarket2021,YueAlgorithmicTradingMarketImpact2025}, while others document adverse selection, short-term volatility, and fairness concerns \citep{hanLiquidityHighFrequencyMarket2014,kirilenkoFlashCrashHighFrequency2017,brogaardHighFrequencyTrading2018,korajczykHighFrequencyMarketMaking2019,shkilkoEveryCloudHas2020,hoffmannDynamicLimitOrder2014,yaoWhyTradingSpeed2018,goldsteinHighFrequencyTradingStrategies2023}. Faster information processing can reduce market-making costs and improve liquidity, but it can also expose slower liquidity providers to adverse selection and alter their incentives to supply liquidity \citep{bongaertsCompetitionLiquidityProviders2021}. The resulting effects on market quality therefore depend on traders' strategic responses and market conditions \citep{brogaardStrategicLiquidityProvision2025}.

    A central empirical challenge in studying these effects is identification. Direct HFT labels are proprietary, rarely available to researchers, and often limited in coverage \citep{brogaardHighFrequencyTradingPrice2014}. Public-data proxies based on message traffic, rapid cancellations \citep{hasbrouckLowlatencyTrading2013}, and immediate-or-cancel orders \citep{aquilinaQuantifyingHighFrequencyTrading2021} are more broadly applicable, but may capture heterogeneous forms of fast or intensive trading rather than HFT-specific behavior. More importantly, these measures do not directly capture the behavioral implication of low-latency information processing that makes HFT economically distinctive. Precise price placement offers a way to identify trading behavior that systematically relies on timely order-book information.

    We propose a new order-level proxy for HFT activity based on exact matching of the prevailing best quote. Active exact-match orders are marketable limit orders submitted at the best opposite-side quote, while passive exact-match orders are posted at the best same-side quote. When quotes change multiple times per second, repeatedly matching the current best quote requires timely observation and processing of the order book. Exact matching therefore provides a measurable signature of order placement based on current market information. The economic logic of the proxy is closely related to the thin-margin nature of low-latency trading. \citet{aquilinaQuantifyingHighFrequencyTrading2021} document that the average race is worth just over half a tick, so even small deviations from the best available price can materially affect trading outcomes. Exact price placement is thus informative not simply because it reflects fast reaction, but because it reveals the ability to process real-time order-book information and select the precise contemporaneous best quote. This focus is also consistent with theoretical work showing that competition between HFTs and slower algorithms depends critically on tick-size constraints, particularly when the bid--ask spread is binding at one tick \citep{liWhoProvidesLiquidity2021}. Compared with approaches that rely primarily on speed-based signatures \citep{chakrabartyIdentifyingHighFrequency2023}, our method captures a more specific behavioral outcome of latency-sensitive information processing.

    China's centralized, order-driven equity market lets us observe trading within a single exchange order book, without the fragmentation across venues found in some developed markets. Commonly available quotes provide coarse snapshots, while millisecond-level information requires reconstructing the book from detailed Level-2 order messages. Traders who reconstruct the book can therefore act on information that is not yet visible to those relying on lower-frequency updates. Our Level-2 data record each order's submission time, price, quantity and direction, along with subsequent executions and cancellations. Comparing submitted prices with reconstructed quotes lets us distinguish exact matching from more aggressive or less precise price placement and test whether it reveals latency-sensitive behavior. Section~\ref{sec:Data and Methodology} describes the institutional setting and data.

    We validate the measure through order-level behavior and return predictability. Exact-match orders respond faster to quote updates, are canceled sooner and are more common when depth imbalance favors the order direction and bid--ask spreads are narrow. These patterns are consistent with trading based on timely order-book information \citep{yangBackRunningSeekingHiding2020,hirscheyHighFrequencyTradersAnticipate2021,goldsteinHighFrequencyTradingStrategies2023}. Active exact-match orders also predict direction-adjusted returns from milliseconds through ten minutes. Among marketable limit orders, those submitted exactly at the best opposite-side quote predict returns more strongly than orders submitted at more aggressive prices. Price-placement precision thus adds information beyond order direction and aggressiveness. This interpretation is consistent with evidence that fast traders anticipate order flow and price movements \citep{hirscheyHighFrequencyTradersAnticipate2021,ait-sahaliaHowWhenAre2025}.

    We then exploit plausibly exogenous cross-channel variation in order-message dissemination latency at the Shenzhen Stock Exchange (SZSE) market open to study the effects of latency-sensitive HFT activity on market quality. Stocks are assigned to different trading channels, and when opening order flow congests message delivery, dissemination delays differ across channels. We use the fast group's first crossing below three seconds as a common event and compare outcomes after that crossing but before the slow group crosses. Active exact matching and intrasecond price variation increase, while the effects on displayed liquidity and transaction costs are mixed. Faster access to order-level information allows latency-sensitive traders to adjust order placement and order management more quickly. Consistent with this mechanism, active exact-match shares increase after the latency improvement. The accompanying changes in liquidity, price variation, and transaction costs provide evidence on the market-quality consequences of the induced increase in latency-sensitive HFT activity. This within-market variation in information timeliness complements studies of exchange automation, system upgrades, co-location, and other low-latency technologies.

    How speed affects market quality may depend on market design. Where fast traders can recycle inventory throughout the day and continuously revise two-sided quotes, lower latency can support liquidity provision through faster quote adjustment and inventory management. China's A-share market instead restricts the same-day resale of newly purchased shares, lacks maker-taker rebates, and imposes different execution rules and constraints on rapid order submission and cancellation. In this environment, HFT activity tilts toward directional liquidity taking rather than continuous market making, accounting for the clearest activity response in active exact matching and for market-quality effects that are more pronounced for short-horizon price variation and adverse selection than for average spread measures.

    This paper makes four contributions. First, we introduce an order-level quote-matching measure of latency-sensitive trading activity. Whether a trader submits an explicitly priced order at the prevailing best quote captures a specific behavioral manifestation of timely order-book processing, complementing measures based primarily on speed or message intensity. Second, we show that price-placement precision contains predictive information beyond trade direction and order aggressiveness. Among marketable limit orders, exact matches at the best opposite-side quote predict short-horizon returns more strongly than orders priced more aggressively. An order's position relative to the prevailing quote thus provides information about the timeliness and informativeness of the trading decision. This evidence complements studies of informed marketable orders \citep{barardehiDetectingInformedTrading2024} and retail limit orders \citep{anandRetailLimitOrders2026}. Third, we introduce a novel quasi-experimental design to study latency-sensitive HFT activity and its effects on market quality. The design exploits plausibly exogenous differences in order-message dissemination latency across trading channels caused by congestion at the market open. Improved information timeliness increases active exact-match shares, providing treatment-response evidence that the shock affects our HFT proxy. We then examine how the latency improvement affects liquidity, short-horizon price variation, transaction costs, and pricing error. This design complements research using exchange automation, system upgrades, co-location, and other low-latency technologies or proxies \citep{hendershottDoesAlgorithmicTrading2011,conradHighfrequencyQuotingTrading2015,boehmerAlgorithmicTradingMarket2021,hasbrouckLowlatencyTrading2013,boehmerCompetitiveLandscapeHighFrequency2018,malcenieceHighFrequencyTrading2019}. Finally, we contribute to the limited evidence on HFT in emerging markets by documenting latency-sensitive trading in China's retail-dominated centralized limit order market. Our strongest evidence concerns active exact matching and directional liquidity taking, extending the evidence beyond settings in which HFT is closely associated with automated market making. These findings broaden the evidence on how speed and real-time information are used across different market settings.

    This paper is organized as follows. Section~\ref{sec:Data and Methodology} presents the data and measure; Section~\ref{sec:Validation Tests} validates the proxy; Section~\ref{sec:HFT Order Return Predictability} examines return predictability; Section~\ref{sec:High-frequency Trading and Market Quality} studies the latency experiment; and Section~\ref{sec:Conclusion} concludes.

    \section{Data and Methodology}\label{sec:Data and Methodology}

    \subsection{Institutional Background of HFT in China}\label{sec:institutional_background}
    China's A-share market differs sharply from many U.S. and European equity markets in the institutional environment faced by HFTs. Ordinary A-share trading takes place in a centralized, order-driven limit order market without designated market makers or maker-taker rebates. Unlike many developed markets, the SZSE does not offer a standard price-protected IOC limit-order instruction that allows traders to execute only at a specified top-of-book price and automatically cancel any unexecuted remainder. Its immediate-or-cancel-style market orders may instead execute across multiple price levels, making them less suitable for latency-sensitive strategies that seek precise top-of-book execution. In addition, Chinese exchanges impose strict surveillance on program trading, particularly high cancellation ratios and rapid order submissions within short time intervals. These institutional features raise the cost of exploratory order submission, rapid cancellation, and continuous quote revision.

    The T+1 trading rule also affects the relative economics of liquidity provision and liquidity taking. Shares acquired through an executed buy order generally cannot be resold until the following trading day \citep[Articles 3.1.4--3.1.5]{szseTradingRules2023}. This restriction limits the rapid intraday recycling of inventory that supports continuous, low-inventory, two-sided market making. Continuous low-inventory quoting and frequent inventory adjustment therefore become more demanding in terms of inventory commitment and capital. Directional liquidity taking relies less on the same inventory-recycling mechanism, because it does not require repeated offsetting trades on both sides of the book. T+1 therefore constrains strategies that depend on rapid intraday inventory turnover more than directional trading, making continuous liquidity provision relatively less attractive.

    As a result, HFT activity in China is less naturally expressed through market-maker-style liquidity provision.
    In U.S. markets, quote updates account for about 95\% of order-book events, reflecting intensive quote revision and liquidity provision \citep{zhouQueuingUncertaintyLimit2025}.
    In our Shenzhen sample, valid best quotes update 1.327 to 7.053 times per second across size groups during the first hour of trading, and 51.52\%--64.90\% of updates occur within 100 milliseconds of the preceding update (Table~\ref{tab:panel_average_number_of_quote_updates_in_size_quintiles.}). \citet{zhouQueuingUncertaintyLimit2025} reports the composition of order-book events, whereas our statistics measure the frequency of best-quote updates.
    Frequent quote changes make precise contemporaneous price placement economically meaningful and motivate our focus on active exact matching by liquidity takers.

	\subsection {Data}
    We use Level-2 data for China's A-share market from specialized market-data services. The Level-2 data contain three streams.
    Tick data consist of three-second snapshots of the order book state. Each tick data observation includes the top ten bid prices and ask prices, along with exchange timestamps, corresponding volumes, cumulative turnover, and other market statistics. Order data record each submission, including its submission timestamp, quantity, price, and direction (buy or sell). Transaction data record execution time, price, size and direction for each trade, as well as order cancellations.
    Access to these streams differs across investor types.
    % Retail investors typically access market information through trading applications that display order book snapshots and transaction prices derived from tick data, providing them with a relatively static view of market conditions updated every three seconds\footnote{Some public data streams are even slower, for example, Level-1 snapshot data updating every six seconds.}.
    Retail investors typically access market information through standard trading applications, which display order book snapshots updated every three to six seconds depending on the platform. This provides a relatively static and coarse view of market conditions, regardless of the exact refresh interval.
    In contrast, HFT firms possess both the access rights and technical capabilities to reconstruct real-time order books using the complete order data. This enables institutional traders, particularly HFTs, to observe order book dynamics at millisecond frequencies and respond to market changes as they occur.
    Table \ref{tab:panel_examples_of_information_asymmetry_for_ticker_symbol_300750_on_october_10,_2025} illustrates the information asymmetry for ticker symbol 300750 (CATL) on October 10, 2025. Panel A presents Level-2 snapshot data sampled at a three-second frequency, reflecting the latency-constrained information typically available to retail investors via standard trading platforms. In contrast, Panel B displays reconstructed high-frequency quotes, 22 limit-order-book updates within a 50-millisecond window. While we report only the top five levels for conciseness, institutional traders can reconstruct the full depth of the order book. The reconstruction procedure is described in Section~\ref{sec:order_book_reconstruction}.

    Our baseline sample spans July 1, 2023, through October 31, 2025, focusing on securities traded on the Shenzhen Stock Exchange. {We filter the sample by excluding stocks in the bottom 30\% of the market capitalization distribution, measured using prior-day market values. This screening is consistent with common practices in the market microstructure literature, which aim to mitigate microcap-related distortions. For example, \citet{conradHighfrequencyQuotingTrading2015} restrict their sample to stocks with prices above \$1 and market capitalizations exceeding \$100 million, while \citet{hasbrouckLowlatencyTrading2013} focus on the 500 largest stocks by market capitalization. In the context of the Chinese equity market, \citet{liuSizeValueChina2019} document that stocks in the bottom 30\% of the size distribution often embed significant shell value, which may contaminate trading-based analyses. Motivated by these considerations, we exclude the smallest 30\% of stocks from our sample.}
    The reconstructed continuous-trading core contains 47,418,470,233 orders. We reconstruct the limit order book from the complete, unsampled sequence of submissions, executions, and cancellations. For computationally intensive order-level analyses, we use systematic sampling with a nominal overall rate of 0.1\%. Order characteristics for sampled observations use quotes from the fully reconstructed book, so sampling does not alter the quote history used to identify exact matches. The sampling rule does not use order characteristics and retains observations across stocks and trading days. Stock-size and order-type screens are applied separately from sampling. After these screens, including the exclusion of market orders without resampling, the final order-level sample contains 42,506,216 limit orders, 2,760 stocks, and 566 trading days.

    Table~\ref{tab:panel_average_number_of_quote_updates_in_size_quintiles.} measures valid best-quote updates using the full event sequence. Across size quintiles and the two half-hour intervals from 09:30 to 10:30, average update rates range from 1.327 to 7.053 per second, and 51.52\%--64.90\% occur within 100 milliseconds of the preceding valid update. This frequency motivates examining precise quote matching as a signature of timely information processing.

    In the market quality section (Section~\ref{sec:High-frequency Trading and Market Quality}), the difference-in-differences analysis uses an additional proprietary exchange-colocation data feed covering October 10 to November 30, 2025.
    Its coverage partly overlaps the July 2023--October 2025 baseline window and extends into November 2025. The different window reflects the availability of this exchange-provided feed, which we obtain separately from standard Level-2 data. Its local receipt timestamps let us measure channel-level dissemination latency and determine when order data become timely relative to three-second tick updates.
    We restrict the DID analysis to dates covered by this feed. As in the order-level analysis, activity measures use explicitly priced limit orders and exclude opposite-side-best and same-side-best market instructions. The DID analysis in Section~\ref{sec:High-frequency Trading and Market Quality} constructs latency, exact-match activity, and market-quality measures from the complete message flow, subject to their respective eligibility requirements, without applying the order-level sampling procedure.

    \subsection{Methodology of Identification}
    We reconstruct the limit order book from order-level data to recover the best bid and ask available at each instant. We then identify orders submitted at those quotes and use their prevalence to proxy latency-sensitive trading activity. Section \ref{sec:Identification_Proxy} details the economic logic underlying this construction and why exact price matching is more attractive and feasible for latency-sensitive traders than for slower participants.
    We assess the proxy through behavior and return predictability. The behavioral tests examine response times and cancellations, both commonly associated with HFT. The return tests ask whether exact-match orders predict subsequent price movements. If exact price matching captures latency-sensitive information processing, these orders should exhibit stronger short-horizon return predictability relative to alternative pricing strategies and other order-characteristic indicators. The measure captures latency-sensitive trading activity that relies on contemporaneous order-book information.

    \subsubsection{{Order Book Reconstruction}} \label{sec:order_book_reconstruction}
    We focus on stocks traded on the Shenzhen Stock Exchange, which operates as a centralized, order-driven market without dark pools. Using detailed order-level data, we reconstruct the real-time limit order book by sequentially processing order submissions, executions, and cancellations according to the exchange’s trading rules. During the opening call auction (09:15–09:25), orders are collected and matched at the exchange-determined clearing price that maximizes executable volume; we take the auction price directly from the data and reconstruct the full post-auction order book at 09:25, but exclude this period from proxy construction. During continuous trading (09:30–11:30 and 13:00–14:57), limit orders execute immediately if they cross standing quotes, with traded volume removed following price–time priority and any unexecuted remainder added to the book; market orders execute against the best available prices; cancellations remove the remaining order volume. We explicitly account for price limit and price cage constraints, excluding invalid orders and enforcing execution bounds in the reconstruction. From the evolving order book, we continuously compute snapshots of the best ten bid and ask quotes. The snapshot example in Table~\ref{tab:panel_examples_of_information_asymmetry_for_ticker_symbol_300750_on_october_10,_2025} illustrates the difference in temporal resolution.

	\subsubsection{Identification Proxy} \label{sec:Identification_Proxy}
    Our identification process begins with the reconstruction of the real-time order book from Level-2 data by sequentially processing order submissions, executions, and cancellations according to the exchange's trading rules.
    We refer to this reconstructed order book as ``reconstructed quotes'' (\(RQuote\)), which represent the full market state observable only to traders with access to millisecond-level data, such as high-frequency traders (HFTs).
    We use these quotes to distinguish orders placed using different information sets.

    Two features motivate exact matching. Low-latency infrastructure lets traders observe and process quote changes at millisecond frequency.
    As documented in Table~\ref{tab:panel_average_number_of_quote_updates_in_size_quintiles.}, quotes update multiple times per second, and a substantial fraction of updates occur within very short intervals.
    HFT strategies also operate on thin margins. Small deviations from the best available price can eliminate expected profits or impose execution and adverse-selection costs.
    Consequently, exact price matching should be more attractive and more feasible for latency-sensitive traders whose strategies depend on precise contemporaneous order-book information.
    Traders relying on lower-frequency information are less likely to consistently submit orders that coincide with the prevailing best quotes at the millisecond level.

\begin{figure}[htbp]
\centering
\resizebox{\linewidth}{!}{%
\begin{tikzpicture}[
    x=1cm,y=1cm,>=Latex,font=\normalsize,
    timeline/.style={line width=0.8pt,-{Latex[length=2.5mm]}},
    event/.style={circle,draw=black,line width=0.7pt,minimum size=4.5mm,inner sep=0pt},
    hftbox/.style={rectangle,draw=blue!70!black,fill=blue!4,line width=0.8pt,minimum height=1.3cm,minimum width=2.6cm,text width=2.3cm,align=center},
    nonhftbox/.style={rectangle,draw=red!70!black,fill=red!4,line width=0.8pt,minimum height=1.4cm,minimum width=3.3cm,text width=3.0cm,align=center},
    hftarrow/.style={-{Latex[length=2mm]},draw=blue!70!black,line width=0.8pt},
    nonhftarrow/.style={-{Latex[length=2mm]},draw=red!70!black,line width=0.8pt},
    guideblue/.style={dashed,draw=blue!70!black,line width=0.8pt,-{Latex[length=2mm]}},
    guidered/.style={dashed,draw=red!70!black,line width=0.8pt,-{Latex[length=2mm]}}
]
\node[align=center,font=\bfseries] at (-0.55,6.15) {Order flow};
\draw[line width=0.8pt] (0.6,6.15)--(11.7,6.15);
\draw[timeline] (13.3,6.15)--(17.6,6.15);
\node[right] at (17.65,6.15) {Time};
\node[event,fill=blue!75] (t1) at (2.0,6.15) {};
\node[above=2mm of t1,font=\large] {$t_1$};
\node[below=4mm of t1,align=center,fill=white] {New quote\\\textbf{best price} $=P_1$};
\node[event,fill=red!85] (t2) at (9.0,6.15) {};
\node[above=2mm of t2,font=\large] {$t_2$};
\node[below=4mm of t2,align=center,fill=white] {Low-latency trader's limit order\\\ifdefined\slidesdiagramonly\textcolor{red}{\textbf{exact match}}\else\textbf{exact match}\fi\ $P_1$};
\node at (12.5,6.15) {$\cdots$};
\node[event,fill=gray!45] (tk) at (16.0,6.15) {};
\node[above=2mm of tk,font=\large] {$t_k$};
\node[below=4mm of tk,align=center,fill=white] {\textbf{Best price} $=P_2$\\Higher-latency trader's limit order $P_1\neq P_2$\ifdefined\slidesdiagramonly\\\textcolor{red}{\textbf{not-exact match}}\fi};
\node[align=center,text=blue!70!black,font=\bfseries\small] at (-0.55,3.45) {Low-latency\\trader};
\node[hftbox] (hftreceive) at (2.0,3.45) {Receive\\$P_1$};
\node[hftbox] (hftprocess) at (5.5,3.45) {Process\\signal};
\node[hftbox] (hftsubmit) at (9.0,3.45) {Submit\\at $P_1$};
\draw[hftarrow] (hftreceive.east)--(hftprocess.west);
\draw[hftarrow] (hftprocess.east)--(hftsubmit.west);
\draw[blue!70!black,<->,line width=0.7pt]
    ($(hftreceive.south)+(0,-0.35)$)--($(hftsubmit.south)+(0,-0.35)$);
\node[text=blue!70!black,fill=white,inner sep=2pt] at (5.5,1.95) {\large Microseconds};
\node[align=center,text=red!70!black,font=\bfseries\small] at (-0.55,0.75) {Higher-latency\\trader};
\node[nonhftbox] (nonreceive) at (3.0,0.75) {Receive $P_1$\\with high latency};
\node[nonhftbox] (nonprocess) at (9.5,0.75) {Process\\signal};
\node[nonhftbox] (nonsubmit) at (16.0,0.75) {Submit\\at $P_1$};
\draw[nonhftarrow] (nonreceive.east)--(nonprocess.west);
\draw[nonhftarrow] (nonprocess.east)--(nonsubmit.west);
\begin{pgfonlayer}{background}
\draw[guideblue] (t1.south)--(hftreceive.north);
\draw[guideblue] (hftsubmit.north)--(t2.south);
% Connect delayed receipt directly to the Non-HFT receive box.
\draw[guidered] (t1.south)--(nonreceive.north);
\draw[guidered] (nonsubmit.north)--(tk.south);
\end{pgfonlayer}
\draw[red!70!black,<->,line width=0.7pt]
    ($(nonreceive.south)+(0,-0.35)$)--($(nonsubmit.south)+(0,-0.35)$);
\node[text=red!70!black,fill=white,inner sep=2pt] at (9.5,-0.8) {\large Tens to hundreds of milliseconds};
\end{tikzpicture}%
}
\ifdefined\slidesdiagramonly
\else
\caption{Information Processing and Exact-Match Limit Orders.}
\label{fig:exact_match_mechanism}
\begin{minipage}{\textwidth}\small
This figure illustrates how differences in information-processing speed affect whether limit orders match the contemporaneous best quote. Following new order-book information at $t_1$, a low-latency trader receives and processes the quote quickly enough to submit a limit order at the still-prevailing best price $P_1$ at $t_2$. A higher-latency trader instead submits at the stale price $P_1$ after the best quote has changed to $P_2$. The diagram illustrates why timely access to order-book information makes exact matching more likely. The reference price may correspond to the opposite-side quote for active matching or the same-side quote for passive matching.
\end{minipage}
\fi
\end{figure}
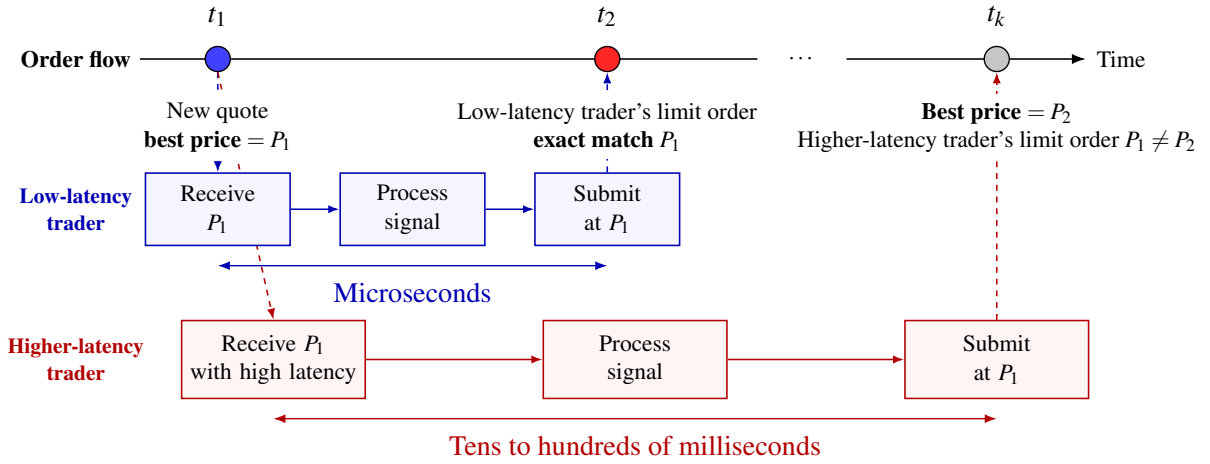

    We therefore compare submitted prices with the contemporaneous best quotes.
    Figure~\ref{fig:exact_match_mechanism} illustrates the mechanism: a low-latency trader can submit a limit order while the observed best quote is still current, whereas a delayed response may reach the market after that quote has changed.
    $$[\text{Insert Figure~\ref{fig:exact_match_mechanism} here.}]$$
    We construct the following indicator variables based on millisecond-level reconstructed quotes (high-frequency information):

    \begin{align}
    \mathit{RQuoteMatchActive} &= \mathbf{1}\{\text{order price} = \text{best price on the opposite side of reconstructed quotes}\},\\
    \mathit{RQuoteMatchPassive} &= \mathbf{1}\{\text{order price} = \text{best price on the same side of reconstructed quotes}\}.
    \end{align}
    % For comparison, we also define analogous indicators using lower-frequency, exchange-published three-second tick data:
    % \begin{align}
    % &\mathit{TickQuoteMatchActive} \notag\\&= \mathbf{1}\{\text{order price} = \text{best price on the opposite side of the most recent tick snapshot}\},\\
    % &\mathit{TickQuoteMatchPassive} \notag\\&= \mathbf{1}\{\text{order price} = \text{best price on the same side of the most recent tick snapshot}\},
    % \end{align}
    Here ``order price'' means the submitted price of a limit order. The indicator $\mathbf{1}\{\cdot\}$ equals one when the condition holds.

    We treat active quote matching as our primary HFT proxy. An active exact-match limit order is submitted at the best opposite-side quote and is marketable at the current top-of-book price.
    We also define passive quote matching as an auxiliary measure. A passive exact-match order is submitted at the best same-side quote. We treat passive matching as an auxiliary measure because posting at the same-side best quote does not require immediate execution and can reflect a broader set of liquidity-provision decisions. These indicators measure order-placement behavior consistent with latency-sensitive trading, not trader identities. Their observed shares, individually or combined, therefore do not measure HFT's share of market activity.

    The regressions include the following control variables. $BuyOrder$ is an indicator for buy orders. $OneCentLarger$ and $MultiCentLarger$ capture orders priced one cent and more than one cent, respectively, more aggressively than the best opposite quote. $NonRoundVol$ equals one if the submitted order size is not a multiple of 500 shares. \citet{liTrackingRetailInstitutional2023} document a greater tendency for retail investors in China to trade in multiples of 500 shares. $NonRoundPrice$ equals one if the submitted limit price is not an integer-yuan price. Prior studies document stronger round-number price clustering among individual investors \citep{chiaoPriceClustering2009,kuoCognitiveLimitationInvestment2015}. These variables capture order discreteness associated in prior work with retail trading behavior. Finally, $LargeOrder$ equals one if the submitted order notional exceeds RMB 200,000.

    $$[\text{Insert Table \ref{tab:panel_descriptive_statistics} here.}]$$

    Table~\ref{tab:panel_descriptive_statistics} presents descriptive statistics for our main variables. Panel A reports the mean values of indicator variables, which represent the percentage of orders for which each condition equals one. Approximately 30\% of orders match the opposite best price in reconstructed quotes ($RQuoteMatchActive$), while more than 30\% match the same-side best price ($RQuoteMatchPassive$).
    % In comparison, less than 30\% of orders match the best price in three-second tick data.
    Panel B provides an initial descriptive comparison. Active exact-match orders have higher subsequent direction-adjusted returns than non-matching orders. Section~\ref{sec:HFT Order Return Predictability} formally examines this relation using multivariate regressions. Panel C reports order types before market orders are excluded. Of the original 42,644,479 selected orders, 138,263 (0.3242\%) are market orders: 132,422 opposite-best and 5,841 same-side-best instructions. These instructions select a reference quote without requiring a numerical limit price, so they are excluded from the exact-match measures and main order-level regressions. The retained sample contains 42,506,216 limit orders.

	\section{Validation Tests} \label{sec:Validation Tests}
	We assess the behavioral validity of the exact-match proxy using order response times, cancellations, and associations with order imbalance and quoted spreads.
    These tests examine whether exact-match orders exhibit characteristics consistent with established HFT trading patterns documented in the literature.
    \subsection{Response Time}
    HFT orders should respond quickly to market updates. We test whether exact-match orders share this behavior. We define \textit{response time} as the elapsed time from the latest prior change in best-bid or best-ask price or quantity to order submission.
    Slower traders may act on earlier quotes, so measuring response from the latest update understates their response times and makes the comparison conservative. Exact-match orders nevertheless have significantly shorter measured response times.

    Table~\ref{tab:panel_descriptive_statistics_of_order_response_time} reports response-time statistics for all sampled limit orders. Panel A shows median response times declining from 0.11 seconds in the smallest size quintile to 0.05 seconds in the largest. Panel B1 reports medians of 0.04 seconds for both active and passive matches in the smallest quintile, and 0.02 and 0.03 seconds, respectively, in the largest. The corresponding non-matching medians are 0.15--0.16 seconds in the smallest quintile and 0.06 seconds in the largest. Panel B2 shows that the smallest-quintile medians for $OneCentLarger$ and $MultiCentLarger$ are 0.12 and 0.18 seconds, compared with 0.06 and 0.08 seconds in the largest quintile. Other order characteristics also correlate with response time, but precise quote matching is associated with especially short response intervals.

    % We also observe that orders matching three-second tick-level quotes exhibit relatively short response times in the descriptive statistics; however, this pattern does not persist once we control for firm and time fixed effects in the regression analysis. In contrast, orders classified as $RQuoteMatchPassive$ consistently display significantly shorter response times.
    % We also observe that orders classified as $RQuoteMatchPassive$ consistently display significantly shorter response times.
    % A plausible interpretation is that these orders reflect {race failures} in high-frequency trading. When multiple HFTs simultaneously attempt to execute against the best opposite-side quote, only the fastest order is successfully filled. Orders arriving marginally later fail to access the opposite-side liquidity and instead rest or execute at the same-side best price. Consequently, such orders appear passive ex post, despite originating from aggressive trading intentions. This mechanism is consistent with the speed-sensitive nature of HFT competition and further supports our identification strategy.

    $$[\text{Insert Table \ref{tab:order_response_time} here.}]$$

    Table~\ref{tab:order_response_time} reports panel regression results examining the relationship between order response time and our identification proxies.
    \begin{equation}
    \label{eq:response_time_reg}
    \begin{aligned}
    ResponseIn100ms_{i,t}
    = \beta_1 RQuoteMatchActive_{i,t}
    + \beta_2 RQuoteMatchPassive_{i,t}
    + \boldsymbol{\gamma}' \mathbf{X}_{i,t}
    + \alpha_i
    + \delta_t
    + \varepsilon_{i,t},
    \end{aligned}
    \end{equation}
    Equation~(\ref{eq:response_time_reg}) relates short-horizon order response to the informational content of contemporaneous quotes. The dependent variable, $ResponseIn100ms_{i,t}$, is an indicator equal to one if order $t$ for stock $i$ is submitted less than 100 milliseconds after the latest prior best-quote price or quantity change. The variables $RQuoteMatchActive_{i,t}$ and $RQuoteMatchPassive_{i,t}$ indicate whether the order price matches the best available price on the opposite or same side of the reconstructed order book at submission, respectively. The control vector $\mathbf{X}_{i,t}$ includes order direction and lagged stock-level characteristics. All specifications include stock fixed effects $\alpha_i$ and date fixed effects $\delta_t$, which absorb time-invariant stock heterogeneity and common date-level conditions. Standard errors are clustered at the stock level. In the fully controlled specification of Table~\ref{tab:order_response_time}, active and passive quote matching have the two largest positive coefficients among the included order-characteristic indicators. They are associated with 19.2 and 17.7 percentage-point higher probabilities of responding within 100 milliseconds, respectively.

	\subsection{Cancellation Time}
    Rapid cancellation allows HFTs to update quotes and manage adverse-selection risk. \citet{hasbrouckLowlatencyTrading2013} and \citet{aquilinaQuantifyingHighFrequencyTrading2021} document that HFTs frequently cancel orders within extremely short horizons to avoid being executed against toxic order flow as market conditions evolve.
    We test whether exact-match orders have shorter lifespans conditional on cancellation. Fully executed orders are excluded so that the measured interval reflects cancellation decisions and monitoring speed.

    Table~\ref{tab:panel_descriptive_statistics_of_order_cancellation_time} describes canceled limit orders. Median cancellation times range from 35.55 to 45.10 seconds across size quintiles. In the largest quintile, active exact-match orders have a median of 3.01 seconds versus 36.04 seconds for other orders; passive matches have a median of 15.01 seconds versus 82.34 seconds for non-passive matches. The medians for $OneCentLarger$, $MultiCentLarger$, $NonRoundVol$, and $NonRoundPrice$ are 6.56, 10.36, 26.76, and 32.95 seconds, respectively.
    The difference is economically pronounced in cancellation durations: in the largest size quintile, the median for active exact-match orders is less than one-tenth that for non-active orders.

    \begin{equation}
    \label{eq:cancel_time_reg}
    \begin{aligned}
    CancelIn100ms_{i,t}
    = \beta_1 RQuoteMatchActive_{i,t}
    + \beta_2 RQuoteMatchPassive_{i,t}
    + \boldsymbol{\gamma}' \mathbf{X}_{i,t}
    + \alpha_i
    + \delta_t
    + \varepsilon_{i,t},
    \end{aligned}
    \end{equation}

    Equation~\eqref{eq:cancel_time_reg} examines whether reconstructed-quote matching predicts rapid order cancellation. The dependent variable, $CancelIn100ms_{i,t}$, is an indicator equal to one if order $t$ for stock $i$ is canceled within 100 milliseconds of submission. The variables $RQuoteMatchActive_{i,t}$ and $RQuoteMatchPassive_{i,t}$ capture active and passive matches to the reconstructed order book, respectively. The control vector $\mathbf{X}_{i,t}$ includes order-level characteristics and prevailing market conditions. All specifications include stock fixed effects $\alpha_i$ and date fixed effects $\delta_t$, which absorb time-invariant stock heterogeneity and common date-level shocks. Standard errors are clustered at the stock level.

    $$[\text{Insert Table \ref{tab:order_cancellation_time} here.}]$$

    Table~\ref{tab:order_cancellation_time} reports three specifications for cancellation within 100 milliseconds, conditional on cancellation. Active quote matching is strongly associated with rapid cancellation. In the fully controlled specification, its coefficient is 0.05406, corresponding to a 5.4 percentage-point higher probability of cancellation within 100 milliseconds among canceled orders. This coefficient exceeds those for passive matching and most other order-characteristic indicators, although $OneCentLarger$ has a larger coefficient of 0.06500.

    \subsection{Order Imbalance and Bid-Ask Spread}

    We ask whether exact-match submissions coincide with the conditions in which HFTs tend to trade: pronounced directional depth imbalance and narrow bid--ask spreads, which limit transaction costs \citep{goldsteinHighFrequencyTradingStrategies2023}.

    Order imbalance ($OrderImbalance$) is defined as the signed relative depth imbalance at the best quotes prior to the order submission:
    \begin{equation}
    OrderImbalance_{i,t}
    = side_{i,t} \times \frac{Bid1Depth_{i,t} - Ask1Depth_{i,t}}{Bid1Depth_{i,t} + Ask1Depth_{i,t}},
    \end{equation}
    where $side_{i,t}$ equals $+1$ for buy orders and $-1$ for sell orders. This measure captures whether available depth favors the direction of the submitted order.

    In the order-level regressions, $Spread$ is the relative quoted spread expressed in percent:
    \begin{equation}
    Spread_{i,t}
    = 100\times\frac{Ask1_{i,t} - Bid1_{i,t}}{Mid_{i,t}},
    \end{equation}
    where $Mid_{i,t} = (Ask1_{i,t} + Bid1_{i,t})/2$.

    To examine the market conditions associated with exact-match activity, we estimate the following panel regression:
    \begin{equation}
    \label{eq:hft_microstructure_reg}
    \begin{aligned}
    Y_{i,t}
    = \beta_1 OrderImbalance_{i,t}
    + \beta_2 Spread_{i,t}
    + \boldsymbol{\gamma}' \mathbf{X}_{i,t}
    + \alpha_i
    + \delta_t
    + \varepsilon_{i,t},
    \end{aligned}
    \end{equation}
    where the dependent variable $Y_{i,t}$ is alternatively defined as an indicator for active quote matching ($RQuoteMatchActive$), passive quote matching ($RQuoteMatchPassive$), or other order characteristics related to price aggressiveness, order size, and discreteness. The vector $\mathbf{X}_{i,t}$ includes order direction, lagged market capitalization, lagged turnover, and lagged inverse price. All specifications include stock fixed effects $\alpha_i$ and date fixed effects $\delta_t$. Standard errors are clustered at the stock level.

    $$[\text{Insert Table \ref{tab:order_imbalance_and_spread_channel} here.}]$$

    Table~\ref{tab:order_imbalance_and_spread_channel} reports the estimation results. Across the included order-characteristic indicators, $RQuoteMatchActive$ has the largest positive loading on directional order imbalance (0.09831) and the most negative loading on spreads ($-0.59515$). Active exact-match orders are therefore disproportionately submitted when displayed depth favors the order direction and quoted trading costs are low.

    In contrast, $RQuoteMatchPassive$ loads negatively on order imbalance and displays a smaller negative loading on spreads, consistent with its auxiliary interpretation as top-of-book competition. The alternative order characteristics in Columns~(3)--(7) also have smaller absolute loadings on both variables than active matching.

    Taken together, the rapid response, rapid order management, and strong sensitivity to favorable microstructure conditions provide complementary behavioral validation of active exact matching as a proxy for latency-sensitive order placement. Section~\ref{sec:HFT Order Return Predictability} examines its return-predictive content.

    \section{Return Predictability of Exact-Match Orders}  \label{sec:HFT Order Return Predictability}

    \subsection{Short-Term Return Predictability}
    To evaluate the informational content of the proxy, we examine whether active exact-match orders predict short-horizon price movements more strongly than comparable orders with different price-placement strategies.
	If exact price matching captures latency-sensitive information processing, these orders should be followed by more favorable short-horizon price movements after controlling for mechanical execution effects and observable order characteristics.

	To reduce the influence of the order's immediate price impact, we measure returns from the first valid reconstructed midpoint following order submission instead of the pre-submission midpoint.  The terminal price is the last valid midpoint at or before the exact submission timestamp plus the specified horizon.

\begin{figure}[H]
    \centering
    \includegraphics[width=\linewidth]{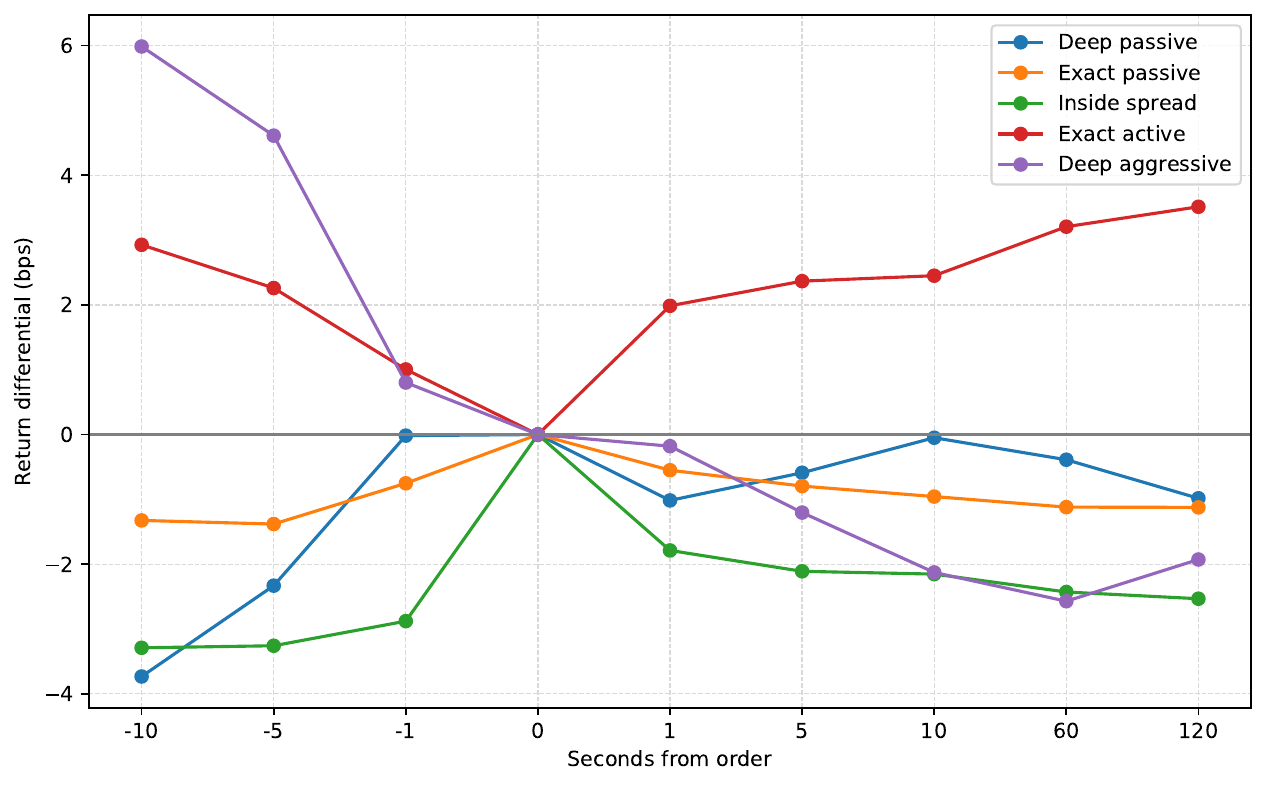}
    \caption{Order Aggressiveness and Return.}
    \label{fig:return_aggressiveness}
    \begin{minipage}{\textwidth}
    \small
    This figure compares short-horizon return patterns across limit-order aggressiveness categories. The sample includes all common stocks listed on the Shenzhen Stock Exchange, excluding the smallest 30\% by previous-day market capitalization, over the period July 1, 2023 to October 31, 2025. Orders are classified according to their limit prices relative to the prevailing best bid and best ask. Deep passive orders are buy (sell) orders priced below the best bid (above the best ask), while deep aggressive orders are buy (sell) orders priced above the best ask (below the best bid). At-the-quote passive (active) orders exactly match the best same-side (opposite-side) quote, and inside-the-spread orders are priced strictly between the best bid and best ask. For each second relative to order arrival, the figure reports the difference between the mean direction-adjusted mid-quote return for each category and the corresponding mean for all other orders. The horizontal axis measures seconds relative to order arrival, and the vertical axis reports category-minus-complement return differentials in basis points. Post-submission returns are measured from the next valid quote after submission to the quote at the specified horizon, excluding observations crossing lunch breaks or the market close. Negative horizons use the corresponding historical quote convention to report pre-submission return differences.
    \end{minipage}
\end{figure}

	We adjust returns for order direction:

	\begin{align}
		Ret_{i,t+\Delta t} = \left(\frac{Mid_{i,t+\Delta t}}{Mid_{i,t}} - 1\right) \times side_{i,t} \times 10000,
	\end{align}

	where $Mid_{i,t+\Delta t}$ represents the mid-price for stock $i$ at time $t+\Delta t$ after order submission, $Mid_{i,t}$ represents the first valid reconstructed midpoint following order submission, and $side$ equals $+1$ for buy orders and $-1$ for sell orders. Returns are expressed in basis points.
	Positive returns indicate price movements in the order direction: increases after buys and decreases after sells. Negative returns indicate adverse movements. Direction adjustment makes buy and sell orders comparable, so higher average returns imply stronger predictive ability.

    $$[\text{Insert Figure~\ref{fig:return_aggressiveness} here.}]$$

    Figure \ref{fig:return_aggressiveness} plots each order category\textquotesingle s direction-adjusted mid-quote return differential relative to its complement group. Positive values mean higher average returns than all other orders, not positive absolute returns.
    Active exact-match orders (\(RQuoteMatchActive\)) have positive post-submission return differentials relative to other orders, consistent with timely directional information. In contrast, orders with less aggressive pricing strategies generate markedly weaker and less persistent return responses, suggesting lower informational content. These categories include deep passive orders placed behind the best same-side quote, passive exact matches at that quote, and orders placed inside the spread. They do not seek immediate execution against the prevailing opposite-side quote and have lower post-submission return differentials than active exact matches. Deep aggressive orders cross beyond the best opposite-side quote, accepting greater price concessions to obtain execution. Their weaker post-submission differentials indicate that greater aggressiveness alone does not imply stronger directional information.
    These findings are consistent with \citet{brogaardHighFrequencyTradingPrice2014}, who document a strong correlation between high-frequency trading activity and returns in the seconds surrounding order submission. In particular, they show that this correlation peaks within one second and remains economically meaningful for up to ten seconds.

    We estimate the order-level regression corresponding to Figure \ref{fig:return_aggressiveness}:
    \begin{equation}
    \begin{aligned}
    Ret10s_{i,t} =\;& \beta_1 RQuoteMatchActive_{i,t}
    + \beta_2 RQuoteMatchPassive_{i,t} \\
    &+ \boldsymbol{\gamma}' \mathbf{X}_{i,t}
    + \alpha_i + \delta_t + \varepsilon_{i,t},
    \end{aligned}
    \end{equation}
    where the dependent variable \(Ret10s_{i,t}\) is the direction-adjusted mid-quote return measured from the first mid-quote observation following order submission to the mid-quote ten seconds after submission for order \(t\) and stock \(i\). The variables \(RQuoteMatchActive_{i,t}\) and \(RQuoteMatchPassive_{i,t}\) indicate whether the order price exactly matches the best prevailing price on the opposite or same side of the millisecond-level reconstructed order book, respectively. The vector \(\mathbf{X}_{i,t}\) includes order-level characteristics and lagged stock controls. All specifications include stock fixed effects \(\alpha_i\) and date fixed effects \(\delta_t\), absorbing time-invariant stock characteristics and common date-level shocks. Standard errors are clustered at the stock level.

    $$[\text{Insert Table \ref{tab:matched_orders_predict_future_price} here.}]$$

    Table \ref{tab:matched_orders_predict_future_price} reports the short-horizon return regressions with the matching proxies and controls.
    The coefficient on \(RQuoteMatchActive\) is positive and highly significant in every specification. In Column (1), the active coefficient is 2.732 basis points ($t=67.70$); in the fully controlled Column (3), it is 2.157 basis points ($t=39.20$). The association remains economically large after conditioning on aggressiveness and other order characteristics. Prices tend to move in the direction of exact-match limit orders.

    To assess the economic magnitude of these returns, we average the ten-second direction-adjusted mid-quote returns of active exact-match orders within each stock-day and then equal-weight across stocks. The resulting average daily return is 4.125 basis points. Motivated by China's T+1 trading restriction, we consider one representative trading opportunity per stock per day. Compounding the daily returns gives an average monthly return of 0.84\% and an annualized return of 10.95\%. These figures illustrate the magnitude of the predictive signal rather than realizable trading profits. They do not account for bid-ask spreads, commissions, transaction taxes, short-selling constraints, or the restrictions on execution imposed by T+1 trading. Investors with existing shares available for sale can sell those holdings and repurchase them later the same day, even though newly purchased shares generally cannot be resold until the next trading day. Short-horizon predictability can therefore inform execution timing and the management of inventory and adverse-selection risk under T+1.

    The coefficient on \(RQuoteMatchPassive\) is positive and significant in Column (1), then becomes negative and significant with the additional aggressiveness controls in Columns (2) and (3). The initial positive association thus depends on correlated price-placement features. Passive matches rest at the best same-side quote and do not immediately trade against the opposite side.
    This pattern is consistent with the auxiliary interpretation of passive matching as top-of-book competition.

    More aggressive prices have weaker predictive content after controlling for exact matching. Both \(OneCentLarger\) and \(MultiCentLarger\) enter with negative and highly significant coefficients in Columns (2) and (3). The positive coefficient for exact matching therefore cannot be explained by a general relation between aggressive prices and stronger information.

    The controls $NonRoundVol$ and $LargeOrder$ have positive significant coefficients in the fully controlled model. $NonRoundPrice$ is positive but insignificant (0.031 basis points, $t=0.62$). Their coefficients are smaller than that of \(RQuoteMatchActive\), indicating that these alternative characteristics do not account for the predictive association of active exact matching.

    \subsection{Predictability Across Horizons}
    We extend the return horizon from 100 milliseconds to ten minutes using a common sample of 39,509,310 limit orders across 2,760 stocks and 566 trading days. Each regression includes the full control set in Table~\ref{tab:matched_orders_predict_future_price}, stock and date fixed effects, and stock-clustered standard errors. Returns use the same precise-time definition as the short-horizon regressions: the starting mid is the first valid quote after submission, and the endpoint is the last valid quote at or before the exact submission timestamp plus the horizon.

\begin{figure}[H]\centering
\includegraphics[width=\linewidth]{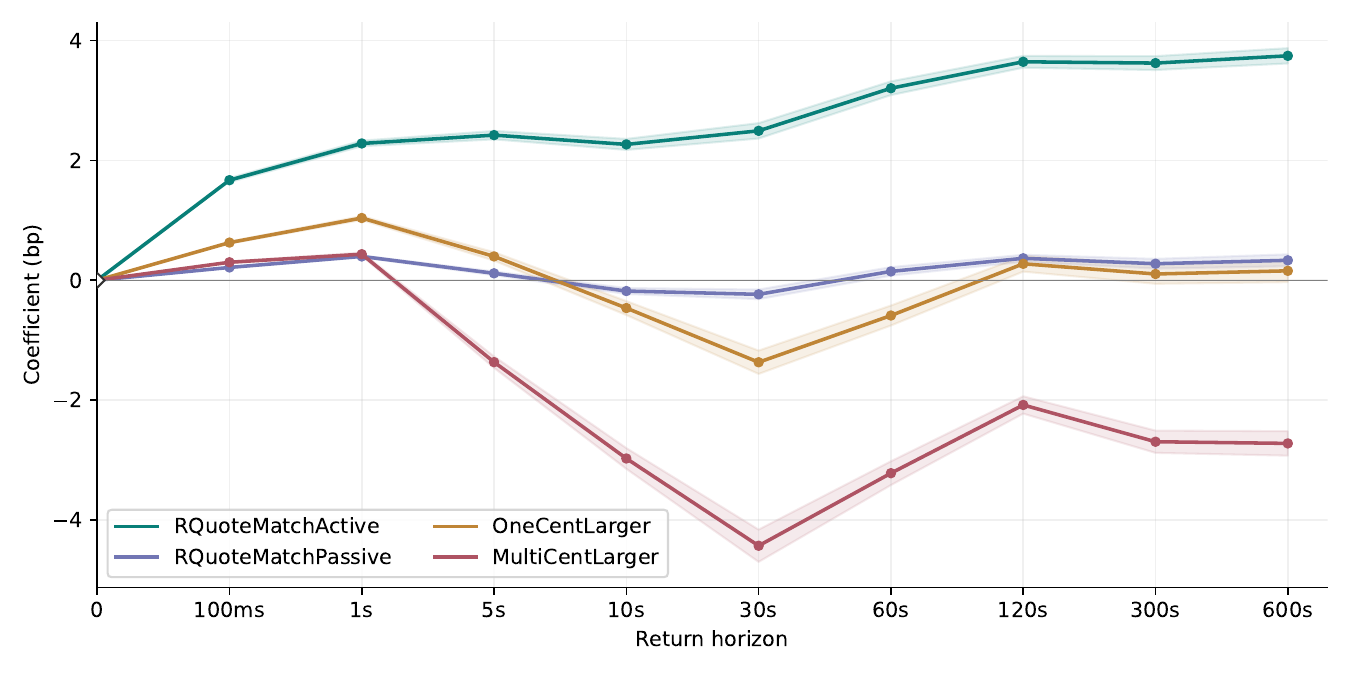}
\caption{Return Predictability from 100 Milliseconds to Ten Minutes.}\label{fig:order_horizons}
\begin{minipage}{\textwidth}\small
This figure reports the return predictability of exact-match orders across horizons from 100 milliseconds to ten minutes. The sample includes all common stocks listed on the Shenzhen Stock Exchange, excluding the smallest 30\% of stocks by previous-day market capitalization, over the period July 1, 2023 to October 31, 2025. The analysis uses 39,509,310 limit orders submitted during continuous trading hours after excluding market orders. The dependent variable is the direction-adjusted mid-quote return measured from the first valid midpoint after order submission to the last valid midpoint at or before the specified horizon. RQuoteMatchActive (RQuoteMatchPassive) indicates whether an order is submitted at the best opposite-side (same-side) quote reconstructed from millisecond-level order-book data. All regressions include full controls, stock and date fixed effects, and stock-clustered standard errors. Coefficients are reported in basis points, and bars represent 90\% pointwise confidence intervals. Horizontal positions represent discrete horizon categories rather than proportional elapsed time.
\end{minipage}
\end{figure}

    Figure~\ref{fig:order_horizons} and Table~\ref{tab:horizons} show that the active exact-match coefficient is positive at all nine horizons: 1.671 bp at 100ms, 2.268 bp at 10s and 3.747 bp at 600s. The persistence of predictive content through ten minutes suggests that active exact matching captures information about price movements over longer intraday horizons, beyond the immediate response to an order. This interpretation accords with \citet{brogaardHighFrequencyTradingPrice2014}, who find that HFTs trade in the direction of permanent price changes, and with \citet{hirscheyHighFrequencyTradersAnticipate2021}, who shows that HFTs use patterns in past trades and orders to anticipate other investors' order flow. Our results provide complementary evidence that latency-sensitive price placement predicts subsequent price movements over horizons extending well beyond milliseconds.

    $$[\text{Insert Figure~\ref{fig:order_horizons} and Table~\ref{tab:horizons} here.}]$$

    Internet Appendix Section~\ref{app:buy_sell} reports separate regressions for buy and sell orders. Active exact matching predicts returns in both directions, with stronger buy-side predictability at 100 milliseconds and stronger sell-side predictability at several longer horizons. This asymmetry may reflect China's T+1 trading restriction and short-selling constraints, which limit investors' ability to trade on negative information and may delay its incorporation into prices relative to positive information.

    \subsection{Return Predictability of Exact-Match Orders Conditional on Order Quantity}

    We next ask whether order quantity relative to displayed depth adds predictive content. High-frequency traders monitor quote updates at millisecond frequencies, which involve adjustments to both prices and quantities. To limit adverse-selection risk and inventory exposure, HFTs have incentives to submit aggressive orders whose sizes do not exceed the available depth at the best opposite quote.

    To test this implication, we construct an indicator for quantity matching and its interaction with active price matching:
    \begin{align}
    RQuoteVolLess_{i,t}
    &= \mathbf{1}\{Vol_{i,t} < BestOppVol_{i,t}\}, \\
    RQuoteMatchActiveVolLess_{i,t}
    &= RQuoteMatchActive_{i,t} \times RQuoteVolLess_{i,t}, \\
    RQuoteVolMatch_{i,t}
    &= \mathbf{1}\{Vol_{i,t} = BestOppVol_{i,t}\}, \\
    RQuoteMatchActiveVolMatch_{i,t}
    &= RQuoteMatchActive_{i,t} \times RQuoteVolMatch_{i,t}, \\
    RQuoteVolMore_{i,t}
    &= \mathbf{1}\{Vol_{i,t} > BestOppVol_{i,t}\}, \\
    RQuoteMatchActiveVolMore_{i,t}
    &= RQuoteMatchActive_{i,t} \times RQuoteVolMore_{i,t},
    \end{align}
    where $Vol_{i,t}$ denotes the submitted order size and $BestOppVol_{i,t}$ is the displayed depth at the best opposite quote immediately prior to order submission.

    We estimate the following panel regression:
    \begin{equation}
    \label{eq:price_quantity_match_reg}
    \begin{aligned}
    Ret10s_{i,t}
    &= \beta_1 RQuoteMatchActiveVolMatch_{i,t}\\
    &+ \beta_2 RQuoteMatchActiveVolLess_{i,t}\\
    &+ \beta_3 RQuoteMatchActiveVolMore_{i,t}\\
    &+ \beta_4 RQuoteMatchPassive_{i,t}
    + \boldsymbol{\gamma}' \mathbf{X}_{i,t}
    + \alpha_i
    + \delta_t
    + \varepsilon_{i,t},
    \end{aligned}
    \end{equation}
    where $Ret10s_{i,t}$ is the mid-quote return over the 10-second window following order submission. The vector $\mathbf{X}_{i,t}$ includes order direction and lagged stock-level controls. All specifications include stock fixed effects $\alpha_i$ and date fixed effects $\delta_t$, and standard errors are clustered at the stock level.

    \medskip
    $$[\text{Insert Table \ref{tab:quantity_matched_orders_predict_future_price} here.}]$$

    Table~\ref{tab:quantity_matched_orders_predict_future_price} separates orders by quantity relative to the best opposite depth. Orders that match or fall below that depth ($RQuoteMatchActiveVolMatch_{i,t}$ and $RQuoteMatchActiveVolLess_{i,t}$) have strong positive return predictability; orders exceeding it ($RQuoteMatchActiveVolMore_{i,t}$) are associated with negative subsequent returns.
    In the fully controlled specification, the coefficients are 2.711 basis points for orders below the displayed opposite volume, 1.818 for equal-volume orders, and $-4.408$ for orders exceeding it.
    Because HFT strategies typically emphasize inventory and execution risk control, such traders are less likely to submit orders larger than the available depth at the best opposite quote.
    Return predictability is strongest for orders below displayed depth, remains positive for exact quantity matches, and reverses for orders exceeding displayed depth.

    Internet Appendix Sections~\ref{app:intraday} and \ref{app:regimes} report intraday patterns and regime dependence in exact-match activity and return predictability. Internet Appendix Section~\ref{app:cross_section} examines the cross-sectional determinants and industry patterns of exact-match activity (Table~\ref{tab:fmb_hft_market_quality} and Figure~\ref{fig:hft_industry_market_share}). Internet Appendix Section~\ref{app:robustness} presents robustness tests using Fama--MacBeth regressions, alternative return horizons, lower-frequency snapshot information, and response-speed controls (Tables~\ref{tab:FMB_MatchedOrders_Ret10s} and \ref{tab:matched_orders_predict_future_price_robustness}). The active exact-match coefficient remains positive across these robustness specifications.

    \section{Latency, Exact-Match Activity, and Market Quality}
    \label{sec:High-frequency Trading and Market Quality}

    Having identified and validated active exact matching as a proxy for latency-sensitive HFT activity, we exploit plausibly exogenous cross-channel latency variation to generate exogenous variation in this activity and study its consequences for market quality. At the market open, stocks assigned to different trading channels face different dissemination delays. Faster order-message delivery allows traders to act on more current order-book information. We first describe the latency-crossing design and examine whether the latency improvement increases active exact-match activity. We then estimate the corresponding responses of liquidity, short-horizon price variation, transaction costs, and pricing error.

    \subsection{Research Design}\label{sec:did_research_design}

    The SZSE assigns each stock to one of ten trading channels for the duration of a trading day. Assignments are set by the exchange, fixed within a day, and may vary across days.\footnote{The SZSE Binary market-data specification identifies tick-by-tick order and execution messages by MsgType 30xx92 and 30xx91, respectively. These messages carry a channel identifier (ChannelNo) and a message sequence number (ApplSeqNum); see \citet{szseBinarySpec}. Differences in message traffic across channels can generate different delivery queues during the opening surge. In our sample, channels cross the same latency threshold at different times.} We apply k-means to the channels' per-second median latency paths from 10 to 60 seconds after the open, forming two groups. The group that first crosses below three seconds is designated fast, and the other slow. Throughout the DID analysis, treatment refers to stocks in the fast group and control to stocks in the slow group.

    We measure latency in the order-message feed. At the market open, surges in order submissions can congest data delivery and increase dissemination latency. The volume of continuously streamed order messages can delay delivery by more than ten seconds during peak traffic. Tick data, by contrast, are compact and refreshed at fixed three-second intervals, and are largely unaffected by this congestion. Because HFT strategies depend on order-level data while retail traders rely mainly on tick-level price updates, dissemination delays affect the two groups asymmetrically. Stocks in a channel share the same dissemination infrastructure. We use group-median latency to define the event.

\begin{figure}[H]
    \centering
    \includegraphics[width=\linewidth]{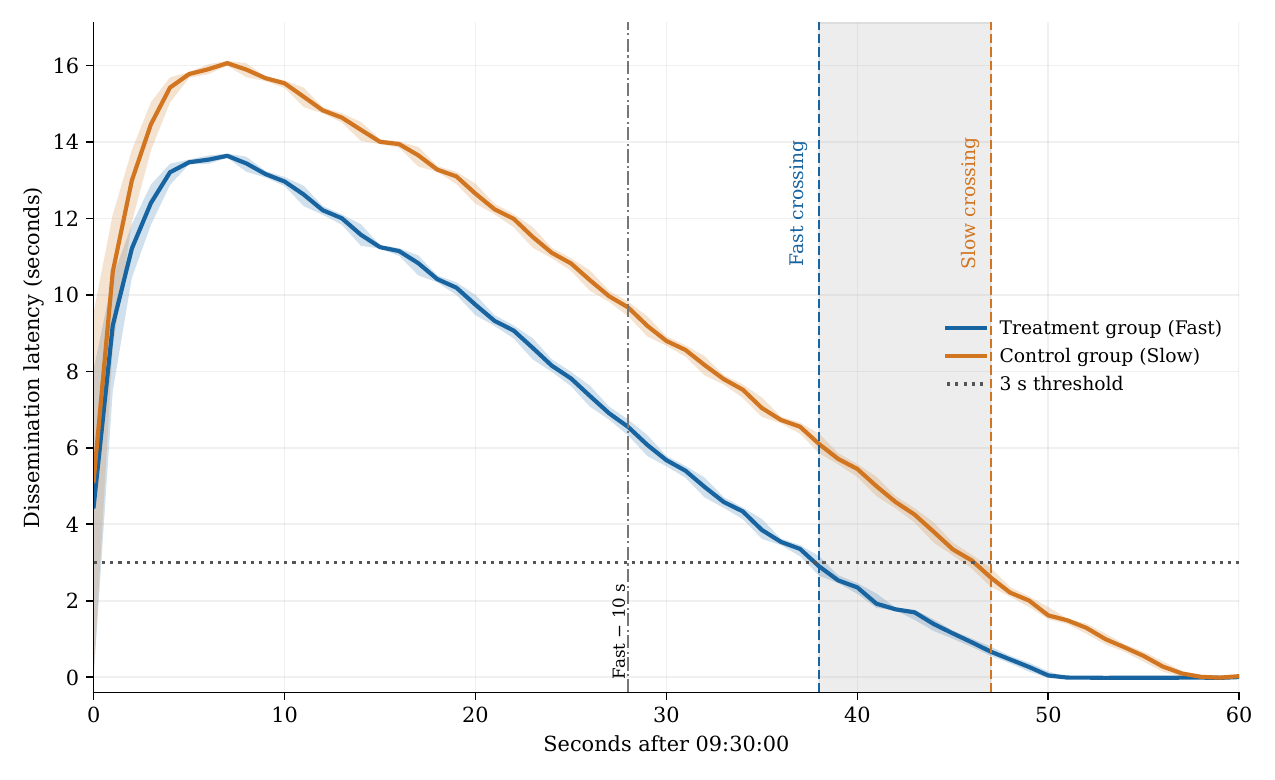}
    \caption{Latency Convergence and Event Windows at Market Open on October 20, 2025.}
    \label{fig:latency_timeseries}
    \begin{minipage}{\textwidth}
    \small
    This figure illustrates the cross-channel variation in dissemination latency underlying the difference-in-differences design. Solid curves show group-median dissemination latency for valid matched order messages. Shaded bands show the observed minimum and maximum latency within each group and exchange second. The fast and slow group medians first cross below three seconds at 38 and 47 seconds after the market open, respectively. The estimation window is $28 \leq s < 47$ seconds after 09:30:00, including the pre-event interval before the fast crossing and ending before the slow group crosses. Between 38 and 47 seconds, the fast group has experienced the latency improvement while the slow group has not yet experienced the same improvement. Gray shading marks the interval between crossings; vertical lines indicate the two crossings and the Fast--10 boundary.
    \end{minipage}
\end{figure}

    Latency typically remains elevated for 30 to 90 seconds before declining. We use three seconds as a timing benchmark because it is the tick-feed refresh interval. A crossing marks an improvement in the timeliness of order data and their usefulness to latency-sensitive strategies. The empirical strategy compares stocks assigned to different channels within the same second and trading day, absorbing common time-series variation.

    $$[\text{Insert Figure~\ref{fig:latency_timeseries} here.}]$$

    Figure~\ref{fig:latency_timeseries} plots group-median dissemination latency on October 20, 2025, together with each group's per-second minimum--maximum band across valid matched order messages. The groups are clearly separated during the estimation window: the fast group's maximum latency is below the slow group's minimum in every retained second. Their median latencies cross below three seconds at 38 and 47 seconds after the open, respectively.

    The DID data window is October 10 to November 30, 2025. Its dates reflect the availability of an additional proprietary colocation feed containing the local receipt timestamps needed to measure channel-level dissemination latency. This window partly overlaps and extends beyond the baseline order-level sample ending October 31. Of the 36 target trading days, two lack the required colocation data and five have simultaneous fast and slow crossings, leaving 29 positive-gap candidates. We require a crossing-time gap strictly greater than one second; five additional days fail this requirement. The final estimation sample contains 24 trading days.

    For day $t$ and calendar second $s$ measured from 09:30:00, define the fast and slow groups' first crossings below three seconds, searching from ten seconds after the open:
    \begin{align}
    G^F_t &= \min \{ s\ge 10 : Latency_{F,t,s} < 3 \},\\
    G^S_t &= \min \{ s\ge 10 : Latency_{S,t,s} < 3 \}.
    \end{align}
    Thus, $G^F_t$ and $G^S_t$ mark the two groups' first crossings. We truncate the sample at the slow group's crossing, $G^S_t$, which prevents the slow group's improvement in data timeliness from contaminating the estimated treatment coefficient.

    Let $Fast_{i,t}$ indicate assignment to the fast group and define $Post_{t,s}=1\{s\ge G^F_t\}$. At the stock--date--second level, the static DID specification is
    \begin{equation}
    Y_{i,t,s} =
    \beta_1 Fast_{i,t}\times Post_{t,s}
    + \alpha_{i,t}
    + \gamma_{t,s}
    + \varepsilon_{i,t,s}.
    \end{equation}
    Stock $\times$ date fixed effects $\alpha_{i,t}$ absorb stock-day heterogeneity, and date $\times$ calendar-second fixed effects $\gamma_{t,s}$ absorb common second-level shocks. Identification comes from the change in the fast--slow outcome difference around the fast crossing, while the slow group has not yet crossed.

    We interpret cross-channel latency variation as an encouragement to latency-sensitive trading. The increase in active exact-match share provides evidence that improved information timeliness raises proxy-measured HFT activity. A causal interpretation requires parallel trends and the absence of concurrent channel-specific shocks. Interpreting the market-quality responses as effects of HFT activity additionally requires an exclusion restriction: within the narrow censored event window, cross-channel latency affects market quality through the induced change in latency-sensitive trading, with no separate effect through other channels. Under these assumptions, the reduced-form estimates provide causal evidence on the market-quality consequences of the induced increase in HFT activity.

    We examine pre-event differences and post-event dynamics using the same groups, common clock and censored window:
    \begin{equation*}
    Y_{i,t,s} =
    \sum_{\ell\in\mathcal K\setminus\{-1\}}
    \theta_{\ell} Fast_{i,t}\times 1\{s-G^F_t=\ell\}
    +\alpha_{i,t}+\gamma_{t,s}+\varepsilon_{i,t,s}.
    \end{equation*}
    Here $\mathcal K=\{-10,-9,\ldots,10\}$ spans the event-time interval $[-10,10]$, with $s-G^F_t=-1$ as the reference period. Event seconds at or after the slow crossing are excluded. The transaction-cost outcomes use $s-G^F_t=-2$ for the forward-window reason described below. We use two-way clustering by stock and date $\times$ calendar-second for the main tables and event-study figures. Internet Appendix Section~\ref{app:did_clustering} reports the corresponding results with stock $\times$ date clustering.

    We measure exact-match activity with three proxies. Active count is the number of active exact-match limit orders in a stock-second. Active share and passive share divide the corresponding matching-order counts by the number of limit orders. These outcomes measure the response of latency-sensitive order placement to lower latency.

    Market-quality proxies comprise relative quoted spread (Spread), best-quote notional depth (LogDepth), intrasecond mid-quote range (Range), effective spread (ES), one-second adverse-selection exposure (AS1s), one-second realized spread (RS1s), wide-quote incidence (WideSpread), and transaction-based pricing error (PEbps). Price measures are in basis points, depth is in log units, and binary and share outcomes are in percentage points.

    Let $\mathcal Q_{i,t,s}$ denote valid reconstructed quote updates in a second, with best ask $A_q$, best bid $B_q$, midpoint $M_q=(A_q+B_q)/2$, and corresponding displayed quantities $V^A_q$ and $V^B_q$. All prices below are in yuan; $C_{i,t-1}$ is the previous closing price. The quote-based measures are
    \begin{align*}
    Spread_{i,t,s}
      &=\frac{10{,}000}{|\mathcal Q_{i,t,s}|}
        \sum_{q\in\mathcal Q_{i,t,s}}\frac{A_q-B_q}{M_q},\\
    LogDepth_{i,t,s}
      &=\log\left(1+\frac{1}{|\mathcal Q_{i,t,s}|}
        \sum_{q\in\mathcal Q_{i,t,s}}(A_qV^A_q+B_qV^B_q)\right),\\
    Range_{i,t,s}
      &=10{,}000\,\frac{\max_{q\in\mathcal Q_{i,t,s}}M_q
                 -\min_{q\in\mathcal Q_{i,t,s}}M_q}{C_{i,t-1}}.
    \end{align*}
    Spread and depth use equal weights across valid quote updates. When a second has no valid update, the construction carries forward depth; Spread and Range remain missing.

    For a transaction with aggressor direction $D\in\{-1,1\}$, price $P$, pre-trade midpoint $M_-$ and midpoint one second later $M_{+1}$, define
    \begin{align}
    ES &= 2D(P-M_-)/M_-\times10{,}000,\\
    AS1s &= D(M_{+1}-M_-)/M_-\times10{,}000,\\
    RS1s &= 2D(P-M_{+1})/M_-\times10{,}000=ES-2AS1s.
    \end{align}
    We aggregate these measures using identical transactions and transaction-notional weights within each stock-second. Future quotes must be strictly before the slow crossing and, for pre-event trades, before the fast crossing. This removes $s-G^F_t=-1$ entirely, so these three event studies use $s-G^F_t=-2$ as the reference and jointly test the eight pre-event coefficients at $s-G^F_t=-10,\ldots,-3$.

    For tick size $\tau$, we define WideSpread as
    \begin{equation*}
    WideSpread_{i,t,s}
      =100\,1\{\exists q\in\mathcal Q_{i,t,s}: A_q-B_q>\tau\}.
    \end{equation*}
    For this indicator, $\mathcal Q_{i,t,s}$ also includes the valid quote state at the start of the second. Seconds without a valid state are missing. Internet Appendix Section~\ref{app:wide_share} reports the distinct fixed-grid WideSpreadShare measure.

    Pricing error is the average absolute temporary price deviation across trades in a stock-second, expressed in basis points:
    \begin{equation*}
    PEbps_{i,t,s}
      =\frac{10{,}000}{N^{PE}_{i,t,s}}
        \sum_{j=1}^{N^{PE}_{i,t,s}}|\widehat e_{i,t,s,j}|.
    \end{equation*}
    Here $N^{PE}_{i,t,s}$ is the number of trades in that second with a pricing-error estimate after model eligibility and warm-up requirements, and $\widehat e_{i,t,s,j}$ is the temporary log-price component for the $j$th such trade. We estimate this component using fixed historical VAR parameters and a finite pricing-error filter, then average its absolute value with equal trade weights. Internet Appendix Section~\ref{app:pe} describes its construction and coverage.

    $$[\text{Insert Table~\ref{tab:balance_smd} here.}]$$

    As a covariate-balance check, Table~\ref{tab:balance_smd} compares the 48,188 stock-days in the final censored active-count sample. Absolute standardized differences are below 0.02 for all reported covariates.

    \subsection{Exact-Match Activity}
\begin{figure}[H]\centering
\includegraphics[width=.92\linewidth]{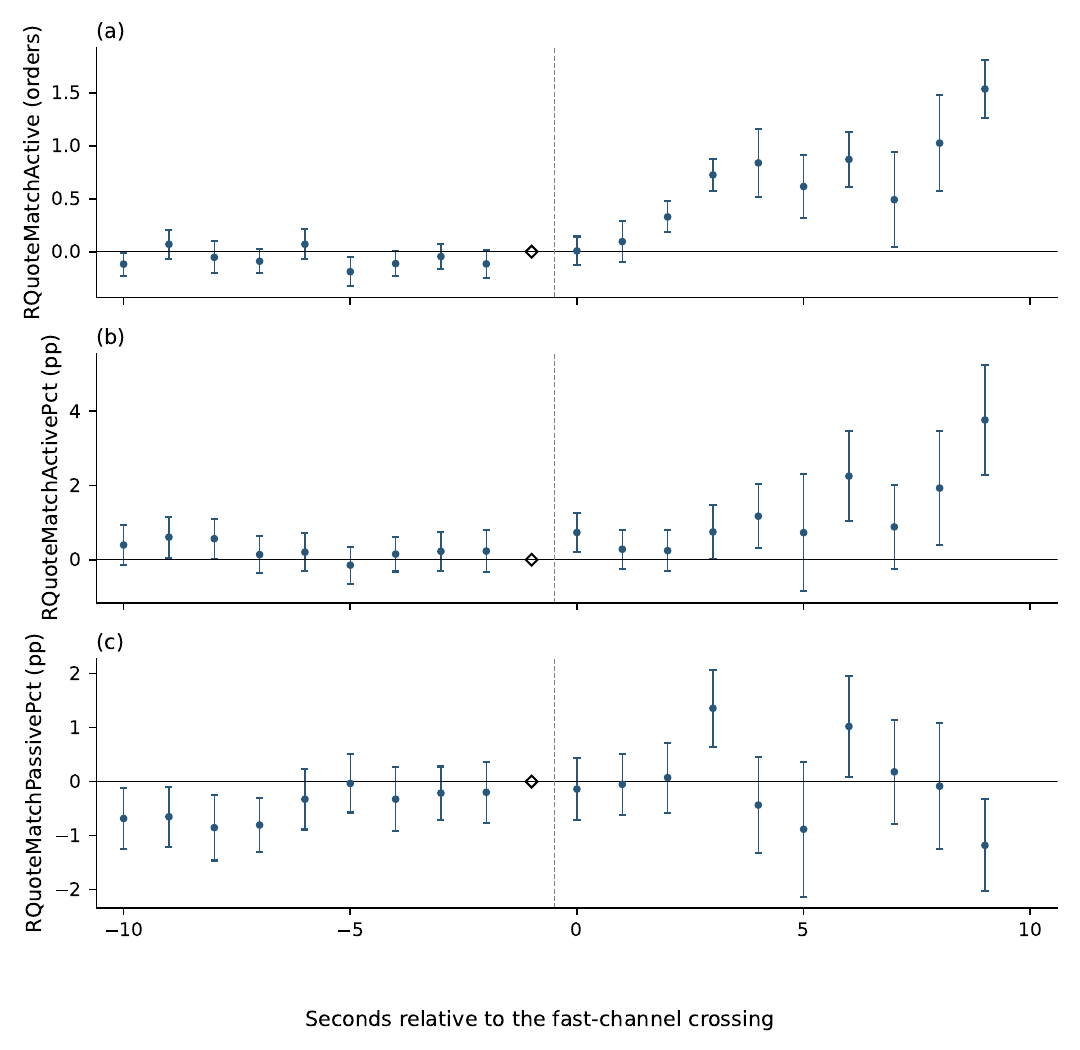}
\caption{Event Studies: Exact-Match Activity.}\label{fig:did_hft_panel}
\begin{minipage}{\textwidth}\small This figure presents event-time estimates of exact-match activity around the fast-group latency crossing. The sample includes 24 trading days and uses the censored window $G^F_t-10 \leq s < G^S_t$ with $G^S_t-G^F_t>1$, excluding the smallest 30\% of stocks by prior-day capitalization. Panels (a)--(c) report active exact-match counts, active exact-match shares, and passive exact-match shares. Counts include limit orders only; shares are computed from limit orders and reported in percentage points (pp). All specifications include stock $\times$ date and date $\times$ calendar-second fixed effects, with two-way clustering by stock and date $\times$ calendar-second. Bars indicate 90\% pointwise confidence intervals; open diamonds mark the omitted period $s-G^F_t=-1$. Only supported event-time bins within $\pm10$ seconds are displayed.\end{minipage}
\end{figure}

    $$[\text{Insert Figure~\ref{fig:did_hft_panel} here.}]$$

    Figure~\ref{fig:did_hft_panel}(a) shows active exact-match counts rising after the fast group's latency-threshold crossing. Seven of the nine pre-event coefficients have 90\% confidence intervals that include zero, and the estimates show no clear monotonic pattern. However, the joint pre-event test rejects the null that all pre-event coefficients equal zero ($p=0.028$). We therefore treat the count result as supporting evidence.

    To separate matching intensity from overall order activity, we examine active exact-match shares among limit orders. Figure~\ref{fig:did_hft_panel}(b) shows no clear pre-event pattern, and the joint pre-event test does not reject ($p=0.228$). Post-event coefficients are generally positive. Active share provides the clearest treatment-response evidence: active matching becomes more prevalent after the latency improvement, beyond changes in total order activity.

    Figure~\ref{fig:did_hft_panel}(c) shows a less distinct post-event pattern for passive quote-matching shares. The static coefficient is significantly positive, but most individual post-event estimates are insignificant and the joint pre-event test rejects ($p=0.038$). The static estimate compares the full pre-event and post-event windows, whereas the event study uses the second immediately before crossing as its reference. We treat passive matching as supporting evidence.

    $$[\text{Insert Table~\ref{tab:did_latency_in_3s} here.}]$$

    Table~\ref{tab:did_latency_in_3s} reports the static activity estimates: active matching increases by 0.447 orders per stock-second ($t=6.95$), active share by 0.505 percentage points ($t=3.03$), and passive share by 0.515 percentage points ($t=3.07$), with two-way clustering. All three static activity estimates are significantly positive, but their pre-event diagnostics differ. We place the greatest weight on active share as evidence that the latency shock increases proxy-measured HFT activity.

    Internet Appendix Section~\ref{app:exclusion_tests} examines market orders and more aggressively priced limit orders under the same latency-crossing design. Orders priced more than one tick beyond the best opposite-side quote become less prevalent after the latency improvement. Together with the increase in active exact matching, this finding supports a shift toward more precise price placement.

    \subsection{Liquidity and Price Variation}
\begin{figure}[H]\centering
\includegraphics[width=\linewidth]{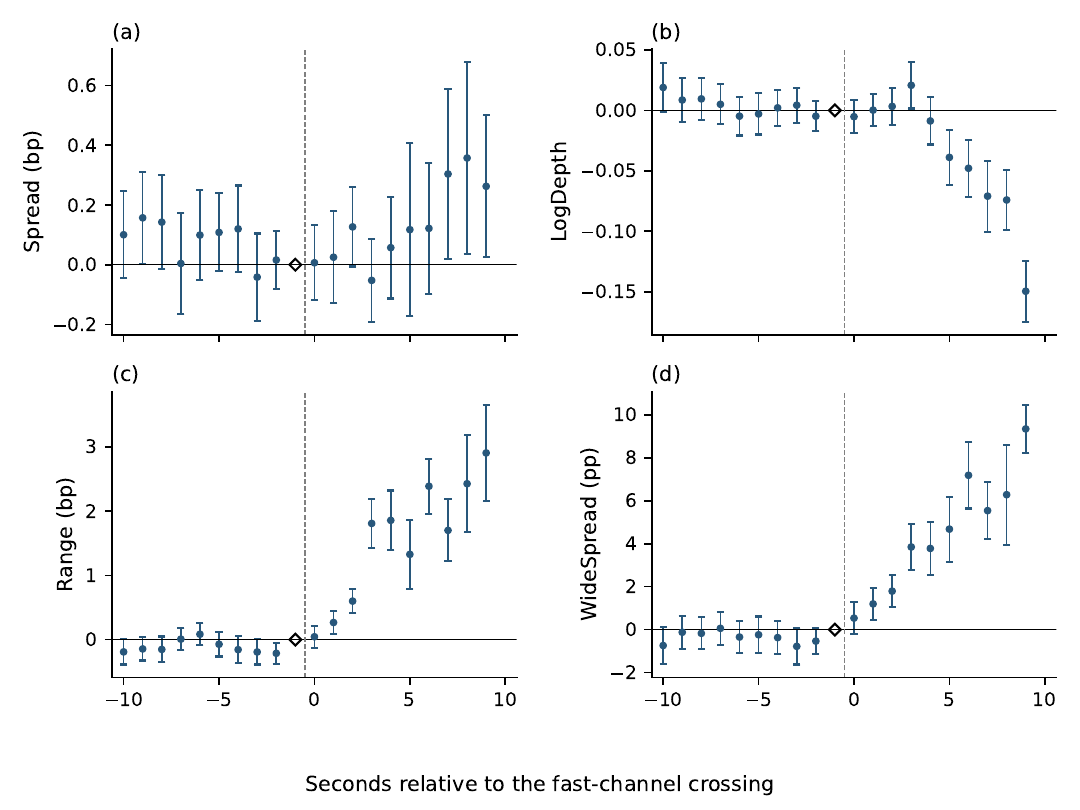}
\caption{Event Studies: Liquidity and Price Variation.}\label{fig:did_market_panel}
\begin{minipage}{\textwidth}\small This figure presents event-time estimates of liquidity and price-variation outcomes around the fast-group latency crossing. The sample includes 24 trading days and follows the censored event-time design described for Figure~\ref{fig:did_hft_panel}, with $G^F_t-10 \leq s < G^S_t$, $G^S_t-G^F_t>1$, and the smallest 30\% of stocks by prior-day capitalization excluded. Panels (a)--(d) report Spread, LogDepth, Range, and WideSpread. The specifications include stock $\times$ date and date $\times$ calendar-second fixed effects with two-way clustering by stock and date $\times$ calendar-second. Bars indicate 90\% pointwise confidence intervals; open diamonds mark the omitted period $s-G^F_t=-1$. WideSpread indicates whether a valid wide quote occurs within the second, including an initial carried-forward quote state. Axes use outcome-specific scales; bp denotes basis points and pp denotes percentage points.\end{minipage}
\end{figure}

    $$[\text{Insert Figure~\ref{fig:did_market_panel} and Table~\ref{tab:did_market_expanded} here.}]$$

    Figure~\ref{fig:did_market_panel} and Table~\ref{tab:did_market_expanded}, Panel A, report quoted liquidity and short-horizon price variation.

    The average Spread estimate is close to zero and insignificant ($p=0.960$), while its pre-event joint test rejects ($p=0.004$). LogDepth, measured as $\log(1+\text{best-quote notional depth in yuan})$, declines weakly.

    Range increases by 1.029 basis points ($t=10.36$), with a pre-event joint $p=0.082$. WideSpread increases by 2.964 percentage points ($p<0.001$), with a pre-event joint $p=0.720$. Wide-quote incidence and average spread width are distinct outcomes: wide quotes become more frequent without a statistically detectable change in average quoted spread.

    \subsection{Transaction Costs and Pricing Error}
\begin{figure}[H]\centering
\includegraphics[width=\linewidth]{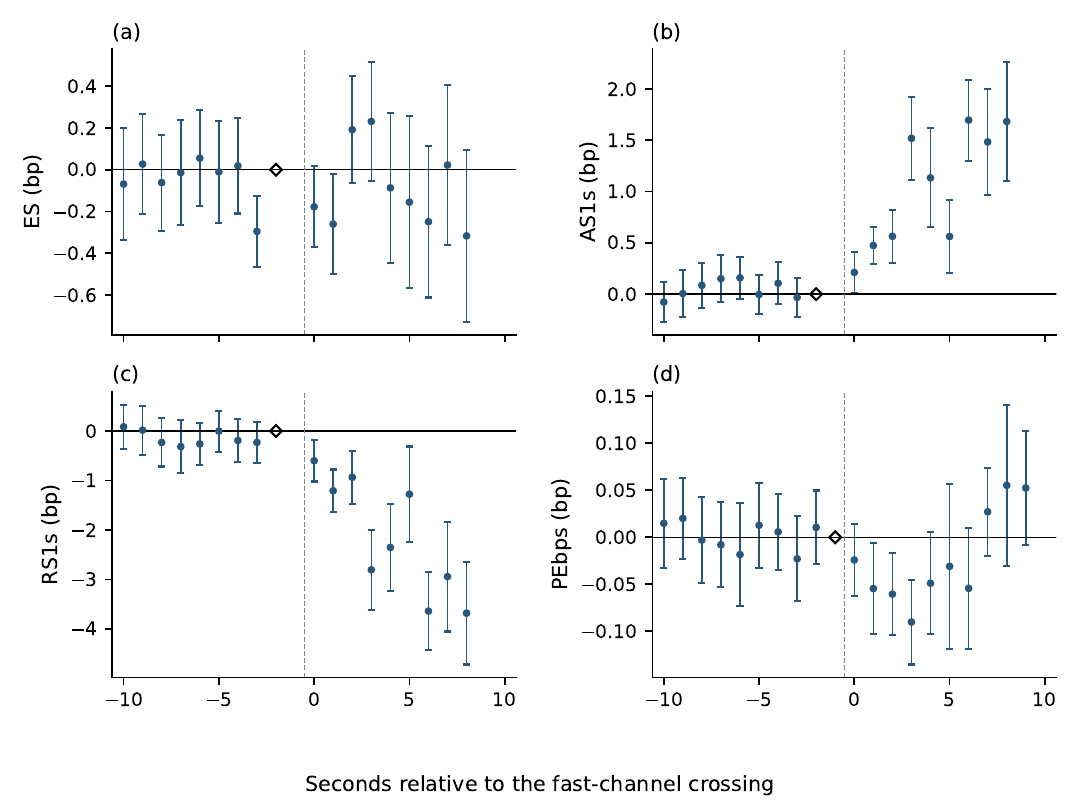}
\caption{Event Studies: Transaction Costs and Pricing Error.}\label{fig:did_transaction_panel}
\begin{minipage}{\textwidth}\small This figure presents event-time estimates of transaction-cost and pricing-error outcomes around the fast-group latency crossing. The sample includes 24 trading days and follows the censored event-time design described for Figure~\ref{fig:did_hft_panel}. Panels (a)--(d) report Effective Spread (ES), Adverse Selection (AS1s), Realized Spread (RS1s), and Pricing Error (PEbps), all measured in basis points. Panels (a)--(c) use the common one-second-valid transaction sample ($N=482{,}510$). Because $s-G^F_t=-1$ is unavailable, these panels use $s-G^F_t=-2$ as the omitted reference period and include eight pre-event leads. Panel (d) uses the PE-available transaction sample ($N=541{,}710$), omits $s-G^F_t=-1$, and includes nine pre-event leads; Internet Appendix Section~\ref{app:pe} describes transaction and warm-up selection. All specifications include stock $\times$ date and date $\times$ calendar-second fixed effects with two-way clustering by stock and date $\times$ calendar-second. Open diamonds mark omitted periods; bars indicate 90\% pointwise confidence intervals.\end{minipage}
\end{figure}

    $$[\text{Insert Figure~\ref{fig:did_transaction_panel} here.}]$$

    Figure~\ref{fig:did_transaction_panel} and Table~\ref{tab:did_market_expanded}, Panel B, report transaction costs and pricing error. ES, AS1s and RS1s share the same one-second-valid transaction sample and use $s-G^F_t=-2$ as the event-study reference. PE instead uses its own available sample and a $s-G^F_t=-1$ reference.

    ES changes by $-0.049$ bp ($p=0.599$), AS1s increases by 0.648 bp, and RS1s decreases by 1.345 bp (both $p<0.001$).

    PEbps declines by 0.04486 bp ($t=-3.23$, $p=0.00134$), with a pre-event joint $p=0.800$. The event-time path is non-monotonic. Because this model-based measure is available only for transaction-seconds satisfying the estimation and warm-up requirements, we report it as complementary evidence.

    \subsection{Discussion}

    The activity results indicate that lower latency changes the composition of order placement. Active exact matching provides the clearest treatment-response evidence, consistent with the earlier behavioral and return-predictability results showing that these orders rely on contemporaneous order-book information. Passive matching responds less clearly in event time, providing weaker evidence of an immediate adjustment in liquidity provision.

    The market-quality results point to greater short-horizon exposure for liquidity providers. Prices move further in the aggressor's direction, wide-spread states become more frequent, and displayed depth declines weakly, even though average quoted and effective spreads change little. The decline in pricing error captures a different dimension of price behavior: it reflects smaller temporary transaction-price deviations rather than lower intrasecond price variation. The pricing-error result is therefore complementary to, rather than inconsistent with, the increase in short-horizon price variation. Under the identifying assumptions in Section~\ref{sec:did_research_design}, these patterns provide causal evidence on the market-quality consequences of latency-induced HFT activity.

    The institutional environment described in Section~\ref{sec:institutional_background} provides a possible interpretation. The T+1 trading rule limits rapid inventory recycling, while the absence of a standard price-protected IOC limit-order instruction, the lack of maker-taker incentives, and constraints on rapid order submission and cancellation reduce the relative attractiveness of continuous liquidity provision. Improved information timeliness may therefore be used more intensively for directional liquidity taking than for continuous market making. These institutional features do not generate the identifying variation; rather, they characterize the environment in which the latency shock operates and help explain the form of the trading response.

    Taken together, the results suggest that the market-quality consequences of faster information processing depend on how speed is deployed. In our setting, the trading response is clearest in active exact matching, while the strongest market-quality responses appear in short-horizon price variation and adverse selection rather than in average spread measures.

    \section{Conclusion} \label{sec:Conclusion}
    This paper develops an order-level proxy for latency-sensitive HFT activity based on active exact matching: limit orders submitted precisely at the contemporaneous best opposite-side quote. The evidence supports active exact matching as an informative behavioral signature of timely order-book processing. Active exact-match orders are submitted sooner after quote updates, are canceled more rapidly, and cluster in favorable microstructure conditions. They also predict subsequent price movements from 100 milliseconds through ten minutes, with stronger return predictability than orders priced one or multiple cents more aggressively than the prevailing best quote.

    Using plausibly exogenous cross-channel variation in order-message dissemination latency at the SZSE market open, we study the market-quality effects of latency-sensitive HFT activity. Improved information timeliness increases active exact-match shares and is accompanied by greater intrasecond price variation and a higher incidence of wide spreads. Average quoted and effective spreads remain statistically unchanged, while displayed depth declines weakly. At the transaction level, one-second adverse-selection exposure increases and realized spreads decline, while pricing error declines modestly on its available sample. Under the identifying assumptions discussed in Section~\ref{sec:High-frequency Trading and Market Quality}, these joint activity and market-quality responses provide causal evidence on the consequences of latency-induced HFT activity.

    The paper makes four contributions. First, we introduce an order-level proxy for latency-sensitive HFT that captures precise contemporaneous price placement without requiring proprietary trader identifiers or relying solely on response speed.

    Second, we show that active exact-match orders predict subsequent returns more strongly than more aggressively priced orders, indicating that price-placement precision contains predictive information beyond order direction and aggressiveness. This result extends the literature on marketable limit orders by identifying precise targeting of the prevailing best quote as an additional dimension of order informativeness.

    Third, we introduce a novel quasi-experimental design to study the market-quality effects of latency-sensitive HFT activity. The design exploits congestion-driven differences in order-message dissemination latency across trading channels at the market open. The latency improvement increases active exact-match shares. The design provides causal evidence on how latency-induced HFT activity affects liquidity, short-horizon price variation, adverse selection, transaction costs, and pricing error.

    Fourth, we add to the limited evidence on HFT in China and emerging markets by documenting latency-sensitive trading in a market structure that differs substantially from those studied in much of the existing literature. China's equity market features a large retail presence, a T+1 restriction on the same-day resale of newly purchased shares, no maker-taker rebates, different execution rules for marketable orders, and constraints on rapid order submission and cancellation. Our strongest evidence concerns active exact matching and directional liquidity taking, broadening the evidence beyond settings in which HFT is closely associated with continuous automated liquidity provision.

    Future research could link exact-match activity to trader-level identifiers, distinguish among latency-sensitive trading strategies, and examine how the use and market-quality effects of speed vary across trading venues and market designs.
\clearpage

\nocite{}
\bibliographystyle{jfe}
\bibliography{references_hft.bib}
\clearpage

	%\section{References}
	%\bibliographystyle{apalike}
	%\bibliography{references.bib}

\clearpage

\clearpage

\clearpage

% \clearpage

% \begin{figure}[H]
%     \centering
%     \caption{High-Frequency Trading Activity and Market Indices}
%     \begin{minipage}{\textwidth}
%     \small
%     \end{minipage}
% \end{figure}
% \clearpage

\clearpage

\clearpage
\clearpage
\clearpage

% Top-align the table float pages; retain natural table height.
\makeatletter
\setlength{\@fptop}{0pt}
\makeatother
\begin{table}[htbp]\singlespacing\footnotesize\renewcommand{\arraystretch}{0.96}
\setlength{\tabcolsep}{1.5pt}
  \caption{Descriptive Statistics.}\begin{minipage}{\textwidth}
\small
This table summarizes order characteristics, exact-match measures, and their associated short-horizon return patterns in the sample. The sample consists of common stocks listed on the Shenzhen Stock Exchange, excluding the smallest 30\% of stocks by previous-day market capitalization. The sample period is July 1, 2023 to October 31, 2025. Stocks are sorted into five size quintiles based on prior-day market capitalization. Panel A reports the fraction of orders with each characteristic by size quintile. Panel B reports mean ten-second forward mid-quote returns conditional on size quintile and order characteristics. \textit{RQuoteMatchActive} (\textit{RQuoteMatchPassive}) equals one if an order is submitted at the best opposite-side (same-side) quote. \textit{BuyOrder} equals one for buy orders. \textit{OneCentLarger} and \textit{MultiCentLarger} indicate orders priced exactly one tick and more than one tick beyond the best opposite-side quote, respectively. \textit{NonRoundVol} and \textit{NonRoundPrice} indicate non-round order quantities and prices. \textit{LargeOrder} equals one if order notional exceeds RMB 200,000. Returns are direction-adjusted mid-quote returns measured from the first valid midpoint after submission to the last valid midpoint at or before ten seconds after submission. Returns crossing lunch breaks or the session close are excluded rather than assigned zero. Panel A contains 42,506,216 limit orders; Panel B contains 42,465,539 orders with valid ten-second returns. Market orders are excluded from matching indicators and main regressions. Panel C reports excluded order types and limit-order benchmarks before exclusion; shares in Panel C use the original 42,644,479 selected orders. Returns are reported in basis points.
\end{minipage}\par\footnotesize
  \label{tab:panel_descriptive_statistics}
  \centering
  \paperfit{\textwidth}{
  \begin{tabular}{llrrrrrrrr}
    \toprule
    \multicolumn{10}{c}{\textbf{Panel A: Average of Dummy Variables}} \\
    \midrule
     &  & RQuoteMatchActive & RQuoteMatchPassive & BuyOrder & OneCentLarger & MultiCentLarger & NonRoundVol & NonRoundPrice & LargeOrder \\
    \midrule
Small &  & 0.3120 & 0.3565 & 0.5071 & 0.0355 & 0.0410 & 0.7821 & 0.9816 & 0.0117 \\
2 &  & 0.3004 & 0.3435 & 0.5092 & 0.0381 & 0.0452 & 0.7660 & 0.9800 & 0.0133 \\
3 &  & 0.2903 & 0.3350 & 0.5104 & 0.0404 & 0.0500 & 0.7594 & 0.9784 & 0.0147 \\
4 &  & 0.2816 & 0.3286 & 0.5106 & 0.0426 & 0.0570 & 0.7690 & 0.9766 & 0.0164 \\
Large &  & 0.2502 & 0.3011 & 0.5118 & 0.0445 & 0.0748 & 0.7822 & 0.9666 & 0.0248 \\
    \midrule
    \multicolumn{10}{c}{\textbf{Panel B: Forward 10-Second Mid-Quote Returns Conditional on Dummy Values}} \\
    \midrule
     &  & RQuoteMatchActive & RQuoteMatchPassive & BuyOrder & OneCentLarger & MultiCentLarger & NonRoundVol & NonRoundPrice & LargeOrder \\
    \midrule
\multirow{2}{*}{Small} & 0 & 1.7737 & 3.3160 & 2.4357 & 2.7585 & 2.8681 & 1.7324 & 3.0242 & 2.7468 \\
 & 1 & 4.9228 & 1.7452 & 3.0674 & 2.6893 & 0.1346 & 3.0411 & 2.7510 & 3.5412 \\
    \addlinespace[1ex]
\multirow{2}{*}{2} & 0 & 1.7822 & 3.1211 & 2.3484 & 2.6670 & 2.8081 & 1.6861 & 2.7810 & 2.6429 \\
 & 1 & 4.6865 & 1.7627 & 2.9496 & 2.3377 & -0.5938 & 2.9501 & 2.6519 & 3.5127 \\
    \addlinespace[1ex]
\multirow{2}{*}{3} & 0 & 1.7784 & 2.9627 & 2.2777 & 2.5786 & 2.7295 & 1.6506 & 2.6756 & 2.5470 \\
 & 1 & 4.4668 & 1.7568 & 2.8283 & 2.0858 & -0.6905 & 2.8462 & 2.5561 & 3.3421 \\
    \addlinespace[1ex]
\multirow{2}{*}{4} & 0 & 1.6154 & 2.6286 & 2.0602 & 2.3308 & 2.4889 & 1.4764 & 2.3335 & 2.2930 \\
 & 1 & 4.0747 & 1.6523 & 2.5451 & 1.7905 & -0.6924 & 2.5575 & 2.3071 & 3.1907 \\
    \addlinespace[1ex]
\multirow{2}{*}{Large} & 0 & 1.3317 & 1.9570 & 1.5867 & 1.8204 & 1.9787 & 1.1908 & 1.9565 & 1.7658 \\
 & 1 & 3.1379 & 1.3811 & 1.9715 & 0.9926 & -0.6277 & 1.9486 & 1.7776 & 2.4836 \\
    \bottomrule
\end{tabular}
  }
% LIMIT_ONLY_MARKET_PANEL
\par\medskip\textbf{Panel C: Order Types Before Exclusion}\par
\paperfit{\textwidth}{\begin{tabular}{lrrr}\toprule
Order type & Orders & Share (\%) & Response $<100$ms (\%) \\
\midrule
Limit orders (L) & 42,506,216 & 99.6758 & 53.88 \\
Opposite-best market orders (O) & 132,422 & 0.3105 & 31.98 \\
Same-side-best market orders (S) & 5,841 & 0.0137 & 37.44 \\
All market orders (O/S) & 138,263 & 0.3242 & 32.21 \\
Active-matching limit orders & 11,782,436 & 27.6295 & 60.95 \\
Passive-matching limit orders & 13,792,088 & 32.3420 & 59.40 \\
\bottomrule\end{tabular}}
\end{table}

\clearpage

\begin{table}[htbp]\singlespacing\footnotesize\renewcommand{\arraystretch}{0.9}\centering
\def\sym#1{\ifmmode^{#1}\else\(^{#1}\)\fi}
\caption{Order Response Time: Panel Regression.}
\label{tab:order_response_time}
\begin{minipage}{\textwidth}
\small
This table reports panel regressions examining whether exact-match orders are associated with faster responses to changes in the best quote. The sample consists of common stocks listed on the Shenzhen Stock Exchange, excluding the smallest 30\% of stocks by prior-day market capitalization. The sample period is July 1, 2023 to October 31, 2025. The analysis uses an approximately 0.1\% systematic sample of limit orders submitted during continuous trading hours, excluding market orders. The dependent variable equals one if an order is submitted within 100 milliseconds after the latest prior change in best-bid or best-ask price or quantity. \textit{RQuoteMatchActive} (\textit{RQuoteMatchPassive}) equals one if an order is submitted at the best opposite-side (same-side) quote. Other order controls include \textit{BuyOrder}, \textit{OneCentLarger}, \textit{MultiCentLarger}, \textit{NonRoundVol}, \textit{NonRoundPrice}, and \textit{LargeOrder}, as well as lagged stock characteristics. All regressions include stock and date fixed effects, and standard errors are clustered at the stock level. t-statistics are reported in parentheses.
\end{minipage}\par\footnotesize
\paperfit{\textwidth}{
\begin{tabular}{lrrr}\toprule
 & (1) & (2) & (3) \\
 & ResponseIn100ms & ResponseIn100ms & ResponseIn100ms \\
\midrule
RQuoteMatchActive & 0.19399\sym{***} & 0.21583\sym{***} & 0.19167\sym{***} \\
 & (77.50) & (81.32) & (86.03) \\
RQuoteMatchPassive & 0.17807\sym{***} & 0.19987\sym{***} & 0.17650\sym{***} \\
 & (72.20) & (75.79) & (78.25) \\
BuyOrder & -0.01073\sym{***} & -0.01059\sym{***} & -0.01049\sym{***} \\
 & (-40.80) & (-39.65) & (-38.08) \\
LaggedLogSize & 0.08647\sym{***} & 0.08573\sym{***} & 0.07603\sym{***} \\
 & (17.03) & (17.03) & (15.89) \\
LaggedTurnoverRate & 0.89259\sym{***} & 0.87743\sym{***} & 0.88740\sym{***} \\
 & (64.26) & (63.21) & (65.29) \\
LaggedInverseClosePrice & -0.05869 & -0.06498 & -0.09322\sym{*} \\
 & (-0.99) & (-1.11) & (-1.68) \\
OneCentLarger &  & 0.11749\sym{***} & 0.10220\sym{***} \\
 &  & (75.93) & (82.40) \\
MultiCentLarger &  & 0.06656\sym{***} & 0.05170\sym{***} \\
 &  & (42.43) & (32.67) \\
NonRoundVol &  &  & 0.14365\sym{***} \\
 &  &  & (81.69) \\
NonRoundPrice &  &  & 0.00288\sym{***} \\
 &  &  & (2.84) \\
LargeOrder &  &  & -0.05622\sym{***} \\
 &  &  & (-20.43) \\
\midrule
Stock FE & Yes & Yes & Yes \\
Date FE & Yes & Yes & Yes \\
Observations & 42,506,216 & 42,506,216 & 42,506,216 \\
\bottomrule\end{tabular}
}
\par\footnotesize $t$-statistics in parentheses. \sym{*} $p<0.10$, \sym{**} $p<0.05$, \sym{***} $p<0.01$.
\end{table}

\clearpage

\begin{table}[htbp]\singlespacing\footnotesize\renewcommand{\arraystretch}{0.9}\centering
\def\sym#1{\ifmmode^{#1}\else\(^{#1}\)\fi}
\caption{Order Cancellation Time: Panel Regression.}
\label{tab:order_cancellation_time}
\begin{minipage}{\textwidth}
\small
This table reports panel regressions examining whether exact-match orders are associated with faster order cancellations. The sample consists of common stocks listed on the Shenzhen Stock Exchange, excluding the smallest 30\% of stocks by prior-day market capitalization. The sample period is July 1, 2023 to October 31, 2025. The analysis uses an approximately 0.1\% systematic sample of limit orders submitted during continuous trading hours, excluding market orders. Only canceled orders are included in the estimation sample. The dependent variable, \textit{CancelIn100ms}, equals one if an order is canceled within 100 milliseconds after submission and zero otherwise. \textit{RQuoteMatchActive} (\textit{RQuoteMatchPassive}) equals one if an order is submitted at the best opposite-side (same-side) quote. Other order controls include \textit{BuyOrder}, \textit{OneCentLarger}, \textit{MultiCentLarger}, \textit{NonRoundVol}, \textit{NonRoundPrice}, and \textit{LargeOrder}. All specifications include stock and date fixed effects, and standard errors are clustered at the stock level. t-statistics are reported in parentheses.
\end{minipage}\par\footnotesize
\paperfit{\textwidth}{
\begin{tabular}{lrrr}\toprule
 & (1) & (2) & (3) \\
 & CancelIn100ms & CancelIn100ms & CancelIn100ms \\
\midrule
RQuoteMatchActive & 0.05383\sym{***} & 0.05400\sym{***} & 0.05406\sym{***} \\
 & (40.96) & (41.06) & (41.13) \\
RQuoteMatchPassive & 0.01142\sym{***} & 0.01154\sym{***} & 0.00993\sym{***} \\
 & (40.22) & (40.53) & (34.99) \\
BuyOrder & 0.00155\sym{***} & 0.00155\sym{***} & 0.00154\sym{***} \\
 & (10.88) & (10.94) & (10.86) \\
LaggedLogSize & 0.00955\sym{***} & 0.00952\sym{***} & 0.00921\sym{***} \\
 & (6.70) & (6.68) & (6.30) \\
LaggedTurnoverRate & -0.03439\sym{***} & -0.03433\sym{***} & -0.03399\sym{***} \\
 & (-16.30) & (-16.27) & (-16.61) \\
LaggedInverseClosePrice & 0.04057\sym{***} & 0.04045\sym{***} & 0.04018\sym{***} \\
 & (4.54) & (4.52) & (4.46) \\
OneCentLarger &  & 0.06363\sym{***} & 0.06500\sym{***} \\
 &  & (19.07) & (19.43) \\
MultiCentLarger &  & 0.04269\sym{***} & 0.04505\sym{***} \\
 &  & (7.60) & (8.01) \\
NonRoundVol &  &  & 0.00811\sym{***} \\
 &  &  & (39.35) \\
NonRoundPrice &  &  & 0.00587\sym{***} \\
 &  &  & (7.07) \\
LargeOrder &  &  & -0.00283\sym{***} \\
 &  &  & (-6.14) \\
\midrule
Stock FE & Yes & Yes & Yes \\
Date FE & Yes & Yes & Yes \\
Observations & 10,414,541 & 10,414,541 & 10,414,541 \\
\bottomrule\end{tabular}
}
\par\footnotesize $t$-statistics in parentheses. \sym{*} $p<0.10$, \sym{**} $p<0.05$, \sym{***} $p<0.01$.
\end{table}

\clearpage

\begin{table}[htbp]\singlespacing\footnotesize\renewcommand{\arraystretch}{1.08}\centering
\setlength{\tabcolsep}{1.5pt}
\def\sym#1{\ifmmode^{#1}\else\(^{#1}\)\fi}
\caption{Validation Tests: Market Conditions Associated with Exact-Match Activity.}
\label{tab:order_imbalance_and_spread_channel}

\begin{minipage}{\textwidth}\small
This table reports panel regressions examining whether exact-match activity concentrates in market conditions consistent with latency-sensitive trading. The sample consists of common stocks listed on the Shenzhen Stock Exchange, excluding the smallest 30\% of stocks by prior-day market capitalization. The sample period is July 1, 2023 to October 31, 2025. The analysis uses an approximately 0.1\% systematic sample of limit orders submitted during continuous trading hours, excluding market orders. Each column reports a panel regression with the dependent variable indicated in the column header. The main explanatory variables are order imbalance, measured as the normalized difference between \textit{Bid1Depth} and \textit{Ask1Depth} signed by order direction, and the prevailing bid--ask spread immediately before order submission. Exact-match indicators and other order characteristics follow the definitions in Table~\ref{tab:panel_descriptive_statistics}. Control variables include order direction, lagged market capitalization, turnover rate, and inverse price level. All specifications include stock and date fixed effects, and standard errors are clustered at the stock level. t-statistics are reported in parentheses. The order-level Spread regressor is measured in percent.
\end{minipage}\par\footnotesize
\paperfit{\textwidth}{
\begin{tabular}{lrrrrrrr}\toprule
 & (1) & (2) & (3) & (4) & (5) & (6) & (7) \\
 & RQuoteMatchActive & RQuoteMatchPassive & OneCentLarger & MultiCentLarger & NonRoundVol & NonRoundPrice & LargeOrder \\
\midrule
OrderImbalance & 0.09831\sym{***} & -0.08869\sym{***} & 0.01478\sym{***} & 0.00509\sym{***} & -0.00660\sym{***} & 0.00039\sym{***} & -0.00135\sym{***} \\
 & (48.77) & (-48.14) & (62.07) & (22.86) & (-45.56) & (8.26) & (-29.23) \\
Spread & -0.59515\sym{***} & -0.08386\sym{***} & -0.05667\sym{***} & 0.08206\sym{***} & -0.01360\sym{***} & -0.02367\sym{***} & 0.01320\sym{***} \\
 & (-16.16) & (-13.04) & (-15.01) & (10.32) & (-11.66) & (-10.10) & (15.57) \\
BuyOrder & -0.01000\sym{***} & 0.01073\sym{***} & -0.00164\sym{***} & -0.00044\sym{***} & -0.00155\sym{***} & 0.00187\sym{***} & -0.00256\sym{***} \\
 & (-34.91) & (24.67) & (-17.40) & (-3.47) & (-6.78) & (17.14) & (-29.43) \\
LaggedLogSize & -0.06890\sym{***} & -0.08372\sym{***} & 0.00660\sym{***} & 0.05253\sym{***} & 0.05473\sym{***} & -0.03745\sym{***} & 0.01704\sym{***} \\
 & (-22.99) & (-28.32) & (4.90) & (40.89) & (17.39) & (-17.65) & (12.19) \\
LaggedTurnoverRate & -0.38027\sym{***} & -0.53609\sym{***} & 0.14060\sym{***} & 0.31089\sym{***} & -0.17532\sym{***} & -0.04293\sym{***} & 0.02293\sym{***} \\
 & (-37.77) & (-50.44) & (34.10) & (45.33) & (-15.59) & (-13.94) & (13.50) \\
LaggedInverseClosePrice & 0.42868\sym{***} & 0.08303\sym{***} & 0.01726\sym{**} & 0.13375\sym{***} & 0.23045\sym{***} & -0.10620\sym{***} & 0.05449\sym{***} \\
 & (10.30) & (4.67) & (2.21) & (9.61) & (7.62) & (-8.79) & (8.12) \\
\midrule
Stock FE & Yes & Yes & Yes & Yes & Yes & Yes & Yes \\
Date FE & Yes & Yes & Yes & Yes & Yes & Yes & Yes \\
Observations & 42,506,216 & 42,506,216 & 42,506,216 & 42,506,216 & 42,506,216 & 42,506,216 & 42,506,216 \\
\bottomrule\end{tabular}
}
\par\footnotesize $t$-statistics in parentheses. \sym{*} $p<0.10$, \sym{**} $p<0.05$, \sym{***} $p<0.01$.
\end{table}

\clearpage

\begin{table}[htbp]\singlespacing\footnotesize\renewcommand{\arraystretch}{0.9}\centering
\def\sym#1{\ifmmode^{#1}\else\(^{#1}\)\fi}
\caption{Short-Term Return Predictability of Exact-Match Orders.}
\label{tab:matched_orders_predict_future_price}
\begin{minipage}{\textwidth}
\small
This table reports panel regressions examining whether exact-match activity predicts short-horizon price movements. The sample consists of common stocks listed on the Shenzhen Stock Exchange, excluding the smallest 30\% of stocks by prior-day market capitalization. The sample period is July 1, 2023 to October 31, 2025. The analysis uses an approximately 0.1\% systematic sample of limit orders submitted during continuous trading hours, excluding market orders. The dependent variable, \textit{Ret10s}, is the direction-adjusted forward mid-quote return measured from the first valid midpoint after order submission to the last valid midpoint at or before ten seconds after submission. \textit{RQuoteMatchActive} (\textit{RQuoteMatchPassive}) equals one if an order is submitted at the best opposite-side (same-side) quote reconstructed from millisecond-level order-book data. Control variables include standard order characteristics and prevailing stock characteristics at the time of order submission. All specifications include stock and date fixed effects, and standard errors are clustered at the stock level. t-statistics are reported in parentheses.
\end{minipage}\par\footnotesize
\paperfit{\textwidth}{
\begin{tabular}{lrrr}\toprule
 & (1) & (2) & (3) \\
 & Ret10s & Ret10s & Ret10s \\
\midrule
RQuoteMatchActive & 2.73230\sym{***} & 2.31494\sym{***} & 2.15714\sym{***} \\
 & (67.70) & (41.71) & (39.20) \\
RQuoteMatchPassive & 0.35130\sym{***} & -0.06503\sym{**} & -0.20812\sym{***} \\
 & (19.04) & (-2.21) & (-6.83) \\
BuyOrder & 0.47894\sym{***} & 0.47825\sym{***} & 0.48232\sym{***} \\
 & (42.36) & (42.40) & (42.68) \\
LaggedLogSize & 1.26375\sym{***} & 1.34368\sym{***} & 1.25839\sym{***} \\
 & (9.79) & (10.31) & (9.77) \\
LaggedTurnoverRate & 9.82431\sym{***} & 10.32051\sym{***} & 10.36744\sym{***} \\
 & (26.49) & (26.96) & (26.99) \\
LaggedInverseClosePrice & 7.97731\sym{***} & 8.48126\sym{***} & 8.20674\sym{***} \\
 & (10.53) & (10.86) & (10.74) \\
OneCentLarger &  & -0.32212\sym{***} & -0.43477\sym{***} \\
 &  & (-4.62) & (-6.21) \\
MultiCentLarger &  & -2.71343\sym{***} & -2.82843\sym{***} \\
 &  & (-26.35) & (-27.33) \\
NonRoundVol &  &  & 0.95048\sym{***} \\
 &  &  & (74.51) \\
NonRoundPrice &  &  & 0.03053 \\
 &  &  & (0.62) \\
LargeOrder &  &  & 1.00205\sym{***} \\
 &  &  & (30.15) \\
\midrule
Stock FE & Yes & Yes & Yes \\
Date FE & Yes & Yes & Yes \\
Observations & 42,465,539 & 42,465,539 & 42,465,539 \\
\bottomrule\end{tabular}
}
\par\footnotesize $t$-statistics in parentheses. \sym{*} $p<0.10$, \sym{**} $p<0.05$, \sym{***} $p<0.01$.
\end{table}

\clearpage

\begin{table}[p]\centering\singlespacing\footnotesize
\caption{Return Predictability Across Horizons.}\label{tab:horizons}
\begin{minipage}{\textwidth}\small This table reports return predictability of exact-match orders across horizons from 100 milliseconds to ten minutes. All nine horizons use the same sample of 39,509,310 limit orders, 2,760 stocks, and 566 trading days. Returns are direction-adjusted mid-quote returns measured from the first valid midpoint after submission to the last valid midpoint at or before the specified horizon. All specifications include the full set of order and stock controls, stock and date fixed effects, and stock-clustered standard errors. Returns are reported in basis points, and t-statistics are shown in parentheses.\end{minipage}\par\medskip
\setlength{\tabcolsep}{2pt}
\renewcommand{\arraystretch}{1.08}
\paperfit{\linewidth}{\begin{tabular}{lrrrrrrrrr}\toprule
 & 100ms & 1s & 5s & 10s & 30s & 60s & 120s & 300s & 600s \\\midrule
RQuoteMatchActive & 1.671$^{***}$ & 2.284$^{***}$ & 2.423$^{***}$ & 2.268$^{***}$ & 2.495$^{***}$ & 3.205$^{***}$ & 3.647$^{***}$ & 3.625$^{***}$ & 3.747$^{***}$ \\
 & (81.12) & (80.72) & (56.59) & (40.66) & (31.97) & (46.00) & (61.15) & (51.45) & (48.38) \\
RQuoteMatchPassive & 0.214$^{***}$ & 0.398$^{***}$ & 0.116$^{***}$ & -0.180$^{***}$ & -0.235$^{***}$ & 0.148$^{***}$ & 0.367$^{***}$ & 0.276$^{***}$ & 0.333$^{***}$ \\
 & (39.35) & (44.08) & (5.98) & (-5.96) & (-4.82) & (3.42) & (10.24) & (5.60) & (5.49) \\
OneCentLarger & 0.629$^{***}$ & 1.039$^{***}$ & 0.399$^{***}$ & -0.464$^{***}$ & -1.369$^{***}$ & -0.589$^{***}$ & 0.275$^{***}$ & 0.105 & 0.158 \\
 & (62.97) & (52.38) & (9.32) & (-6.54) & (-11.55) & (-5.82) & (3.47) & (1.05) & (1.37) \\
MultiCentLarger & 0.300$^{***}$ & 0.435$^{***}$ & -1.367$^{***}$ & -2.975$^{***}$ & -4.432$^{***}$ & -3.221$^{***}$ & -2.081$^{***}$ & -2.695$^{***}$ & -2.723$^{***}$ \\
 & (37.39) & (25.52) & (-23.52) & (-28.25) & (-27.06) & (-26.65) & (-23.64) & (-23.83) & (-21.92) \\
BuyOrder & 0.070$^{***}$ & 0.245$^{***}$ & 0.476$^{***}$ & 0.495$^{***}$ & 0.587$^{***}$ & 1.259$^{***}$ & 2.005$^{***}$ & 2.996$^{***}$ & 3.649$^{***}$ \\
 & (32.75) & (56.85) & (54.63) & (40.84) & (20.41) & (24.91) & (24.32) & (21.77) & (18.78) \\
NonRoundVol & 0.312$^{***}$ & 0.445$^{***}$ & 0.770$^{***}$ & 0.976$^{***}$ & 1.180$^{***}$ & 1.164$^{***}$ & 1.196$^{***}$ & 1.330$^{***}$ & 1.291$^{***}$ \\
 & (65.53) & (71.55) & (81.31) & (73.90) & (55.69) & (48.74) & (46.01) & (37.39) & (28.76) \\
NonRoundPrice & -0.115$^{***}$ & -0.178$^{***}$ & -0.060$^{**}$ & 0.065 & 0.358$^{***}$ & 0.431$^{***}$ & 0.425$^{***}$ & 0.488$^{***}$ & 0.532$^{***}$ \\
 & (-18.39) & (-16.09) & (-2.15) & (1.31) & (3.71) & (4.96) & (4.87) & (3.97) & (3.52) \\
LargeOrder & -0.037$^{***}$ & 0.074$^{***}$ & 0.537$^{***}$ & 1.070$^{***}$ & 1.791$^{***}$ & 1.720$^{***}$ & 1.534$^{***}$ & 1.340$^{***}$ & 1.170$^{***}$ \\
 & (-3.45) & (4.84) & (21.62) & (30.44) & (27.19) & (21.61) & (16.00) & (10.38) & (7.31) \\
LaggedLogSize & -0.009 & 0.118$^{***}$ & 0.683$^{***}$ & 1.290$^{***}$ & 2.501$^{***}$ & 2.319$^{***}$ & 1.690$^{***}$ & 1.718$^{***}$ & 1.571$^{***}$ \\
 & (-0.63) & (5.02) & (9.42) & (9.59) & (8.28) & (7.50) & (8.46) & (8.21) & (6.44) \\
LaggedTurnoverRate & -1.345$^{***}$ & -0.381$^{***}$ & 5.551$^{***}$ & 10.469$^{***}$ & 13.260$^{***}$ & 6.915$^{***}$ & 2.834$^{***}$ & 5.448$^{***}$ & 3.654$^{***}$ \\
 & (-27.65) & (-5.01) & (26.01) & (26.90) & (15.41) & (8.65) & (4.31) & (6.24) & (3.75) \\
LaggedInverseClosePrice & 1.776$^{***}$ & 2.626$^{***}$ & 5.750$^{***}$ & 8.469$^{***}$ & 13.526$^{***}$ & 14.450$^{***}$ & 13.451$^{***}$ & 14.917$^{***}$ & 16.642$^{***}$ \\
 & (8.79) & (8.50) & (10.97) & (10.66) & (9.88) & (11.10) & (12.62) & (11.37) & (9.87) \\
\midrule
Stock FE & Yes & Yes & Yes & Yes & Yes & Yes & Yes & Yes & Yes \\
Date FE & Yes & Yes & Yes & Yes & Yes & Yes & Yes & Yes & Yes \\
Observations & 39,509,310 & 39,509,310 & 39,509,310 & 39,509,310 & 39,509,310 & 39,509,310 & 39,509,310 & 39,509,310 & 39,509,310 \\
Stocks & 2,760 & 2,760 & 2,760 & 2,760 & 2,760 & 2,760 & 2,760 & 2,760 & 2,760 \\
Trading days & 566 & 566 & 566 & 566 & 566 & 566 & 566 & 566 & 566 \\
\bottomrule\end{tabular}}
\par\smallskip $^{*}p<0.10$, $^{**}p<0.05$, $^{***}p<0.01$.\end{table}

\clearpage

\begin{table}[htbp]\singlespacing\footnotesize\renewcommand{\arraystretch}{0.9}\centering
\def\sym#1{\ifmmode^{#1}\else\(^{#1}\)\fi}
\caption{Return Predictability of Exact-Match Orders by Order Quantity.}
\label{tab:quantity_matched_orders_predict_future_price}
\begin{minipage}{\textwidth}
\small
This table reports panel regressions examining whether the return predictability of active exact-match orders differs by order size relative to displayed depth. The sample consists of common stocks listed on the Shenzhen Stock Exchange, excluding the smallest 30\% of stocks by prior-day market capitalization. The sample period is July 1, 2023 to October 31, 2025. The analysis uses an approximately 0.1\% systematic sample of limit orders submitted during continuous trading hours, excluding market orders. The dependent variable, \textit{Ret10s}, is the direction-adjusted forward mid-quote return measured from the first valid midpoint after submission to the last valid midpoint at or before ten seconds after submission. \textit{RQuoteMatchActiveVolLess}, \textit{RQuoteMatchActiveVolMatch}, and \textit{RQuoteMatchActiveVolMore} indicate active exact-match orders whose size is below, equal to, or above the displayed depth at the matched opposite-side quote, respectively. \textit{RQuoteMatchPassive} equals one if an order is submitted at the best same-side quote. Other order controls follow Table~\ref{tab:matched_orders_predict_future_price}. Stock-level controls include lagged log market capitalization, lagged turnover rate, and lagged inverse closing price. All specifications include stock and date fixed effects, and standard errors are clustered at the stock level. t-statistics are reported in parentheses.
\end{minipage}\par\footnotesize
\paperfit{\textwidth}{
\begin{tabular}{lrr}\toprule
 & (1) & (2) \\
 & Ret10s & Ret10s \\
\midrule
RQuoteMatchActiveVolMatch & 2.41571\sym{***} & 1.81835\sym{***} \\
 & (73.62) & (44.27) \\
RQuoteMatchActiveVolLess & 3.25895\sym{***} & 2.71051\sym{***} \\
 & (71.85) & (45.17) \\
RQuoteMatchActiveVolMore & -3.81442\sym{***} & -4.40834\sym{***} \\
 & (-67.38) & (-68.39) \\
RQuoteMatchPassive & 0.37896\sym{***} & -0.15504\sym{***} \\
 & (20.71) & (-5.02) \\
BuyOrder & 0.47303\sym{***} & 0.47788\sym{***} \\
 & (41.89) & (42.32) \\
LaggedLogSize & 1.27486\sym{***} & 1.27086\sym{***} \\
 & (9.87) & (9.84) \\
LaggedTurnoverRate & 9.95466\sym{***} & 10.48332\sym{***} \\
 & (26.94) & (27.42) \\
LaggedInverseClosePrice & 7.76165\sym{***} & 7.98202\sym{***} \\
 & (10.38) & (10.56) \\
OneCentLarger &  & -0.43101\sym{***} \\
 &  & (-6.14) \\
MultiCentLarger &  & -2.83960\sym{***} \\
 &  & (-27.41) \\
NonRoundVol &  & 0.80777\sym{***} \\
 &  & (64.10) \\
NonRoundPrice &  & 0.08209\sym{*} \\
 &  & (1.69) \\
LargeOrder &  & 1.75957\sym{***} \\
 &  & (46.53) \\
\midrule
Stock FE & Yes & Yes \\
Date FE & Yes & Yes \\
Observations & 42,465,539 & 42,465,539 \\
\bottomrule\end{tabular}
}
\par\footnotesize $t$-statistics in parentheses. \sym{*} $p<0.10$, \sym{**} $p<0.05$, \sym{***} $p<0.01$.
\end{table}

\clearpage

\begin{table}[htbp]\singlespacing\footnotesize\renewcommand{\arraystretch}{0.9}\centering
\caption{Covariate Balance Test.}\label{tab:balance_smd}
\begin{minipage}{\textwidth}\small
This table reports covariate balance between the fast and slow latency groups used in the final DID design. The sample contains 48,188 stock-days from the 24-day censored active-count regression, with one observation per stock-day. Fast and slow denote the latency groups used in the DID estimation. Market variables are measured using the preceding trading day: log size is $\log(1+\text{market capitalization})$, trading amount is measured in billion yuan, turnover is trading amount divided by capitalization, inverse price is the inverse closing price, and 20-day return and volatility measures are based on daily excess returns relative to the value-weighted market. Financial ratios use the most recent disclosures available before each observation. Diff./SD is the fast-minus-slow mean difference scaled by the square root of the average group variance.
\end{minipage}\par\footnotesize
\paperfit{\textwidth}{\begin{tabular}{lrrrrrrr}\toprule
 & Fast mean & Fast median & Slow mean & Slow median & Difference & SE & Diff./SD \\
\midrule
Log(Size) & 23.1255 & 22.9028 & 23.1281 & 22.8953 & -0.0026 & 0.0078 & -0.0030 \\
Trading Amount & 0.5076 & 0.2030 & 0.5231 & 0.2060 & -0.0155 & 0.0107 & -0.0132 \\
Turnover Rate & 0.0327 & 0.0201 & 0.0330 & 0.0200 & -0.0002 & 0.0004 & -0.0057 \\
IVOL (20-Day) & 0.0246 & 0.0209 & 0.0247 & 0.0208 & -0.0001 & 0.0001 & -0.0054 \\
Excess Return (20-Day) & -0.0071 & -0.0246 & -0.0081 & -0.0258 & 0.0010 & 0.0012 & 0.0076 \\
Inverse Price & 0.0871 & 0.0604 & 0.0881 & 0.0602 & -0.0010 & 0.0008 & -0.0121 \\
Gross Margin & 0.2799 & 0.2510 & 0.2765 & 0.2473 & 0.0033\sym{**} & 0.0017 & 0.0182 \\
Revenue Growth & 0.0822 & 0.0437 & 0.0834 & 0.0451 & -0.0012 & 0.0040 & -0.0028 \\
Leverage & 0.4221 & 0.4055 & 0.4246 & 0.4104 & -0.0025 & 0.0019 & -0.0120 \\
\bottomrule\end{tabular}}\par\footnotesize \sym{*} $p<0.10$, \sym{**} $p<0.05$, \sym{***} $p<0.01$.
\end{table}

\clearpage

% \input{tables/did_on_market_quality}
% \clearpage

\begin{table}[p]\centering\singlespacing\footnotesize
\caption{Latency Crossings and Exact-Match Activity.}\label{tab:did_latency_in_3s}
\begin{minipage}{\textwidth}\small This table reports difference-in-differences estimates of the effect of improved information timeliness on exact-match activity. The design contains 24 daily latency-crossing episodes. For each day, stock-second observations are aligned around that day's fast-group latency crossing. The sample uses $G^F_t-10 \leq s < G^S_t$ and $G^S_t-G^F_t>1$, with stock $\times$ date and date $\times$ calendar-second fixed effects. Standard errors are clustered two ways by stock and date $\times$ calendar-second. Outcomes include active exact-match counts and active and passive exact-match shares; shares are reported in percentage points (pp). In the corresponding event studies, the omitted event period is $s-G^F_t=-1$.\end{minipage}\par\medskip
\begin{tabular}{llrrrr}\toprule
Outcome & Unit & Coefficient & $t$ & $N$ & Pre-event $p$ \\\midrule
Active count & orders & 0.44718$^{***}$ & 6.95 & 687,663 & 0.028 \\
Active share & pp & 0.50539$^{***}$ & 3.03 & 685,258 & 0.228 \\
Passive share & pp & 0.51496$^{***}$ & 3.07 & 685,258 & 0.038 \\
\bottomrule\end{tabular}
\par\smallskip\footnotesize $^{*}p<0.10$, $^{**}p<0.05$, $^{***}p<0.01$.
\end{table}

\clearpage

\begin{table}[p]\centering\singlespacing\footnotesize
\caption{Latency Crossings and Market Quality.}\label{tab:did_market_expanded}
\begin{minipage}{\textwidth}\small This table reports difference-in-differences estimates of the effect of improved information timeliness on market quality outcomes. The design contains 24 daily latency-crossing episodes. For each day, stock-second observations are aligned around that day's fast-group latency crossing. The sample uses $G^F_t-10 \leq s < G^S_t$ and $G^S_t-G^F_t>1$, with stock $\times$ date and date $\times$ calendar-second fixed effects. Standard errors are clustered two ways by stock and date $\times$ calendar-second. Outcomes include quoted liquidity (Spread and LogDepth), price variation (Range and WideSpread), transaction costs (ES, AS1s, and RS1s), and pricing error (PEbps). In the corresponding event studies, the omitted event period is $s-G^F_t=-1$ for quote-based outcomes and PEbps. $^{\dagger}$For transaction-based outcomes (ES, AS1s, and RS1s), $s-G^F_t=-2$ is used because the one-second forward window makes $s-G^F_t=-1$ unavailable. ES, AS1s, and RS1s use the same notional-weighted transaction sample. Price measures are reported in basis points (bp), depth is measured as $\log(1+\text{notional depth in yuan})$, and WideSpread is reported in percentage points (pp).\end{minipage}\par\medskip
\begin{tabular}{llrrrr}\toprule
Outcome & Unit & Coefficient & $t$ & $N$ & Pre-event $p$ \\\midrule
\multicolumn{6}{l}{\textbf{Panel A: Liquidity and Price Variation}} \\
Spread & bp & -0.00330 & -0.05 & 685,371 & 0.004 \\
LogDepth & log & -0.01386$^{*}$ & -1.82 & 686,315 & 0.385 \\
Range & bp & 1.02884$^{***}$ & 10.36 & 685,371 & 0.082 \\
WideSpread & pp & 2.96447$^{***}$ & 8.66 & 685,554 & 0.720 \\
\midrule
\multicolumn{6}{l}{\textbf{Panel B: Transaction Costs and Pricing Error}} \\
ES$^{\dagger}$ & bp & -0.04906 & -0.53 & 482,510 & 0.051 \\
AS1s$^{\dagger}$ & bp & 0.64810$^{***}$ & 8.30 & 482,510 & 0.532 \\
RS1s$^{\dagger}$ & bp & -1.34527$^{***}$ & -8.28 & 482,510 & 0.738 \\
PEbps & bp & -0.04486$^{***}$ & -3.23 & 541,710 & 0.800 \\
\bottomrule\end{tabular}
\par\smallskip\footnotesize $^{*}p<0.10$, $^{**}p<0.05$, $^{***}p<0.01$.
\end{table}

\clearpage

% ============================================================
% Internet Appendix
% ============================================================
\newpage
\appendix

\setcounter{table}{0}
\setcounter{figure}{0}
\setcounter{equation}{0}
\renewcommand{\thetable}{IA.\arabic{table}}
\renewcommand{\thefigure}{IA.\arabic{figure}}
\renewcommand{\theequation}{IA.\arabic{equation}}  % 如有公式

\begin{center}
    \vspace*{2cm}
    {\LARGE \textbf{Internet Appendix}} \\[1em]
\end{center}

\renewcommand{\thesection}{IA.\arabic{section}}
\setcounter{section}{0}

\section*{Part I: Appendix Text}
\section{Return Predictability by Order Direction}\label{app:buy_sell}
We repeat the full-control regressions in Table~\ref{tab:horizons} separately for buy and sell orders, excluding BuyOrder from both specifications. Each directional subsample uses a common set of orders across all nine horizons. Table~\ref{tab:buy_sell_horizons} reports the regression estimates, and Table~\ref{tab:buy_sell_differences} tests sell-minus-buy differences in the active exact-match coefficient, allowing for within-stock dependence across order directions.

Active exact matching predicts direction-adjusted returns in both subsamples at every horizon. The buy coefficient is significantly larger at 100 milliseconds and is larger at the 10\% significance level at one second. The sell coefficient is significantly larger at 5, 10, 60, 120, and 300 seconds. The differences at 30 and 600 seconds are not statistically significant. The asymmetry therefore varies with the prediction horizon rather than uniformly favoring one order direction.

One possible interpretation relates to the asymmetric inventory constraints associated with T+1 trading. Purchases do not require an existing position in the stock, whereas sales generally require inventory eligible for sale or access to stock borrowing. These constraints may affect participation on the two sides of the market and the timing of information incorporation. The stronger buy-side coefficient at 100 milliseconds and stronger sell-side coefficients at several longer horizons are consistent with this interpretation.

\begin{center}[Insert Table~\ref{tab:buy_sell_horizons} here.]\end{center}
\begin{center}[Insert Table~\ref{tab:buy_sell_differences} here.]\end{center}

\clearpage

    \section{Intraday Exact-Match Activity and Returns}\label{app:intraday}    
\begin{figure}[H]
    \centering
    \includegraphics[width=0.9\linewidth]{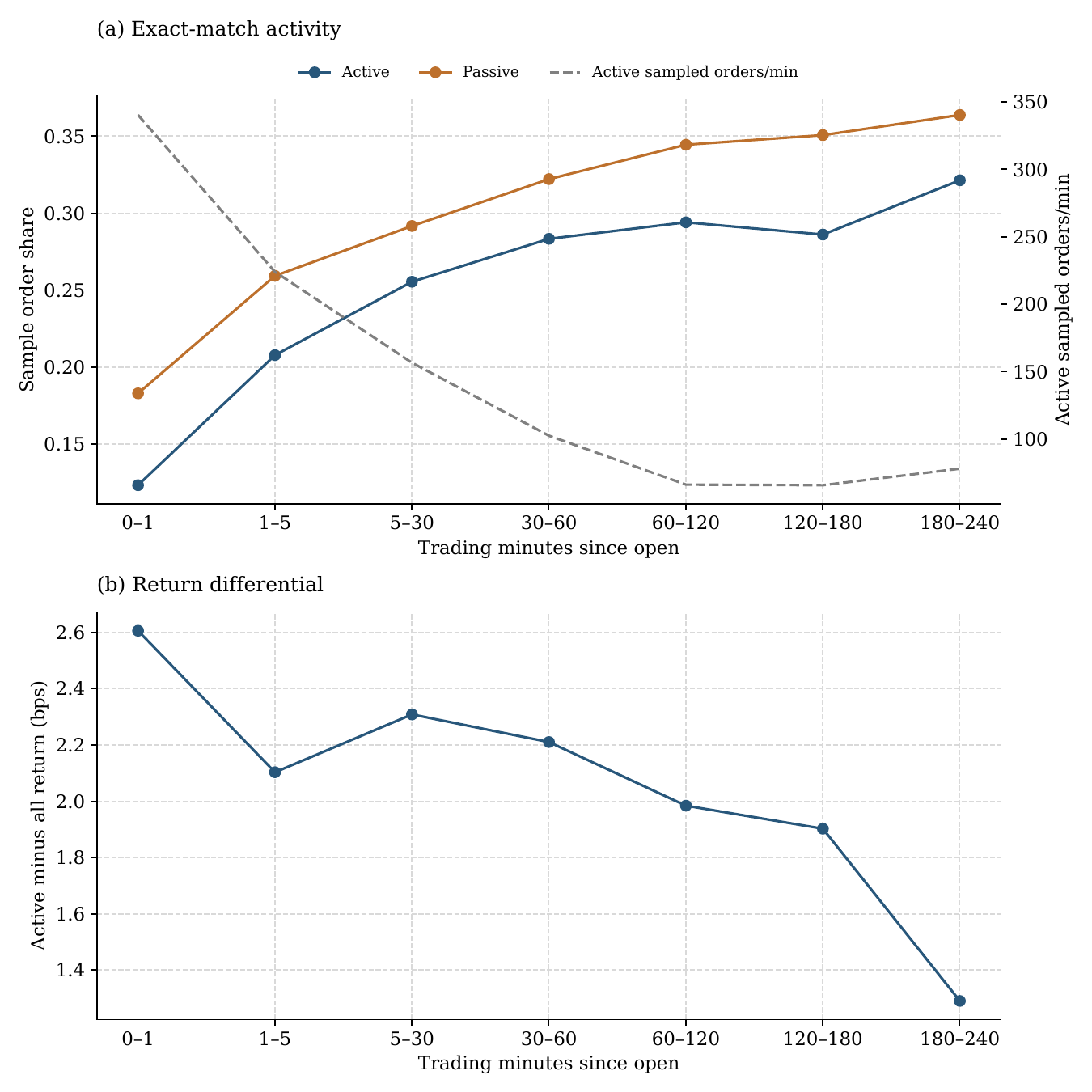}
    \caption{Intraday Exact-Match Activity and Returns}
    \label{fig:market_share_and_return}
    \begin{minipage}{\textwidth}
    \small
    This figure plots intraday order activity, exact-match activity share, and short-horizon returns across trading intervals. The sample includes all common stocks listed on the Shenzhen Stock Exchange, excluding the smallest 30\% by previous-day market capitalization, over the period July 1, 2023 to October 31, 2025. Exact-match orders are classified using their submission prices, and activity share is defined as the fraction of exact-match orders relative to limit-order submissions within each intraday time group.
    Panel~A plots active and passive sample shares, with active sampled orders per minute on the secondary axis. Panel~B plots active-limit-order minus all-limit-order average 10-second returns in basis points. The seven bins are trading minutes since open, with the lunch break removed. The last bin is labeled 180--240 minutes but ends at minute 237 (14:57); its order rate retains a 60-minute denominator. No sampling-rate extrapolation is applied.
    \end{minipage}
\end{figure}

    $$[\text{Insert Figure~\ref{fig:market_share_and_return} here.}]$$
    
    Figure~\ref{fig:market_share_and_return} reports intraday sample shares, observed order rates, and return differentials. In the first minute, there are approximately 340 active exact-match and 505 passive exact-match sampled limit orders per minute; these are not extrapolated full-market counts. Active and passive shares rise from 12.33\% and 18.29\% in the first interval to 32.13\% and 36.37\% in the final interval. The active share is not strictly monotonic over the day.
    
    Panel~B plots the average 10-second return of active exact-match orders minus the average across all sampled limit orders in the same interval. This differential generally declines over the day. The intraday pattern may reflect changes in information arrival, order composition and competitive conditions.

\clearpage

    \section{Return Predictability in Different Regimes}\label{app:regimes}
    
    We examine four sets of market regimes based on recent price dispersion, volatility, and cross-sectional and time-series momentum or reversal. Range and volatility regimes are defined from cross-sectional terciles at each second, whereas the time-series momentum and reversal indicators use fixed direction-adjusted return thresholds.
    
    RngHigh and RngLow use the range of log mid-quotes over the preceding 10 seconds; VolHigh and VolLow use the standard deviation of one-second log returns over a rolling 600-second window, with at least 10 observations. MomCX and RevCX combine terciles of past 10-second returns with order direction to distinguish momentum and reversal orders. MomTS and RevTS use direction-adjusted 10-second return thresholds of $\pm0.3\%$. Quote inputs are lagged one second; missing cross-sectional ranks are assigned to the neutral category. This rolling-range definition differs from the intrasecond range used in the DID analysis.
    
    We estimate regime-dependent return predictability with the following panel regression:
    \begin{equation}
    \label{eq:regime_general}
    \begin{aligned}
    Ret_{i,t+10s}
    = &\ \beta_1 RQuoteMatchActive_{i,t}
    + \beta_2 RQuoteMatchPassive_{i,t} \\
    &+ \sum_{q=1}^{2}\delta_q Regime^{q}_{i,t} \\
    &+ \sum_{q=1}^{2}\theta_q \big(RQuoteMatchActive_{i,t} \times Regime^{q}_{i,t}\big) \\
    &+ \boldsymbol{\gamma}' \mathbf{X}_{i,t}
    + \alpha_i
    + \delta_t
    + \varepsilon_{i,t},
    \end{aligned}
    \end{equation}
    where $Ret_{i,t+10s}$ denotes the 10-second forward mid-quote return. Each specification includes a pair of regime indicators and their interactions with active exact matching: RngHigh and RngLow, VolHigh and VolLow, MomCX and RevCX, or MomTS and RevTS. The reference category consists of observations for which both indicators equal zero. The interaction coefficients measure differences in active exact-match return predictability relative to that category. The vector $\mathbf{X}_{i,t}$ includes order direction, price aggressiveness, order size characteristics, lagged market capitalization, lagged turnover, and lagged inverse price. All regressions include stock fixed effects $\alpha_i$ and date fixed effects $\delta_t$, and standard errors are clustered at the stock level.

    \medskip
    $$[\text{Insert Table \ref{tab:Short-term Return Prediction Power in Regimes} here.}]$$

    Table~\ref{tab:Short-term Return Prediction Power in Regimes}  examines whether the short-term return predictability of quote-matching active orders varies across market regimes. All specifications use 10-second forward mid-quote returns (Ret10s) as the dependent variable and include stock and date fixed effects. The regime variables are dummy indicators, and we interact each regime with \textit{RQuoteMatchActive} to assess whether exact-match return predictability differs across states.
    
    Column (1) focuses on short-term cross-sectional range over the past 10 seconds. High-range periods (RngHigh) are associated with positive subsequent returns, whereas low-range periods (RngLow) predict negative returns. The active quote-matching interaction is larger in low-range regimes (1.887) than in high-range regimes (0.522). This pattern is consistent with stronger predictive differentials when contemporaneous price dispersion is limited.
    
    Column (2) considers cross-sectional volatility over the past 600 seconds. Both high- and low-volatility regimes exhibit significant incremental return predictability for active quote-matching orders, with interaction coefficients of 0.251 and 0.328, respectively. The similar magnitudes suggest that this volatility classification does not sharply differentiate predictive content once other controls are included.
    
    Column (3) evaluates cross-sectional momentum and reversal regimes based on past 10-second returns. MomCX captures trades that follow recent winners (or losers on the sell side), while RevCX captures contrarian trades. The active quote-matching interaction is negative in momentum-following states ($-1.008$) and positive in reversal states (1.208), consistent with stronger predictability when recent price pressure reverses.
    
    Column (4) reports analogous results using time-series thresholds (MomTS and RevTS). The active interaction is negative under momentum continuation ($-1.799$) and positive under reversal (6.482), consistent with the cross-sectional pattern.
        
    The baseline coefficient on \textit{RQuoteMatchActive} is positive and highly significant throughout. The interactions show stronger predictability in reversal states and weaker predictability in momentum states. These patterns are consistent with the informational interpretation of active exact matching.

\clearpage

    \section{Cross-Sectional Determinants and Industry Patterns}\label{app:cross_section}
    
    We provide descriptive evidence on the cross-sectional determinants and industry composition of exact-match activity.
    
    We measure exact-match activity with three stock-day shares. The combined share is \(ExactMatch^{All}_{i,t}\). In the daily formulas below, the matching variables denote order counts rather than individual-order indicators:
    \begin{equation}\label{eq:hft_all}
    ExactMatch^{All}_{i,t}
    = \frac{RQuoteMatchActive_{i,t} + RQuoteMatchPassive_{i,t}}
    {TotalOrderNum_{i,t}} .
    \end{equation}
    
    We separate the share into active and passive components. \(ExactMatch^{Active}_{i,t}\) measures the share of aggressive exact-match orders submitted at the best available opposite-side quote, while \(ExactMatch^{Passive}_{i,t}\) captures the share of passive exact-match orders placed at the best same-side quote:
    \begin{align}\label{eq:hft_active_passive}
    ExactMatch^{Active}_{i,t}
    &= \frac{RQuoteMatchActive_{i,t}}
    {TotalOrderNum_{i,t}}, \\
    ExactMatch^{Passive}_{i,t}
    &= \frac{RQuoteMatchPassive_{i,t}}
    {TotalOrderNum_{i,t}} .
    \end{align}

    To investigate the cross-sectional determinants of exact-match activity, we estimate the following Fama--MacBeth regression at the daily frequency:
    \begin{equation}
    \label{eq:fmb_hft}
    \begin{aligned}
    ExactMatch^{k}_{i,t}
    = \alpha_t
    &+ \beta_{1,t} \log(1+\text{Size}_{i,t-1}) \\
    &+ \beta_{2,t} \text{TradingAmount}_{i,t-1}
    + \beta_{3,t} \text{TurnoverRate}_{i,t-1} \\
    &+ \beta_{4,t} \text{IVOL}_{i,t-1}
    + \beta_{5,t} \text{ExcessReturn}_{i,t-1} \\
    &+ \beta_{6,t} InversePrice_{i,t-1}+\beta_{7,t} GrossMargin_{i,t} \\
    &+ \beta_{8,t}RevenueGrowth_{i,t}+\beta_{9,t}Leverage_{i,t} \\
    &+ \phi_{t}^{A} ExactMatch^{Active}_{i,t-1} \\
    &+ \phi_{t}^{P} ExactMatch^{Passive}_{i,t-1}
    + \varepsilon_{i,t},
    \end{aligned}
    \end{equation}
    where $ExactMatch^{k}_{i,t}$ denotes the daily exact-match activity share for stock $i$ on day $t$, with $k \in \{\text{All}, \text{Active}, \text{Passive}\}$. The explanatory variables include lagged daily market characteristics, inverse price, and the latest financial disclosures published strictly before observation. Industry fixed effects are included in each daily cross section. Log size is computed as $\log(1+\text{market capitalization})$, and IVOL is the standard deviation of daily market-adjusted returns over a 20-trading-day window. ExcessReturn is the sum of daily market-adjusted returns over the same window; both rolling measures are lagged by one trading day. The active and passive lagged shares refer to the preceding selected trading day.
    
    The model is estimated using the standard Fama--MacBeth two-step procedure. In the first step, we run cross-sectional regressions separately for each trading day. In the second step, we report the time-series averages of the daily slope coefficients, and the standard error is the sample standard deviation of daily slopes divided by the square root of 565 valid days, without a Newey--West adjustment.
    
    $$[\text{Insert Table~\ref{tab:fmb_hft_market_quality} here.}]$$
    
\begin{figure}[H]
    \centering
    \includegraphics[width=\linewidth]{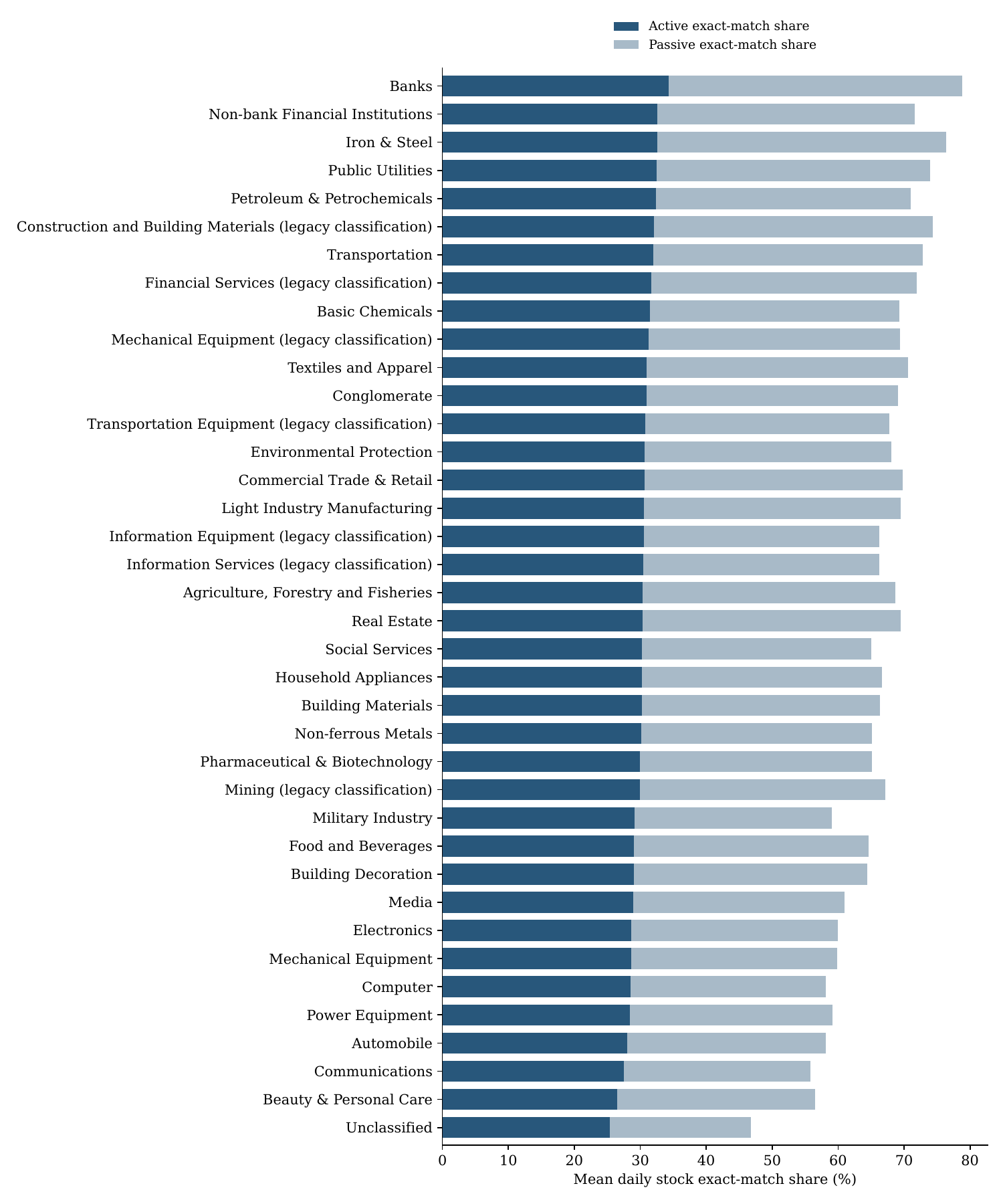}
    \caption{Exact-Match Activity Shares by Industry}
    \label{fig:hft_industry_market_share}
    \begin{minipage}{\textwidth}
    \small
    This figure presents active and passive exact-match shares by industry. Each activity share is first computed at the stock-day level, then averaged to the industry-day level, and finally averaged across days. Industry names follow the source Shenwan dictionary. Retired classification codes remain separate and are labeled as legacy classifications; unclassified observations are also retained.
    \end{minipage}
\end{figure}

    In Table~\ref{tab:fmb_hft_market_quality}, total exact-match share ($ExactMatch^{All}$) and active exact-match share ($ExactMatch^{Active}$) are negatively related to market capitalization, indicating that active exact-match activity is relatively more concentrated in smaller-cap stocks. Passive exact-match activity ($ExactMatch^{Passive}$), by contrast, is positively related to size, consistent with passive quote matching being more common in the deeper, more liquid order books of large-cap stocks. 
    Idiosyncratic volatility ($IVOL$) is strongly negatively associated with exact-match activity across all three measures. This pattern is consistent with stable microstructure conditions favoring exact matching, while volatility increases adverse-selection risk and uncertainty about short-horizon prices. 
    Past 20-day excess returns are positively related to total and passive exact-match share, indicating more exact matching in recent outperformers. This pattern is consistent with more order-flow opportunities in stocks with positive recent returns. 
    Lagged exact-match shares are strongly persistent, consistent with stable stock-level trading patterns. 
    GrossMargin and RevenueGrowth have statistically significant coefficients in all three specifications. Leverage is significant in the all-order and active-share specifications, but not in the passive-share specification. These conditional associations, together with the persistence of matching shares, provide descriptive context for the latency analysis.

    $$[\text{Insert Figure~\ref{fig:hft_industry_market_share} here.}]$$
    
    Figure~\ref{fig:hft_industry_market_share} plots exact-match activity share across industries, decomposing total exact-match activity into active exact-match activity ($ExactMatch^{Active}$) and passive exact-match activity ($ExactMatch^{Passive}$). For each industry, the total length of the stacked bar represents the overall exact-match activity share, while the two segments show the relative contributions of active and passive exact matching.
    
    Exact-match shares differ across industries, and passive quote-matching shares exceed active shares in most classified industries. The pattern complements the daily cross-sectional associations in Table~\ref{tab:fmb_hft_market_quality}. Unclassified industry codes are displayed separately.

\clearpage

    \section{Return Predictability Robustness}\label{app:robustness}
    
    We report four robustness exercises covering the estimation framework, the 120-second native sample, lower-frequency snapshot information, and response speed.
    
    We first re-estimate the baseline with the Fama--MacBeth two-step procedure. 
    We retain the last sampled order in each stock--minute to form the one-minute cross sections. 
    In the first step, we run cross-sectional regressions of future returns on the HFT identification variables and controls for each minute. 
    In the second step, we compute time-series averages of the estimated coefficients and report \textit{t}-statistics using Newey--West covariance with 19 lags across minute regressions. 
    Table~\ref{tab:FMB_MatchedOrders_Ret10s} reports the Fama--MacBeth estimates for 10-second forward mid-quote returns. 
    The coefficient on $RQuoteMatchActive$ remains positive and highly significant across specifications, with magnitudes comparable to those obtained from the order-level panel regressions. 
    $RQuoteMatchPassive$ also loads positively but with substantially smaller economic significance. 
    Active matching therefore predicts short-horizon returns in the minute-level cross sections under this least-squares and covariance specification.
    
    Second, we report 120-second native-sample robustness. Unlike the common-sample horizon analysis, this regression requires a valid return only at 120 seconds and includes 41,861,959 orders rather than 39,509,310. 
    Column (1) of Table~\ref{tab:matched_orders_predict_future_price_robustness} re-estimates the baseline specification using 120-second forward mid-quote returns. 
    The coefficient on $RQuoteMatchActive$ remains positive and highly significant, indicating that active exact-match orders continue to predict favorable price movements over a longer horizon.
    
    Third, we examine whether the predictive power of reconstructed-quote matching simply reflects order placement at prices visible in lower-frequency tick quotes. 
    We define tick-quote matching indicators as follows:
    \begin{align}
    \mathit{TickQuoteMatchActive} 
    &= \mathbf{1}\{\text{order price} = \text{best price on the opposite side of three-second tick quotes}\},\\
    \mathit{TickQuoteMatchPassive} 
    &= \mathbf{1}\{\text{order price} = \text{best price on the same side of three-second tick quotes}\}.
    \end{align}
    We also define indicators for orders that match both reconstructed quotes and three-second tick quotes:
    \begin{align}
    \mathit{RQuoteTickMatchActive} 
    &= \mathit{RQuoteMatchActive} \times \mathit{TickQuoteMatchActive},\\
    \mathit{RQuoteTickMatchPassive} 
    &= \mathit{RQuoteMatchPassive} \times \mathit{TickQuoteMatchPassive}.
    \end{align}
    Column (2) of Table~\ref{tab:matched_orders_predict_future_price_robustness} reports the 10-second return regression with these tick-quote matching variables. 
    The coefficient on $RQuoteMatchActive$ remains positive and highly significant after controlling for tick-quote matching variables. 
    By contrast, the coefficients on $TickQuoteMatchActive$ and $TickQuoteMatchPassive$ are negative, and the coefficients on $RQuoteTickMatchActive$ and $RQuoteTickMatchPassive$ are also negative. 
    The predictive content is concentrated in matches to the millisecond-level reconstructed book, beyond matching prices visible in lower-frequency tick quotes.
    
    Fourth, we test whether active quote matching captures more than raw response speed. 
    We first define $ResponseIn100ms$ as an indicator equal to one if an order is submitted less than 100 milliseconds after the latest prior best-quote price or quantity change.
    We then interact this indicator with reconstructed-quote matching:
    \begin{align}
    \mathit{RQuoteMatchActive100ms} 
    &= \mathit{RQuoteMatchActive} \times \mathit{ResponseIn100ms},\\
    \mathit{RQuoteMatchPassive100ms} 
    &= \mathit{RQuoteMatchPassive} \times \mathit{ResponseIn100ms}.
    \end{align}
    Column (3) controls for whether an order is submitted less than 100 milliseconds after the latest prior best-quote price or quantity change and further separates reconstructed-quote matched orders submitted within this ultra-short response window. 
    The coefficient on $ResponseIn100ms$ is positive and significant. The interaction $RQuoteMatchActive100ms$ is also positive and highly significant after including these controls. 
    Active exact matching adds predictive content beyond response speed, consistent with processing current order-book information and choosing the prevailing best quote.
    
    The estimates support active quote-matching predictability under the alternative estimation framework, longer horizon, and response-speed controls. The snapshot comparison further suggests that this predictive content is not explained solely by matching prices visible in lower-frequency tick quotes. 
    These results support the interpretation of active exact matching as a signature of latency-sensitive processing of the reconstructed book.
\begin{center}[Insert Table~\ref{tab:FMB_MatchedOrders_Ret10s} here.]\end{center}
\begin{center}[Insert Table~\ref{tab:matched_orders_predict_future_price_robustness} here.]\end{center}

\clearpage

\section{DID Results with Alternative Clustering}\label{app:did_clustering}
Table~\ref{tab:did_stock_date} reports the DID estimates with stock $\times$ date clustering. The sample, outcome definitions, fixed effects and coefficient estimates are identical to the main tables; only the clustering rule changes. The main analysis uses two-way clustering by stock and date $\times$ calendar-second.

\begin{center}[Insert Table~\ref{tab:did_stock_date} here.]\end{center}

\clearpage

\section{Exclusion Tests: Market Orders and More Aggressive Limit Orders}\label{app:exclusion_tests}

We examine how market orders and more aggressively priced limit orders respond to the latency improvement. These comparisons help assess whether the increase in active exact matching reflects a broader increase in aggressive order submission. Market orders comprise opposite-side-best and same-side-best instructions, which do not require traders to specify a numerical limit price. OneCentLarger and MultiCentLarger identify limit orders priced exactly one tick and more than one tick beyond the prevailing best opposite-side quote, respectively.

All three measures use the number of limit orders with valid two-sided quotes in the same stock-second as their denominator, consistent with the exact-match measures in the main analysis. MarketOrderShare is therefore a market-to-limit order ratio. OneCentLargerShare and MultiCentLargerShare are fractions of eligible limit orders. We multiply all three measures by 100 for reporting.

We retain the baseline three-second latency-crossing design, including the 24 trading days, fast and slow groups, and censored estimation window. All specifications include stock $\times$ date and date $\times$ calendar-second fixed effects, with two-way clustering by stock and date $\times$ calendar-second.

\begin{center}[Insert Table~\ref{tab:exclusion_tests} here.]\end{center}

Table~\ref{tab:exclusion_tests} shows that MultiCentLargerShare falls by 0.331 percentage points after the latency improvement. The decline remains significant at the 1\% level after Holm adjustment across the three outcomes ($p=0.0014$). OneCentLargerShare increases by 0.062 percentage points, but the estimate is statistically insignificant. MarketOrderShare falls by 0.060 percentage points. This decline is significant at the 10\% level before adjustment ($p=0.0956$), but not after adjustment ($p=0.1913$).

\begin{figure}[H]\centering
\includegraphics[width=.92\linewidth]{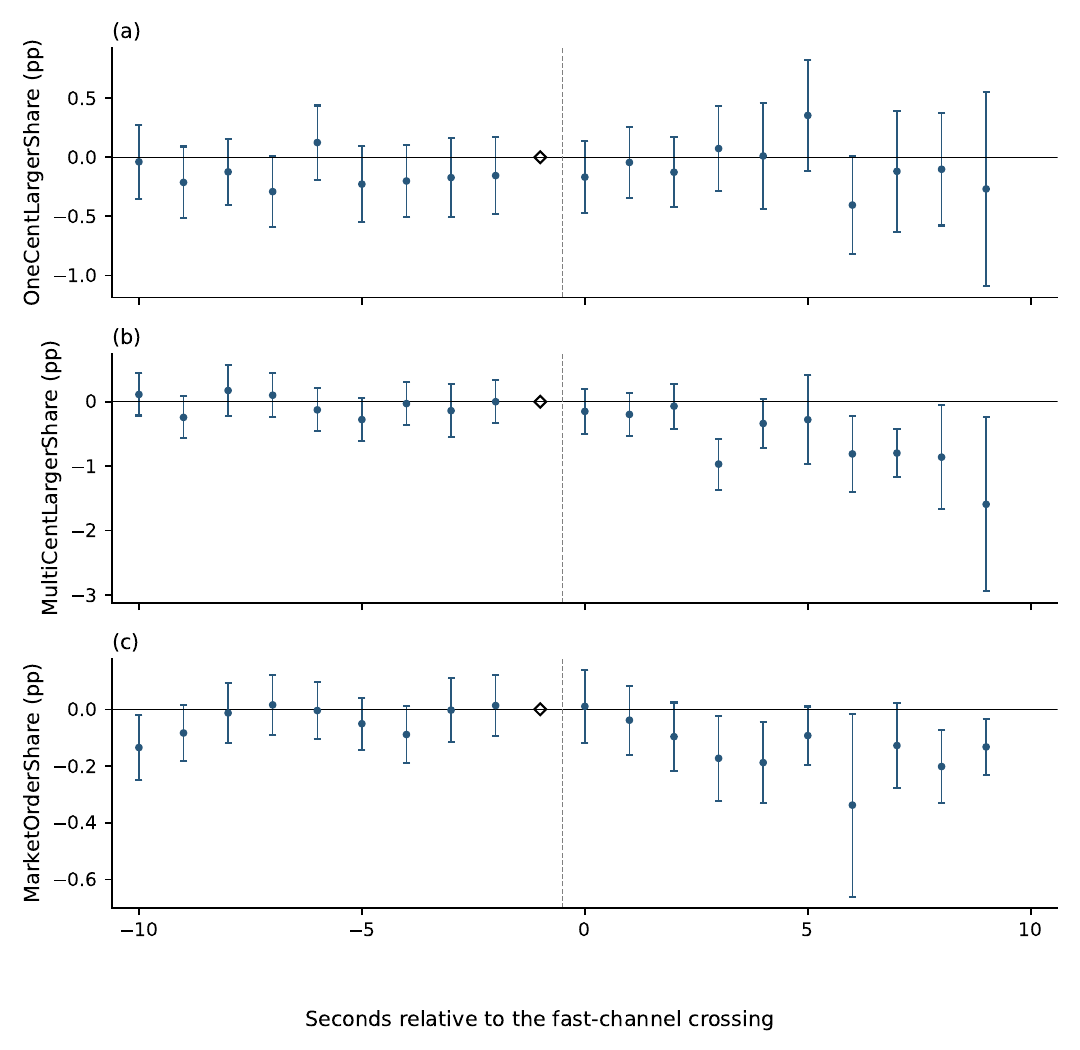}
\caption{Event Studies: Market Orders and More Aggressive Limit Orders.}\label{fig:exclusion_tests}
\begin{minipage}{\textwidth}\small This figure presents event-time estimates around the fast-group latency crossing. Panels (a)--(c) report OneCentLargerShare, MultiCentLargerShare, and MarketOrderShare. Each outcome divides the relevant order count by the number of limit orders with valid two-sided quotes in the same stock-second and multiplies the ratio by 100. MarketOrderShare is a market-to-limit order ratio. The sample contains 685,258 stock-second observations across 24 trading days and uses $G^F_t-10\leq s<G^S_t$, with $G^S_t-G^F_t>1$, excluding the smallest 30\% of stocks by prior-day capitalization. All specifications include stock $\times$ date and date $\times$ calendar-second fixed effects, with two-way clustering by stock and date $\times$ calendar-second. Bars indicate 90\% pointwise confidence intervals; open diamonds mark the omitted period $s-G^F_t=-1$. Only supported event-time bins within $\pm10$ seconds are displayed. Outcomes are reported in percentage points (pp).\end{minipage}
\end{figure}

Figure~\ref{fig:exclusion_tests} presents the corresponding event-time estimates. Joint tests of the nine pre-event coefficients do not reject for any of the three outcomes, with $p$-values ranging from 0.249 to 0.325.

\begin{center}[Insert Figure~\ref{fig:exclusion_tests} here.]\end{center}

The decline in MultiCentLargerShare, together with the increase in active exact-match shares in Table~\ref{tab:did_latency_in_3s}, supports a shift toward more precise price placement. Neither of the other two order categories shows a statistically significant increase. These findings support the proposed latency-sensitive trading channel, while leaving open other channels through which information timeliness may affect market quality.

\clearpage

\section{Fixed-Grid Wide-Spread Share}\label{app:wide_share}
For each stock-second, sample the prevailing best bid and ask at 0, 100, \ldots, 900 milliseconds after its start. Define
\begin{equation}
 WideSpreadShare_{ids}=\frac{1}{10}\sum_{j=0}^{9}1\{Ask_{ids,j}-Bid_{ids,j}>\text{one tick}\}.
\end{equation}
The tick is 0.01 yuan. At each grid point use the last quote at or before the timestamp, resolving equal timestamps by event order; a valid prior-second state may carry forward. All ten states must have positive, finite, noncrossed quotes. Otherwise the outcome is missing; the denominator is never reduced to the number of valid grid points.

This measure records the fraction of fixed grid points with a wide quote and ranges from zero to one. Estimation uses this fraction; coefficients, standard errors, and confidence intervals are multiplied by 100 for reporting in percentage points. The main WideSpread indicator instead records whether a wide state occurred at any time. The measure captures wide-quote incidence on the fixed sampling grid. The candidate panel contains 18,993 observations with WideSpread equal to one but WideSpreadShare equal to zero. No quote-update-count or continuous-duration outcome is added to the DID analysis.

\begin{figure}[H]\centering
\includegraphics[width=.85\linewidth]{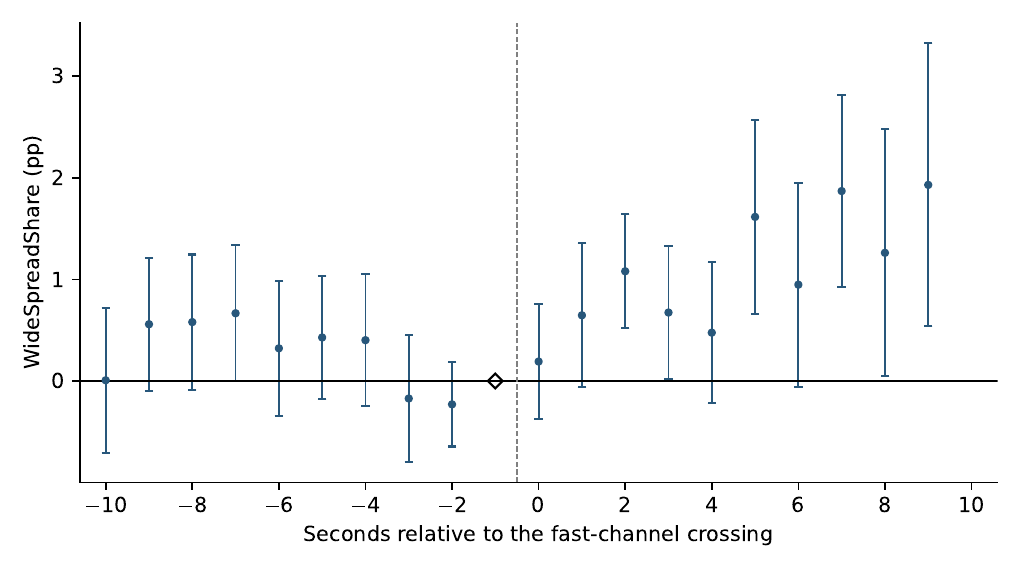}
\caption{Event Study: WideSpreadShare}\label{fig:wide_share}
\begin{minipage}{\textwidth}\small The outcome is in percentage points. Estimates use the same 24-day censored design as the main analysis, with stock $\times$ date and date-by-second fixed effects and two-way clustering by stock and date $\times$ calendar-second. Bars are 90\% pointwise confidence intervals; the omitted second is $-1$. Supported event-time estimates within $[-10,10]$ seconds are displayed. Later bins have fewer supporting dates.\end{minipage}
\end{figure}

The sample, fixed effects and crossing design match the main results. Two-way clustering yields an increase of 0.520 percentage points ($p=0.0527$), with nine-lead pre-event $p=0.275$. The estimate reaches 5\% with stock $\times$ date clustering ($p=0.0437$), but not under the main two-way rule.
\begin{center}[Insert Table~\ref{tab:wide_share} here.]\end{center}
\begin{center}[Insert Figure~\ref{fig:wide_share} here.]\end{center}

\clearpage

\section{Pricing Error Construction and Diagnostics}\label{app:pe}
\subsection{Data, historical estimation and daily boundaries}
We construct a finite-filter pricing-error proxy based on the permanent--temporary decomposition in \citet{hasbrouckQuality1993}. Our transaction-based measure is the second-level mean absolute temporary component. It differs from the pricing-error standard deviation in that paper.

We use the full Shenzhen Stock Exchange transaction stream during $[09{:}30{:}00,09{:}32{:}00)$, excluding auction executions, order messages and cancellations. In the source MetaData schema, transaction records have \textrm{type=0}; \textrm{dataType=2} denotes an aggressive buy and 3 an aggressive sell, whereas 1 denotes cancellation. Prices are in 0.001-yuan units and quantities are shares. Within each security-day, trades are ordered by channel and business sequence, with checks for duplicates, nonpositive prices or volumes, and decreasing exchange timestamps. Split executions are retained separately. The labels use exchange seconds; receipt-clock calibration and group crossings remain those of the existing DID panel.

The training period comprises the 60 trading days from July 10 through October 9, 2025, strictly before the application window of October 10--November 30. We estimate each security's parameters once and hold them fixed throughout the application period, without daily re-estimation. Each training day contributes only its opening two-minute sequence, with returns and lags reset at the start of each day. All 96 training/application dates were available; extraction retained 359,602,103 transactions without trade sampling.

For trade $n\ge2$ in a security-day, let
\begin{align}
 r_n&=\log P_n-\log P_{n-1},\qquad v_n=V_n/1000,\\
 X_n&=(r_n,q_n,q_nv_n,q_n\sqrt{v_n})',\qquad q_n\in\{-1,+1\}.
\end{align}
We use the recorded aggressor direction, so Lee--Ready classification is unnecessary. Volume scaling is identical during training and application. We pool valid historical within-day design rows and estimate, separately for each security,
\begin{equation}
 X_n=c+\sum_{k=1}^{5}A_kX_{n-k}+u_n.
\end{equation}
Exact-match measures are not VAR regressors. The design has 21 columns including the intercept. Daily QR factors of the augmented design/response matrix can be stacked and refactored without changing the pooled least-squares objective; column scaling is reversed after estimation. No lag uses the preceding day's last trade.

We require at least 30 historical days with usable design rows and 210 observations, full design rank, finite coefficients and filter loadings, and a companion-matrix spectral radius below one. Of 3,138 candidate securities, 3,108 pass; 30 lack sufficient history. The median and maximum stability radii are 0.863 and 0.959. These are model-universe counts, not the final DID stock count.

\subsection{Finite BN filter and exact warm-up}
Let $e_1=(1,0,0,0)'$. Recursively compute
\begin{align}
 \Psi_0&=I,\qquad \Psi_h=\sum_{k=1}^{\min(5,h)}A_k\Psi_{h-k},\\
 b_h&=e_1'\Psi_h,\qquad g_j=-\sum_{h=j+1}^{50}b_h,\quad j=0,\ldots,49.
\end{align}
Innovations remain in their original correlated coordinates; there is no Cholesky identification of structural shocks. The 50 lags refer to transactions, not seconds. For an application-day trade, use the frozen parameters to compute
\begin{align}
 \widehat u_n&=X_n-\widehat c-\sum_{k=1}^{5}\widehat A_kX_{n-k},\\
 \widehat s_n&=\sum_{j=0}^{49}\widehat g_j\widehat u_{n-j}.
\end{align}
The first return/vector is at trade 2, the first complete five-lag innovation at trade 7, and the first complete 50-innovation filtered value at trade 56. PE therefore excludes the first 55 trades, not just 50 trades. Initial filter padding is not treated as data; neither prior-day innovations nor future trades initialize the filter. A day with fewer than 56 trades has no valid PE.

For the valid filtered trades $\mathcal V_{ids}$ in exchange second $s$, define
\begin{equation}
 PEbps_{ids}=10{,}000\,|\mathcal V_{ids}|^{-1}\sum_{n\in\mathcal V_{ids}}|\widehat s_{idn}|.
\end{equation}
Trades are equally weighted within a second; stock-seconds receive the usual regression weights rather than trade-count weights. A second with no valid trade has missing PE; we neither fill it with zero nor carry forward an earlier value. A daily trade-weighted descriptive mean is also calculated, but is not repeated into the DID panel, where a stock-day-constant outcome would be absorbed by fixed effects. Across the full extracted universe, 107,947 of 110,769 application security-days have a valid daily mean; the median is 2.979 bp. Because this mean is outside the DID baseline sample, it cannot serve as the denominator for a percentage treatment effect.

\subsection{Truncation and generated-outcome uncertainty}
For stable models, compute the omitted cumulative return loading
\begin{equation}
 a=e_1'\left(I-\sum_{k=1}^{5}A_k\right)^{-1}-\sum_{h=0}^{50}b_h,
 \qquad TailSD=10{,}000\sqrt{a\widehat\Sigma a'},
\end{equation}
where $\widehat\Sigma$ is the historical innovation covariance. TailSD has median 0.000480 bp, 99th percentile 0.11123 bp and maximum 0.47261 bp. TailSD summarizes the omitted cumulative loading; it is not a bound on total truncation error. Model eligibility uses the historical-data and stability criteria above.

PE is a generated transaction-based proxy from a finite filter. Reported inference conditions on the historically estimated parameters, so the standard errors exclude uncertainty from estimating those parameters. The measure includes transaction-price microstructure components.

\subsection{DID coverage and selection}
The original 24-day window contains 687,675 candidate stock-seconds: 541,975 have PE, 129,193 lack a matching transaction-second label, 1,957 have transactions but no valid model, and 14,550 have valid models but incomplete warm-up. Classification is mutually exclusive in that order. Valid PE covers 78.81\% of candidate stock-seconds. Fixed-effect/support cleaning then removes 265 valid rows, giving $N=541,710$.

Pre-event coverage is 77.61\% for Slow and 78.12\% for Fast; post-event coverage is 79.37\% and 81.75\%, respectively. All 24 days retain both groups at the omitted $-1$ second. PE coverage varies with transaction activity and warm-up requirements.

In the PE event study, $s-G^F_t=2$ and $s-G^F_t=3$ are negative at 5\%, but later bins have diminishing date support (only two days at $s-G^F_t=9$).

\clearpage

% Flush each float before accepting the next, preserving exact table order.
\section*{Part II: Appendix Tables}
\begin{table}[htbp]\singlespacing\footnotesize\renewcommand{\arraystretch}{0.9}\centering
\caption{Average Number of Quote Updates by Size Quintile.}\label{tab:panel_average_number_of_quote_updates_in_size_quintiles.}
\begin{minipage}{\textwidth}\small Full valid two-sided L1 event sequence, not sampled orders. Continuous trading ends at 14:57. The sample covers July 2023--October 2025 after the prior-day size screen. Valid best-bid or best-ask price or quantity changes count as updates. Rapid updates have intervals $\le100$ms; lunch breaks and invalid quote states reset the interval. Of 1,127,103 candidate stock-days, 1,125,815 are included; 1,288 unavailable stock-days are excluded rather than zero-filled. The last interval contains 1,620 seconds.
\end{minipage}\par\medskip
\textbf{Panel A: L1 updates per second}
\paperfit{\textwidth}{%
\begin{tabular}{lrrrrrrrr}
\toprule
Quintile & 09:30–10:00 & 10:00–10:30 & 10:30–11:00 & 11:00–11:30 & 13:00–13:30 & 13:30–14:00 & 14:00–14:30 & 14:30–14:57 \\
\midrule
1 & 2.5408 & 1.3269 & 0.9347 & 0.7862 & 0.9412 & 0.8119 & 0.8406 & 1.2734 \\
2 & 3.1110 & 1.6258 & 1.1439 & 0.9501 & 1.1339 & 0.9853 & 1.0222 & 1.5467 \\
3 & 3.7903 & 1.9959 & 1.4002 & 1.1632 & 1.3767 & 1.1996 & 1.2508 & 1.8589 \\
4 & 4.6924 & 2.5115 & 1.7826 & 1.4818 & 1.7471 & 1.5294 & 1.5952 & 2.3101 \\
5 & 7.0527 & 3.9498 & 2.8557 & 2.3976 & 2.8231 & 2.5108 & 2.6112 & 3.6438 \\
\bottomrule
\end{tabular}
}
\par\medskip\textbf{Panel B: updates within 100ms (percent)}
\paperfit{\textwidth}{%
\begin{tabular}{lrrrrrrrr}
\toprule
Quintile & 09:30–10:00 & 10:00–10:30 & 10:30–11:00 & 11:00–11:30 & 13:00–13:30 & 13:30–14:00 & 14:00–14:30 & 14:30–14:57 \\
\midrule
1 & 57.8161 & 51.5225 & 49.4935 & 48.7945 & 51.4058 & 46.8602 & 47.1454 & 46.9241 \\
2 & 58.7377 & 51.9551 & 49.4952 & 48.6793 & 51.3376 & 46.7205 & 47.0428 & 47.1999 \\
3 & 60.3264 & 53.1650 & 50.3276 & 49.3014 & 51.9011 & 47.3616 & 47.6598 & 48.2371 \\
4 & 62.4098 & 55.0219 & 51.7757 & 50.4898 & 53.0046 & 48.5306 & 48.9467 & 50.0701 \\
5 & 64.9023 & 56.6084 & 52.4723 & 50.5934 & 53.4298 & 48.9418 & 49.5136 & 52.5172 \\
\bottomrule
\end{tabular}
}
\end{table}

\clearpage

\begin{landscape}
\begin{table}[htbp]\singlespacing\footnotesize\renewcommand{\arraystretch}{1.1}
\setlength{\tabcolsep}{1pt}
  \caption{Examples of Information Asymmetry for Stock 300750 on October 10, 2025.}
\begin{minipage}{\linewidth}
\small
  This table illustrates information asymmetry for ticker symbol 300750 on October 10, 2025. Panel A reports Level-2 quote data sampled at a three-second frequency. A total of 11 observations are recorded within a 30-second window, which corresponds to the typical quote update frequency displayed in trading applications used by retail investors. Panel A follows the snapshot feed's own three-second slicing convention, so its first displayed quote need not be the first reconstructed quote with the same displayed timestamp. Panel B presents reconstructed high-frequency quotes, where 22 updates occur within only 50 milliseconds. These reconstructed quotes are constructed from detailed order-level messages, which are accessible only to institutional traders. For brevity, only the top five levels are reported in the table, although professional traders can readily reconstruct and process all order-book levels if desired.
\end{minipage}\par\footnotesize
  \label{tab:panel_examples_of_information_asymmetry_for_ticker_symbol_300750_on_october_10,_2025}
  \centering
  \paperfit{\linewidth}{
  \begin{tabular}{rrrrrrrrrrrrrrrrrrrrrr}
    \toprule
    \multicolumn{22}{c}{\textbf{Panel A: Three-second Level2 Quotes}} \\
    \midrule
    \textbf{Time} & \textbf{CumVolume} & \textbf{Bid1} & \textbf{BidDepth1} & \textbf{Bid2} & \textbf{BidDepth2} & \textbf{Bid3} & \textbf{BidDepth3} & \textbf{Bid4} & \textbf{BidDepth4} & \textbf{Bid5} & \textbf{BidDepth5} & \textbf{Ask1} & \textbf{AskDepth1} & \textbf{Ask2} & \textbf{AskDepth2} & \textbf{Ask3} & \textbf{AskDepth3} & \textbf{Ask4} & \textbf{AskDepth4} & \textbf{Ask5} & \textbf{AskDepth5} \\
    \midrule
    93000000 & 845783 & 401.80 & 5000 & 401.71 & 500 & 401.65 & 100 & 401.50 & 200 & 401.33 & 100 & 402.00 & 1400 & 402.05 & 400 & 402.30 & 200 & 402.40 & 400 & 402.47 & 300 \\
    93003000 & 966583 & 401.20 & 100 & 401.16 & 100 & 401.14 & 100 & 401.07 & 100 & 401.06 & 100 & 401.50 & 2100 & 401.99 & 1200 & 402.00 & 1800 & 402.38 & 100 & 402.50 & 100 \\
    93006000 & 1026463 & 402.00 & 920 & 401.54 & 100 & 401.50 & 600 & 401.45 & 100 & 401.42 & 200 & 402.10 & 100 & 402.35 & 200 & 402.38 & 100 & 402.50 & 100 & 402.90 & 100 \\
    93009000 & 1061763 & 402.01 & 700 & 402.00 & 700 & 401.96 & 500 & 401.88 & 100 & 401.80 & 100 & 402.74 & 700 & 402.90 & 300 & 403.00 & 1300 & 403.05 & 1300 & 403.30 & 200 \\
    93012000 & 1078463 & 401.88 & 600 & 401.50 & 200 & 401.21 & 1400 & 401.20 & 3600 & 401.15 & 300 & 402.01 & 2200 & 402.02 & 100 & 402.10 & 100 & 402.15 & 100 & 402.64 & 1500 \\
    93015000 & 1116863 & 401.88 & 100 & 401.80 & 6200 & 401.75 & 100 & 401.68 & 100 & 401.50 & 2000 & 402.00 & 14000 & 402.01 & 100 & 402.17 & 600 & 402.74 & 100 & 402.78 & 2200 \\
    93018000 & 1146663 & 401.88 & 1000 & 401.51 & 100 & 401.50 & 1700 & 401.33 & 200 & 401.22 & 100 & 402.00 & 8000 & 402.01 & 1100 & 402.17 & 600 & 402.73 & 100 & 402.74 & 100 \\
    93021000 & 1171763 & 401.83 & 100 & 401.80 & 200 & 401.60 & 100 & 401.51 & 100 & 401.50 & 100 & 401.99 & 100 & 402.00 & 1500 & 402.01 & 1100 & 402.10 & 100 & 402.61 & 500 \\
    93024000 & 1192663 & 402.00 & 11900 & 401.99 & 200 & 401.90 & 1300 & 401.88 & 200 & 401.83 & 1300 & 402.05 & 800 & 402.38 & 300 & 402.39 & 100 & 402.40 & 300 & 402.41 & 200 \\
    93027000 & 1227663 & 401.99 & 300 & 401.93 & 1800 & 401.91 & 3000 & 401.83 & 500 & 401.68 & 100 & 402.00 & 1500 & 402.01 & 200 & 402.05 & 15200 & 402.10 & 1200 & 402.11 & 100 \\
    93030000 & 1255363 & 401.21 & 300 & 401.20 & 3500 & 401.15 & 100 & 401.13 & 100 & 401.11 & 300 & 401.83 & 100 & 401.88 & 3800 & 401.90 & 100 & 401.91 & 2300 & 401.94 & 200 \\
    \bottomrule
  \end{tabular}}
\end{table}
\clearpage
\begin{table}[htbp]\ContinuedFloat\singlespacing\footnotesize
\renewcommand{\arraystretch}{1.1}
\setlength{\tabcolsep}{1pt}
\centering
\caption[]{Examples of Information Asymmetry for Stock 300750 on October 10, 2025 (continued).}
\paperfit{\linewidth}{\begin{tabular}{rrrrrrrrrrrrrrrrrrrrrr}
    \toprule
    \multicolumn{22}{c}{\textbf{Panel B: Reconstructed Quotes from Detailed Orders}} \\
    \midrule
    \textbf{Time} & \textbf{CumVolume} & \textbf{Bid1} & \textbf{BidDepth1} & \textbf{Bid2} & \textbf{BidDepth2} & \textbf{Bid3} & \textbf{BidDepth3} & \textbf{Bid4} & \textbf{BidDepth4} & \textbf{Bid5} & \textbf{BidDepth5} & \textbf{Ask1} & \textbf{AskDepth1} & \textbf{Ask2} & \textbf{AskDepth2} & \textbf{Ask3} & \textbf{AskDepth3} & \textbf{Ask4} & \textbf{AskDepth4} & \textbf{Ask5} & \textbf{AskDepth5} \\
    \midrule
    93000000 & 613828 & 401.00 & 14172 & 400.99 & 200 & 400.98 & 100 & 400.97 & 100 & 400.96 & 200 & 401.01 & 100 & 401.05 & 200 & 401.08 & 200 & 401.11 & 4500 & 401.12 & 1000 \\
    93000000 & 614228 & 401.00 & 13772 & 400.99 & 200 & 400.98 & 100 & 400.97 & 100 & 400.96 & 200 & 401.01 & 100 & 401.05 & 200 & 401.08 & 200 & 401.11 & 4500 & 401.12 & 1000 \\
    93000010 & 614728 & 401.00 & 13772 & 400.99 & 200 & 400.98 & 100 & 400.97 & 100 & 400.96 & 200 & 401.11 & 4500 & 401.12 & 1000 & 401.15 & 500 & 401.20 & 400 & 401.50 & 2000 \\
    93000010 & 614728 & 401.01 & 2500 & 401.00 & 13772 & 400.99 & 200 & 400.98 & 100 & 400.97 & 100 & 401.11 & 4500 & 401.12 & 1000 & 401.15 & 500 & 401.20 & 400 & 401.50 & 2000 \\
    93000010 & 635328 & 401.01 & 2500 & 401.00 & 13772 & 400.99 & 200 & 400.98 & 100 & 400.97 & 100 & 401.70 & 11500 & 401.80 & 24000 & 401.95 & 600 & 402.00 & 900 & 402.01 & 7000 \\
    93000030 & 635328 & 401.25 & 100 & 401.01 & 2500 & 401.00 & 13772 & 400.99 & 200 & 400.98 & 100 & 401.70 & 11500 & 401.80 & 24000 & 401.95 & 600 & 402.00 & 900 & 402.01 & 7000 \\
    93000040 & 635628 & 401.01 & 4500 & 401.00 & 13772 & 400.99 & 200 & 400.98 & 100 & 400.97 & 100 & 401.70 & 11500 & 401.80 & 24000 & 401.95 & 600 & 402.00 & 900 & 402.01 & 7000 \\
    93000040 & 635628 & 401.01 & 4600 & 401.00 & 13872 & 400.99 & 200 & 400.98 & 100 & 400.97 & 100 & 401.70 & 11500 & 401.80 & 24000 & 401.95 & 600 & 402.00 & 900 & 402.01 & 7000 \\
    93000040 & 635628 & 401.12 & 100 & 401.01 & 4600 & 401.00 & 13872 & 400.99 & 200 & 400.98 & 100 & 401.70 & 11500 & 401.80 & 24000 & 401.95 & 600 & 402.00 & 900 & 402.01 & 7000 \\
    93000040 & 636128 & 401.12 & 100 & 401.01 & 4600 & 401.00 & 13872 & 400.99 & 200 & 400.98 & 100 & 401.70 & 11000 & 401.80 & 24000 & 401.95 & 600 & 402.00 & 900 & 402.01 & 7000 \\
    93000040 & 636628 & 401.01 & 4300 & 401.00 & 13872 & 400.99 & 200 & 400.98 & 100 & 400.97 & 100 & 401.70 & 11000 & 401.80 & 24000 & 401.95 & 600 & 402.00 & 900 & 402.01 & 7000 \\
    93000040 & 636628 & 401.01 & 4300 & 401.00 & 13872 & 400.99 & 200 & 400.98 & 100 & 400.97 & 100 & 401.30 & 300 & 401.70 & 11000 & 401.80 & 24000 & 401.95 & 600 & 402.00 & 900 \\
    93000040 & 636728 & 401.01 & 4300 & 401.00 & 13872 & 400.99 & 200 & 400.98 & 100 & 400.97 & 100 & 401.30 & 200 & 401.40 & 300 & 401.50 & 300 & 401.60 & 300 & 401.70 & 11300 \\
    93000050 & 636828 & 401.01 & 4300 & 401.00 & 13872 & 400.99 & 200 & 400.98 & 100 & 400.97 & 100 & 401.30 & 100 & 401.40 & 300 & 401.50 & 300 & 401.60 & 300 & 401.70 & 11500 \\
    93000050 & 636828 & 401.08 & 1000 & 401.01 & 4300 & 401.00 & 13872 & 400.99 & 200 & 400.98 & 100 & 401.30 & 100 & 401.40 & 300 & 401.50 & 300 & 401.60 & 300 & 401.70 & 11500 \\
    93000050 & 636828 & 401.12 & 400 & 401.08 & 1000 & 401.01 & 4300 & 401.00 & 13872 & 400.99 & 200 & 401.30 & 100 & 401.40 & 300 & 401.50 & 300 & 401.60 & 300 & 401.70 & 11500 \\
    93000050 & 637428 & 401.12 & 300 & 401.08 & 1000 & 401.01 & 4500 & 401.00 & 14172 & 400.99 & 200 & 401.50 & 200 & 401.60 & 300 & 401.70 & 11500 & 401.80 & 24400 & 401.90 & 500 \\
    93000050 & 638028 & 401.08 & 700 & 401.01 & 4500 & 401.00 & 14172 & 400.99 & 200 & 400.98 & 100 & 401.50 & 200 & 401.60 & 300 & 401.70 & 11500 & 401.80 & 24400 & 401.90 & 500 \\
    93000050 & 638128 & 401.08 & 700 & 401.01 & 4500 & 401.00 & 14172 & 400.99 & 200 & 400.98 & 100 & 401.50 & 100 & 401.60 & 300 & 401.70 & 11500 & 401.80 & 24400 & 401.90 & 500 \\
    93000050 & 638228 & 401.08 & 600 & 401.01 & 4800 & 401.00 & 14972 & 400.99 & 200 & 400.98 & 100 & 401.50 & 100 & 401.60 & 300 & 401.70 & 11500 & 401.80 & 24400 & 401.90 & 500 \\
    93000050 & 638328 & 401.08 & 600 & 401.01 & 4800 & 401.00 & 14972 & 400.99 & 200 & 400.98 & 100 & 401.60 & 300 & 401.70 & 11500 & 401.80 & 24400 & 401.90 & 500 & 401.95 & 600 \\
    93000050 & 638328 & 401.58 & 100 & 401.08 & 600 & 401.01 & 5000 & 401.00 & 14972 & 400.99 & 200 & 401.60 & 300 & 401.70 & 11500 & 401.80 & 24400 & 401.90 & 500 & 401.95 & 600 \\
    \bottomrule
  \end{tabular}
  }
\end{table}
\end{landscape}

\clearpage

\begin{table}[htbp]\singlespacing\footnotesize\renewcommand{\arraystretch}{0.9}
\caption{Descriptive Statistics of Order Response Time.}
\label{tab:panel_descriptive_statistics_of_order_response_time}
\centering
\begin{minipage}{\textwidth}
\small
The sample consists of all common stocks listed on the Shenzhen Stock Exchange, excluding the smallest 30\% of stocks by prior-day market capitalization. The analysis covers the period from July 1, 2023 to October 31, 2025, using an approximately 0.1\% systematically sampled subset of orders submitted during continuous trading hours, after excluding market orders. Stocks are classified into five size quintiles based on prior-day market capitalization. Response time is measured in seconds as the elapsed time between order submission and the latest prior change in best-bid or best-ask price or quantity. 
Panel A reports summary statistics of response time by size quintile. Panels B1 and B2 report median response times conditional on quote-matching indicators and order characteristics, respectively. \textit{RQuoteMatchActive} (\textit{RQuoteMatchPassive}) is an indicator equal to one if an order is submitted at the best opposite-side (same-side) quote.  
\textit{BuyOrder} is a dummy variable equal to one for buy orders and zero for sell orders. \textit{OneCentLarger} (\textit{MultiCentLarger}) equals one if the order price is more aggressive than the current best opposite-side quote by exactly one tick (by more than one tick), respectively. \textit{NonRoundVol} is an indicator equal to one if the order size is not a multiple of 500 shares, and \textit{NonRoundPrice} is an indicator equal to one if the submitted limit price is not an integer-yuan price. \textit{LargeOrder} equals one if the submitted order notional exceeds RMB 200,000. Rows labeled ``0'' and ``1'' correspond to indicator values.
Only limit orders are included, and matching uses their submitted numerical prices; market orders (O/S) are excluded without resampling. \end{minipage}\par\footnotesize
\par\medskip\textbf{Panel A: Distribution (seconds)}\par
\paperfit{\textwidth}{\begin{tabular}{lrrrrrrrr}\toprule
Size & Count & Mean & Std & Min & 25\% & Median & 75\% & Max \\
Small & 4,510,567 & 1.00 & 2.92 & 0.00 & 0.00 & 0.11 & 0.84 & 599.80 \\
2 & 5,675,052 & 0.81 & 2.33 & 0.00 & 0.01 & 0.10 & 0.69 & 598.74 \\
3 & 7,259,682 & 0.64 & 1.82 & 0.00 & 0.01 & 0.09 & 0.54 & 345.19 \\
4 & 9,328,279 & 0.50 & 1.45 & 0.00 & 0.01 & 0.07 & 0.43 & 598.88 \\
Large & 15,732,636 & 0.31 & 0.81 & 0.00 & 0.01 & 0.05 & 0.27 & 227.93 \\
\bottomrule\end{tabular}}
\par\medskip\textbf{Panel B1: Conditional medians, quote matching}\par\paperfit{\textwidth}{\begin{tabular}{llrr}\toprule
Size & Value & RQuoteMatchActive & RQuoteMatchPassive \\
\multirow{2}{*}{Small} & 0 & 0.15 & 0.16 \\
 & 1 & 0.04 & 0.04 \\
\multirow{2}{*}{2} & 0 & 0.13 & 0.14 \\
 & 1 & 0.04 & 0.04 \\
\multirow{2}{*}{3} & 0 & 0.11 & 0.11 \\
 & 1 & 0.03 & 0.04 \\
\multirow{2}{*}{4} & 0 & 0.09 & 0.09 \\
 & 1 & 0.03 & 0.03 \\
\multirow{2}{*}{Large} & 0 & 0.06 & 0.06 \\
 & 1 & 0.02 & 0.03 \\
\bottomrule\end{tabular}}
\par\medskip\textbf{Panel B2: Conditional medians, other order characteristics}\par\paperfit{\textwidth}{\begin{tabular}{llrrrrrr}\toprule
Size & Value & BuyOrder & OneCentLarger & MultiCentLarger & NonRoundVol & NonRoundPrice & LargeOrder \\
\multirow{2}{*}{Small} & 0 & 0.11 & 0.11 & 0.10 & 0.36 & 0.20 & 0.11 \\
 & 1 & 0.11 & 0.12 & 0.18 & 0.06 & 0.11 & 0.42 \\
\multirow{2}{*}{2} & 0 & 0.10 & 0.10 & 0.10 & 0.28 & 0.17 & 0.10 \\
 & 1 & 0.11 & 0.10 & 0.14 & 0.06 & 0.10 & 0.30 \\
\multirow{2}{*}{3} & 0 & 0.08 & 0.09 & 0.08 & 0.20 & 0.13 & 0.08 \\
 & 1 & 0.09 & 0.08 & 0.11 & 0.05 & 0.08 & 0.22 \\
\multirow{2}{*}{4} & 0 & 0.07 & 0.07 & 0.07 & 0.16 & 0.11 & 0.07 \\
 & 1 & 0.07 & 0.08 & 0.11 & 0.05 & 0.07 & 0.15 \\
\multirow{2}{*}{Large} & 0 & 0.05 & 0.05 & 0.05 & 0.09 & 0.07 & 0.05 \\
 & 1 & 0.06 & 0.06 & 0.08 & 0.04 & 0.05 & 0.07 \\
\bottomrule\end{tabular}}
\end{table}

\clearpage

\begin{table}[htbp]\singlespacing\footnotesize\renewcommand{\arraystretch}{0.9}
\caption{Descriptive Statistics of Order Cancellation Time.}
\label{tab:panel_descriptive_statistics_of_order_cancellation_time}
\centering
\begin{minipage}{\textwidth}
\small
The sample consists of all common stocks listed on the Shenzhen Stock Exchange, excluding the smallest 30\% of stocks by prior-day market capitalization. The analysis covers the period from July 1, 2023 to October 31, 2025, using an approximately 0.1\% systematically sampled subset of orders submitted during continuous trading hours, after excluding market orders. Stocks are classified into five size quintiles based on prior-day market capitalization. Only canceled orders enter this table. Cancel time is measured in seconds as the elapsed time between order submission and cancellation. Panel A reports summary statistics of cancellation time by size quintile. Panels B1 and B2 report median cancellation times conditional on quote-matching indicators and order characteristics, respectively.
\textit{RQuoteMatchActive} (\textit{RQuoteMatchPassive}) is an indicator equal to one if an order is submitted at the best opposite-side (same-side) quote.  
\textit{BuyOrder} is a dummy variable equal to one for buy orders and zero for sell orders. \textit{OneCentLarger} (\textit{MultiCentLarger}) equals one if the order price is more aggressive than the current best opposite-side quote by exactly one tick (by more than one tick), respectively. \textit{NonRoundVol} is an indicator equal to one if the order size is not a multiple of 500 shares, and \textit{NonRoundPrice} is an indicator equal to one if the submitted limit price is not an integer-yuan price. \textit{LargeOrder} equals one if the submitted order notional exceeds RMB 200,000. 
Rows labeled ``0'' and ``1'' correspond to indicator values.
Only limit orders are included, and matching uses their submitted numerical prices; market orders (O/S) are excluded without resampling. \end{minipage}\par\footnotesize
\par\medskip\textbf{Panel A: Distribution (seconds)}\par
\paperfit{\textwidth}{\begin{tabular}{lrrrrrrrr}\toprule
Size & Count & Mean & Std & Min & 25\% & Median & 75\% & Max \\
Small & 1,179,676 & 594.00 & 1772.68 & 0.00 & 8.93 & 41.96 & 244.25 & 14216.95 \\
2 & 1,452,574 & 650.90 & 1858.03 & 0.00 & 9.00 & 45.10 & 277.24 & 14219.09 \\
3 & 1,821,740 & 672.84 & 1888.31 & 0.00 & 8.98 & 44.83 & 286.69 & 14219.75 \\
4 & 2,279,628 & 660.29 & 1883.30 & 0.00 & 7.85 & 38.94 & 257.26 & 14218.22 \\
Large & 3,680,924 & 711.30 & 1975.17 & 0.00 & 6.83 & 35.55 & 270.04 & 14219.50 \\
\bottomrule\end{tabular}}
\par\medskip\textbf{Panel B1: Conditional medians, quote matching}\par\paperfit{\textwidth}{\begin{tabular}{llrr}\toprule
Size & Value & RQuoteMatchActive & RQuoteMatchPassive \\
\multirow{2}{*}{Small} & 0 & 43.20 & 97.38 \\
 & 1 & 8.00 & 24.93 \\
\multirow{2}{*}{2} & 0 & 46.87 & 111.10 \\
 & 1 & 6.54 & 24.01 \\
\multirow{2}{*}{3} & 0 & 45.72 & 113.82 \\
 & 1 & 5.88 & 21.05 \\
\multirow{2}{*}{4} & 0 & 39.68 & 100.16 \\
 & 1 & 5.00 & 18.01 \\
\multirow{2}{*}{Large} & 0 & 36.04 & 82.34 \\
 & 1 & 3.01 & 15.01 \\
\bottomrule\end{tabular}}
\par\medskip\textbf{Panel B2: Conditional medians, other order characteristics}\par\paperfit{\textwidth}{\begin{tabular}{llrrrrrr}\toprule
Size & Value & BuyOrder & OneCentLarger & MultiCentLarger & NonRoundVol & NonRoundPrice & LargeOrder \\
\multirow{2}{*}{Small} & 0 & 43.93 & 41.97 & 41.96 & 128.35 & 141.46 & 41.64 \\
 & 1 & 39.84 & 12.85 & 18.88 & 31.16 & 41.00 & 65.78 \\
\multirow{2}{*}{2} & 0 & 47.99 & 45.18 & 45.14 & 136.04 & 178.80 & 45.00 \\
 & 1 & 43.01 & 13.82 & 16.95 & 32.91 & 44.29 & 67.76 \\
\multirow{2}{*}{3} & 0 & 47.34 & 44.87 & 44.85 & 136.61 & 191.52 & 44.50 \\
 & 1 & 42.11 & 13.33 & 17.91 & 31.58 & 43.16 & 60.89 \\
\multirow{2}{*}{4} & 0 & 40.98 & 38.96 & 38.95 & 126.75 & 215.31 & 38.64 \\
 & 1 & 36.78 & 9.90 & 16.58 & 28.28 & 37.09 & 56.16 \\
\multirow{2}{*}{Large} & 0 & 36.04 & 35.60 & 35.58 & 122.32 & 292.87 & 35.48 \\
 & 1 & 34.79 & 6.56 & 10.36 & 26.76 & 32.95 & 37.61 \\
\bottomrule\end{tabular}}
\end{table}

\clearpage

\begin{table}[p]\centering\singlespacing\footnotesize
\setlength{\tabcolsep}{2pt}
\renewcommand{\arraystretch}{1.08}
\caption{Return Predictability Across Horizons by Order Direction.}\label{tab:buy_sell_horizons}
\begin{minipage}{\textwidth}\small
This table repeats the full-control specification in Table~\ref{tab:horizons} separately for buy and sell limit orders. The Shenzhen sample spans July 1, 2023 to October 31, 2025 and excludes the smallest 30\% of stocks by prior-day market capitalization. Each directional subsample uses a common set of orders across all nine horizons. Returns are direction-adjusted mid-quote returns, in basis points, from the first valid midpoint after submission to the last valid midpoint at or before the specified horizon. BuyOrder is excluded because direction is constant within each subsample. All other order and stock controls are retained. Stock and date fixed effects are estimated separately by direction; standard errors are clustered by stock. Parentheses report $t$-statistics. Two singleton observations are removed from the buy subsample.
\end{minipage}\par\medskip
\paperfit{\linewidth}{\begin{tabular}{lrrrrrrrrr}\toprule
 & 100ms & 1s & 5s & 10s & 30s & 60s & 120s & 300s & 600s \\
\midrule
\multicolumn{10}{l}{\textbf{Panel A: Buy Orders}} \\
RQuoteMatchActive & 1.691$^{***}$ & 2.291$^{***}$ & 2.355$^{***}$ & 2.211$^{***}$ & 2.500$^{***}$ & 2.912$^{***}$ & 3.281$^{***}$ & 3.121$^{***}$ & 3.126$^{***}$ \\
 & (81.02) & (79.03) & (49.54) & (33.83) & (24.45) & (30.37) & (37.09) & (27.38) & (23.80) \\
RQuoteMatchPassive & 0.231$^{***}$ & 0.379$^{***}$ & 0.008 & -0.264$^{***}$ & -0.212$^{***}$ & -0.088 & 0.094 & -0.070 & -0.038 \\
 & (38.71) & (35.81) & (0.31) & (-6.57) & (-2.99) & (-1.31) & (1.47) & (-0.81) & (-0.37) \\
OneCentLarger & 0.566$^{***}$ & 0.973$^{***}$ & 0.221$^{***}$ & -0.710$^{***}$ & -1.728$^{***}$ & -1.158$^{***}$ & -0.177 & -0.521$^{***}$ & -0.636$^{***}$ \\
 & (54.61) & (44.70) & (4.30) & (-8.24) & (-11.45) & (-8.26) & (-1.50) & (-3.48) & (-3.65) \\
MultiCentLarger & 0.356$^{***}$ & 0.598$^{***}$ & -1.217$^{***}$ & -2.929$^{***}$ & -4.612$^{***}$ & -2.904$^{***}$ & -1.149$^{***}$ & -1.551$^{***}$ & -1.714$^{***}$ \\
 & (32.31) & (24.78) & (-18.51) & (-24.74) & (-24.55) & (-19.53) & (-8.79) & (-9.77) & (-8.78) \\
NonRoundVol & 0.314$^{***}$ & 0.458$^{***}$ & 0.812$^{***}$ & 1.051$^{***}$ & 1.353$^{***}$ & 1.272$^{***}$ & 1.279$^{***}$ & 1.358$^{***}$ & 1.260$^{***}$ \\
 & (61.66) & (64.62) & (68.82) & (62.37) & (46.90) & (38.92) & (33.99) & (26.18) & (19.11) \\
NonRoundPrice & -0.157$^{***}$ & -0.276$^{***}$ & -0.234$^{***}$ & -0.141$^{**}$ & 0.153 & 0.194 & 0.098 & -0.001 & -0.057 \\
 & (-17.04) & (-16.45) & (-6.03) & (-2.05) & (1.03) & (1.61) & (0.76) & (-0.01) & (-0.24) \\
LargeOrder & 0.226$^{***}$ & 0.545$^{***}$ & 1.349$^{***}$ & 2.059$^{***}$ & 3.040$^{***}$ & 3.010$^{***}$ & 3.009$^{***}$ & 2.985$^{***}$ & 2.596$^{***}$ \\
 & (15.96) & (23.98) & (35.57) & (37.96) & (32.23) & (25.77) & (21.72) & (15.85) & (11.13) \\
LaggedLogSize & -0.048$^{***}$ & 0.047 & 0.540$^{***}$ & 1.133$^{***}$ & 2.343$^{***}$ & 1.653$^{***}$ & 0.233 & -1.510$^{***}$ & -4.413$^{***}$ \\
 & (-2.99) & (1.57) & (6.30) & (6.88) & (5.88) & (3.96) & (0.79) & (-3.49) & (-7.39) \\
LaggedTurnoverRate & -1.065$^{***}$ & 0.148 & 5.705$^{***}$ & 9.075$^{***}$ & 11.054$^{***}$ & 11.798$^{***}$ & 16.336$^{***}$ & 33.134$^{***}$ & 43.501$^{***}$ \\
 & (-19.49) & (1.59) & (21.85) & (19.03) & (10.11) & (10.28) & (10.95) & (13.14) & (12.35) \\
LaggedInverseClosePrice & 1.525$^{***}$ & 2.483$^{***}$ & 5.843$^{***}$ & 8.834$^{***}$ & 14.890$^{***}$ & 16.263$^{***}$ & 17.916$^{***}$ & 26.474$^{***}$ & 35.751$^{***}$ \\
 & (6.93) & (6.84) & (10.05) & (10.13) & (8.35) & (8.75) & (8.96) & (7.15) & (6.15) \\
\midrule
Stock FE & Yes & Yes & Yes & Yes & Yes & Yes & Yes & Yes & Yes \\
Date FE & Yes & Yes & Yes & Yes & Yes & Yes & Yes & Yes & Yes \\
Observations & 20,140,046 & 20,140,046 & 20,140,046 & 20,140,046 & 20,140,046 & 20,140,046 & 20,140,046 & 20,140,046 & 20,140,046 \\
Stocks & 2,758 & 2,758 & 2,758 & 2,758 & 2,758 & 2,758 & 2,758 & 2,758 & 2,758 \\
Trading days & 566 & 566 & 566 & 566 & 566 & 566 & 566 & 566 & 566 \\
\bottomrule\end{tabular}}
\end{table}
\clearpage
\begin{table}[p]\ContinuedFloat\centering\singlespacing\footnotesize
\setlength{\tabcolsep}{2pt}
\renewcommand{\arraystretch}{1.08}
\caption[]{Return Predictability Across Horizons by Order Direction (continued).}
\paperfit{\linewidth}{\begin{tabular}{lrrrrrrrrr}\toprule
 & 100ms & 1s & 5s & 10s & 30s & 60s & 120s & 300s & 600s \\
\midrule
\multicolumn{10}{l}{\textbf{Panel B: Sell Orders}} \\
RQuoteMatchActive & 1.653$^{***}$ & 2.274$^{***}$ & 2.473$^{***}$ & 2.287$^{***}$ & 2.406$^{***}$ & 3.371$^{***}$ & 3.791$^{***}$ & 3.613$^{***}$ & 3.335$^{***}$ \\
 & (79.78) & (80.22) & (61.69) & (46.50) & (39.06) & (57.49) & (62.32) & (49.05) & (37.47) \\
RQuoteMatchPassive & 0.197$^{***}$ & 0.415$^{***}$ & 0.211$^{***}$ & -0.124$^{***}$ & -0.322$^{***}$ & 0.292$^{***}$ & 0.476$^{***}$ & 0.245$^{***}$ & -0.069 \\
 & (35.08) & (45.23) & (12.87) & (-5.06) & (-8.27) & (7.34) & (10.00) & (3.77) & (-0.82) \\
OneCentLarger & 0.695$^{***}$ & 1.105$^{***}$ & 0.564$^{***}$ & -0.248$^{***}$ & -1.081$^{***}$ & -0.127 & 0.530$^{***}$ & 0.262$^{**}$ & -0.009 \\
 & (58.53) & (52.41) & (14.64) & (-4.07) & (-11.00) & (-1.44) & (5.73) & (2.17) & (-0.06) \\
MultiCentLarger & 0.246$^{***}$ & 0.265$^{***}$ & -1.541$^{***}$ & -3.067$^{***}$ & -4.339$^{***}$ & -3.678$^{***}$ & -3.262$^{***}$ & -4.379$^{***}$ & -4.749$^{***}$ \\
 & (28.66) & (16.79) & (-27.53) & (-31.08) & (-27.71) & (-27.82) & (-26.51) & (-23.83) & (-22.16) \\
NonRoundVol & 0.309$^{***}$ & 0.427$^{***}$ & 0.710$^{***}$ & 0.871$^{***}$ & 0.949$^{***}$ & 0.973$^{***}$ & 0.982$^{***}$ & 1.026$^{***}$ & 0.807$^{***}$ \\
 & (57.36) & (59.66) & (66.09) & (60.29) & (40.70) & (31.51) & (27.34) & (20.62) & (12.31) \\
NonRoundPrice & -0.075$^{***}$ & -0.088$^{***}$ & 0.101$^{***}$ & 0.256$^{***}$ & 0.548$^{***}$ & 0.622$^{***}$ & 0.669$^{***}$ & 0.815$^{***}$ & 0.857$^{***}$ \\
 & (-9.24) & (-6.40) & (3.39) & (5.28) & (6.09) & (6.10) & (5.55) & (5.42) & (4.11) \\
LargeOrder & -0.275$^{***}$ & -0.353$^{***}$ & -0.207$^{***}$ & 0.153$^{***}$ & 0.617$^{***}$ & 0.529$^{***}$ & 0.166 & -0.270 & -0.493$^{**}$ \\
 & (-17.42) & (-15.80) & (-5.83) & (3.36) & (7.19) & (4.75) & (1.22) & (-1.57) & (-2.39) \\
LaggedLogSize & 0.034$^{**}$ & 0.192$^{***}$ & 0.829$^{***}$ & 1.447$^{***}$ & 2.642$^{***}$ & 2.964$^{***}$ & 3.134$^{***}$ & 4.994$^{***}$ & 7.797$^{***}$ \\
 & (2.34) & (9.04) & (12.53) & (12.83) & (11.66) & (11.69) & (10.86) & (12.70) & (12.64) \\
LaggedTurnoverRate & -1.648$^{***}$ & -0.978$^{***}$ & 5.327$^{***}$ & 11.961$^{***}$ & 15.602$^{***}$ & 1.124 & -12.683$^{***}$ & -26.141$^{***}$ & -41.712$^{***}$ \\
 & (-28.58) & (-10.53) & (23.86) & (31.34) & (18.72) & (1.12) & (-9.23) & (-11.97) & (-12.30) \\
LaggedInverseClosePrice & 2.048$^{***}$ & 2.763$^{***}$ & 5.561$^{***}$ & 7.925$^{***}$ & 11.664$^{***}$ & 11.860$^{***}$ & 7.486$^{***}$ & 0.027 & -7.980 \\
 & (9.77) & (9.59) & (10.70) & (9.71) & (8.38) & (9.65) & (4.36) & (0.01) & (-1.62) \\
\midrule
Stock FE & Yes & Yes & Yes & Yes & Yes & Yes & Yes & Yes & Yes \\
Date FE & Yes & Yes & Yes & Yes & Yes & Yes & Yes & Yes & Yes \\
Observations & 19,369,262 & 19,369,262 & 19,369,262 & 19,369,262 & 19,369,262 & 19,369,262 & 19,369,262 & 19,369,262 & 19,369,262 \\
Stocks & 2,760 & 2,760 & 2,760 & 2,760 & 2,760 & 2,760 & 2,760 & 2,760 & 2,760 \\
Trading days & 566 & 566 & 566 & 566 & 566 & 566 & 566 & 566 & 566 \\
\bottomrule\end{tabular}}
\par\smallskip\begin{minipage}{\textwidth}\footnotesize $^{*}p<0.10$, $^{**}p<0.05$, $^{***}p<0.01$. All tests are two-sided.\end{minipage}
\end{table}

\clearpage

\begin{table}[p]\centering\singlespacing\small
\caption{Differences in Return Predictability Between Sell and Buy Orders.}\label{tab:buy_sell_differences}
\begin{minipage}{\textwidth}\small
This table compares the active exact-match coefficients in Table~\ref{tab:buy_sell_horizons}. The difference is the sell coefficient minus the buy coefficient, in basis points. Difference tests use a direction-interacted model with direction-specific controls and stock and date fixed effects. The joint stock-clustered covariance retains dependence between buy and sell orders in the same stock. The $t$-statistics and $p$-values test a zero sell-minus-buy difference; stars refer to unadjusted two-sided $p$-values.
\end{minipage}\par\medskip
\begin{tabular}{lrrrrr}\toprule
Horizon & Buy & Sell & Sell$-$Buy & Difference $t$ & Difference $p$ \\\midrule
100ms & 1.691 & 1.653 & -0.037$^{***}$ & -7.02 & $<0.0001$ \\
1s & 2.291 & 2.274 & -0.016$^{*}$ & -1.82 & 0.0686 \\
5s & 2.355 & 2.473 & 0.118$^{***}$ & 5.79 & $<0.0001$ \\
10s & 2.211 & 2.287 & 0.076$^{**}$ & 2.42 & 0.0157 \\
30s & 2.500 & 2.406 & -0.094 & -1.43 & 0.1534 \\
60s & 2.912 & 3.371 & 0.459$^{***}$ & 5.84 & $<0.0001$ \\
120s & 3.281 & 3.791 & 0.509$^{***}$ & 5.32 & $<0.0001$ \\
300s & 3.121 & 3.613 & 0.492$^{***}$ & 3.72 & 0.0002 \\
600s & 3.126 & 3.335 & 0.209 & 1.28 & 0.1997 \\
\bottomrule\end{tabular}
\par\smallskip\begin{minipage}{\textwidth}\footnotesize $^{*}p<0.10$, $^{**}p<0.05$, $^{***}p<0.01$.\end{minipage}
\end{table}

\clearpage

\begin{table}[htbp]\singlespacing\footnotesize\renewcommand{\arraystretch}{0.88}\centering
\def\sym#1{\ifmmode^{#1}\else\(^{#1}\)\fi}
\caption{Return Predictability of Exact-Match Orders in Different Regimes.}
\label{tab:Short-term Return Prediction Power in Regimes}
\begin{minipage}{\textwidth}\small
    This table reports panel regression results on the short-term return predictability of quote-matching activity across different market regimes.
    The sample includes all common stocks listed on the Shenzhen Stock Exchange, excluding the smallest 30\% by prior-day market capitalization, from July 1, 2023 to October 31, 2025, based on an approximately 0.1\% systematically sampled subset of orders during continuous trading hours.
    The dependent variable is the 10-second ahead mid-quote return, $Ret10s$. \textit{RQuoteMatchActive} (\textit{RQuoteMatchPassive}) equals one if an order is submitted at the best opposite-side (same-side) quote. The specifications interact active quote matching with regime indicators capturing high/low trailing 10-second log-mid range (\textit{RngHigh}, \textit{RngLow}), high/low volatility (\textit{VolHigh}, \textit{VolLow}), cross-sectional momentum/reversal (\textit{MomCX}, \textit{RevCX}), time-series momentum/reversal (\textit{MomTS}, \textit{RevTS}).
    Controls include \textit{BuyOrder}, \textit{OneCentLarger}, \textit{MultiCentLarger}, \textit{NonRoundVol}, \textit{NonRoundPrice}, \textit{LargeOrder}, and stock characteristics (lagged log size, turnover rate, and inverse closing price). All regressions include stock and date fixed effects. Standard errors are clustered at the stock level. $t$-statistics are reported in parentheses.
Only limit orders are included, and matching uses their submitted numerical prices; market orders (O/S) are excluded without resampling.  Returns follow the precise-time convention in Section~\ref{sec:HFT Order Return Predictability}; unavailable or cross-session endpoints are excluded. In columns (1)--(2), High/Low denote range or volatility terciles. In columns (3)--(4), the same rows denote Momentum/Reversal. Active interaction rows use the corresponding regime. All coefficients and controls are reported.\end{minipage}\par\smallskip
\paperfit{\textwidth}{
\begin{tabular}{lrrrr}\toprule
 & (1) & (2) & (3) & (4) \\
 & Range & Volatility & Cross-sectional & Time-series \\
\midrule
High / Momentum & 0.13544\sym{***} & 0.08663\sym{***} & 2.23056\sym{***} & 10.51587\sym{***} \\
 & (4.88) & (3.46) & (33.02) & (31.66) \\
Low / Reversal & -0.73432\sym{***} & -0.44173\sym{***} & -1.42424\sym{***} & -2.14943\sym{***} \\
 & (-32.39) & (-25.08) & (-27.23) & (-9.74) \\
Active x High / Momentum & 0.52248\sym{***} & 0.25079\sym{***} & -1.00790\sym{***} & -1.79932\sym{***} \\
 & (16.07) & (6.96) & (-17.10) & (-6.95) \\
Active x Low / Reversal & 1.88739\sym{***} & 0.32829\sym{***} & 1.20786\sym{***} & 6.48190\sym{***} \\
 & (46.33) & (8.76) & (29.50) & (31.43) \\
RQuoteMatchActive & 1.42008\sym{***} & 1.98493\sym{***} & 2.42060\sym{***} & 2.52151\sym{***} \\
 & (36.74) & (44.21) & (49.77) & (61.56) \\
RQuoteMatchPassive & -0.14892\sym{***} & -0.19208\sym{***} & -0.15318\sym{***} & 0.08069\sym{***} \\
 & (-5.12) & (-6.43) & (-5.41) & (3.64) \\
BuyOrder & 0.48574\sym{***} & 0.48627\sym{***} & 0.48043\sym{***} & 0.40132\sym{***} \\
 & (43.03) & (42.67) & (43.14) & (42.23) \\
LaggedLogSize & 1.17053\sym{***} & 1.18660\sym{***} & 1.05590\sym{***} & 1.07958\sym{***} \\
 & (9.21) & (9.33) & (8.54) & (9.35) \\
LaggedTurnoverRate & 9.70974\sym{***} & 9.59857\sym{***} & 9.00635\sym{***} & 5.89252\sym{***} \\
 & (25.04) & (25.39) & (24.07) & (18.09) \\
LaggedInverseClosePrice & 8.18688\sym{***} & 8.02638\sym{***} & 8.02503\sym{***} & 5.62243\sym{***} \\
 & (10.80) & (10.65) & (10.62) & (9.02) \\
OneCentLarger & -0.46663\sym{***} & -0.43456\sym{***} & 0.19610\sym{***} & 0.03869 \\
 & (-6.56) & (-6.17) & (4.05) & (0.76) \\
MultiCentLarger & -2.85514\sym{***} & -2.83286\sym{***} & -2.19739\sym{***} & -2.34200\sym{***} \\
 & (-27.37) & (-27.26) & (-27.30) & (-30.82) \\
NonRoundVol & 0.92661\sym{***} & 0.95261\sym{***} & 0.90015\sym{***} & 0.82794\sym{***} \\
 & (74.88) & (74.43) & (74.49) & (78.92) \\
NonRoundPrice & 0.02539 & 0.03427 & -0.08847\sym{*} & 0.00341 \\
 & (0.52) & (0.70) & (-1.87) & (0.07) \\
LargeOrder & 1.00020\sym{***} & 0.99476\sym{***} & 0.94267\sym{***} & 0.86439\sym{***} \\
 & (30.12) & (29.99) & (28.83) & (27.44) \\
\midrule
Stock FE & Yes & Yes & Yes & Yes \\
Date FE & Yes & Yes & Yes & Yes \\
Observations & 42,465,539 & 42,465,539 & 42,465,539 & 42,465,539 \\
\bottomrule\end{tabular}
}
\par\smallskip\begin{minipage}{\textwidth}\footnotesize \sym{*} $p<0.10$, \sym{**} $p<0.05$, \sym{***} $p<0.01$.\end{minipage}
\end{table}

\clearpage

\begin{table}[htbp]\singlespacing\footnotesize\renewcommand{\arraystretch}{0.9}\centering
\def\sym#1{\ifmmode^{#1}\else\(^{#1}\)\fi}
\caption{Cross-Sectional Determinants of Exact-Match Activity.}
\label{tab:fmb_hft_market_quality}
\begin{minipage}{\textwidth}\small
This table reports Fama-MacBeth regression results on the cross-sectional determinants of exact-match activity. The sample includes all common stocks listed on the Shenzhen Stock Exchange, excluding the smallest 30\% by prior-day market capitalization, from July 1, 2023 to October 31, 2025. The dependent variables are daily exact-match shares, defined as the fraction of total order submissions attributable to all exact-match orders ($ExactMatch^{All}$), active exact-match orders ($ExactMatch^{Active}$), and passive exact-match orders ($ExactMatch^{Passive}$). In the first stage, we estimate daily cross-sectional regressions of exact-match share on stock characteristics and recent market conditions. In the second stage, we report the time-series averages of the daily slope coefficients. Explanatory variables include market capitalization ($LogSize=\log(1+Size)$), idiosyncratic volatility ($IVOL$, computed as the standard deviation of daily excess returns over the prior 20 trading days), 20-day cumulative excess returns, lagged exact-match shares (active and passive) over the preceding selected trading day, gross profit margin, revenue growth, and leverage. All daily market characteristics are lagged one trading day to avoid look-ahead bias; financial characteristics use disclosures published strictly before the observation date. Industry fixed effects (Shenwan Level-1 classification) are included in the first-stage regressions.
Only limit orders are included, and matching uses their submitted numerical prices; market orders (O/S) are excluded without resampling.  Both active and passive shares use limit-order counts and a limit-order denominator. Market orders are excluded. % LIMIT_ONLY_DENOMINATOR
 Standard errors are the ordinary time-series standard deviation of daily slopes divided by the square root of the number of days; no Newey--West adjustment is applied.\end{minipage}\par\smallskip\footnotesize
\paperfit{\textwidth}{
\begin{tabular}{lrrr}\toprule
 & (1) & (2) & (3) \\
 & ExactMatch$^{All}$ & ExactMatch$^{Active}$ & ExactMatch$^{Passive}$ \\
\midrule
Log(Size) & -0.00308\sym{***} & -0.00577\sym{***} & 0.00270\sym{***} \\
 & (-16.74) & (-39.16) & (15.73) \\
TradingAmount & -0.00236\sym{***} & -0.00015 & -0.00220\sym{***} \\
 & (-10.34) & (-0.89) & (-10.27) \\
TurnoverRate & 0.00279 & -0.01707\sym{***} & 0.01986\sym{***} \\
 & (0.64) & (-6.57) & (6.15) \\
IVOL(20Day) & -0.29359\sym{***} & -0.11014\sym{***} & -0.18345\sym{***} \\
 & (-26.44) & (-15.73) & (-20.44) \\
ExcessReturn(20Day) & 0.00633\sym{***} & -0.00107 & 0.00740\sym{***} \\
 & (5.68) & (-1.58) & (8.90) \\
InversePrice & 0.06877\sym{***} & -0.02694\sym{***} & 0.09571\sym{***} \\
 & (40.83) & (-21.66) & (64.34) \\
ExactMatch$^{Passive}$(t-1) & 0.86251\sym{***} & 0.14494\sym{***} & 0.71758\sym{***} \\
 & (467.40) & (107.28) & (334.38) \\
ExactMatch$^{Active}$(t-1) & 0.89085\sym{***} & 0.66815\sym{***} & 0.22270\sym{***} \\
 & (449.95) & (355.35) & (111.69) \\
GrossMargin & -0.00393\sym{***} & -0.00287\sym{***} & -0.00106\sym{***} \\
 & (-10.89) & (-11.44) & (-3.55) \\
RevenueGrowth & -0.00045\sym{***} & -0.00023\sym{***} & -0.00022\sym{***} \\
 & (-4.71) & (-3.44) & (-2.60) \\
Leverage & 0.00092\sym{***} & 0.00059\sym{**} & 0.00032 \\
 & (2.92) & (2.52) & (1.22) \\
\midrule
Trading days & 565 & 565 & 565 \\
\bottomrule\end{tabular}
}
\par\smallskip\begin{minipage}{\textwidth}\footnotesize $t$-statistics in parentheses. \sym{*} $p<0.10$, \sym{**} $p<0.05$, \sym{***} $p<0.01$.\end{minipage}
\end{table}

\clearpage

\begin{table}[htbp]\singlespacing\footnotesize\renewcommand{\arraystretch}{0.9}\centering
\def\sym#1{\ifmmode^{#1}\else\(^{#1}\)\fi}
\caption{Fama--MacBeth Regression: Matched Orders Predict Future Returns (Ret10s).}
\label{tab:FMB_MatchedOrders_Ret10s}
\begin{minipage}{\textwidth}
    \small
This table reports Fama--MacBeth cross-sectional regressions examining whether matched order activity predicts short-horizon returns.
The sample includes all common stocks listed on the Shenzhen Stock Exchange, excluding the smallest 30\% by prior-day market capitalization, from July 1, 2023 to October 31, 2025.
The dependent variable is the forward mid-quote return measured from the first mid-quote observation following order submission to the mid-quote ten seconds after submission ($Ret10s$). The data are aligned at the one-minute frequency by retaining the last sampled order in each stock--minute. In the first stage, we estimate cross-sectional regressions for each minute; in the second stage, we report the time-series averages of the slope coefficients. Newey--West standard errors are computed with a fixed lag length of 19 minutes.
Column (1) includes baseline measures of active and passive matched order intensity and standard stock-level controls. Column (2) adds the price-aggressiveness indicators OneCentLarger and MultiCentLarger. Column (3) further adds non-round order quantity and price indicators and the large-order indicator. $t$-statistics are reported in parentheses.
Only limit orders are included, and matching uses their submitted numerical prices; market orders (O/S) are excluded without resampling.  Forward returns use the first valid midpoint after submission and the last valid midpoint at or before the exact submission timestamp plus the horizon. Endpoints crossing lunch or the session close are excluded; unavailable returns remain missing. Inference uses Newey--West covariance with 19 lags across minute regressions. Rank-deficient cross sections use the minimum-norm least-squares solution; this convention can affect individual coefficients.\end{minipage}\par\smallskip\footnotesize
\paperfit{\textwidth}{
\begin{tabular}{lrrr}\toprule
 & (1) & (2) & (3) \\
 & Ret10s & Ret10s & Ret10s \\
\midrule
RQuoteMatchActive & 2.77186\sym{***} & 2.89162\sym{***} & 2.80957\sym{***} \\
 & (294.18) & (303.81) & (296.47) \\
RQuoteMatchPassive & 0.30525\sym{***} & 0.42473\sym{***} & 0.34029\sym{***} \\
 & (54.09) & (77.37) & (61.83) \\
BuyOrder & 0.28197\sym{***} & 0.28758\sym{***} & 0.28529\sym{***} \\
 & (29.57) & (30.34) & (30.28) \\
LaggedLogSize & -0.23706\sym{***} & -0.23490\sym{***} & -0.23668\sym{***} \\
 & (-116.57) & (-115.19) & (-115.83) \\
LaggedTurnoverRate & 4.79345\sym{***} & 4.93380\sym{***} & 5.13545\sym{***} \\
 & (61.62) & (62.89) & (64.92) \\
LaggedInverseClosePrice & -0.06644\sym{**} & -0.08751\sym{***} & 0.30818\sym{***} \\
 & (-2.01) & (-2.61) & (9.30) \\
OneCentLarger &  & 1.05215\sym{***} & 1.00502\sym{***} \\
 &  & (74.96) & (71.17) \\
MultiCentLarger &  & -0.04958\sym{***} & -0.11197\sym{***} \\
 &  & (-3.34) & (-7.51) \\
NonRoundVol &  &  & 0.61411\sym{***} \\
 &  &  & (114.86) \\
NonRoundPrice &  &  & -0.16666\sym{***} \\
 &  &  & (-10.13) \\
LargeOrder &  &  & 0.46587\sym{***} \\
 &  &  & (22.41) \\
\midrule
Observations & 38,701,898 & 38,701,898 & 38,701,898 \\
Minutes & 134,142 & 134,142 & 134,142 \\
\bottomrule\end{tabular}
}
\par\smallskip\begin{minipage}{\textwidth}\footnotesize $t$-statistics in parentheses. \sym{*} $p<0.10$, \sym{**} $p<0.05$, \sym{***} $p<0.01$.\end{minipage}
\end{table}

\clearpage

\begin{table}[htbp]\singlespacing\footnotesize\renewcommand{\arraystretch}{0.9}\centering
\def\sym#1{\ifmmode^{#1}\else\(^{#1}\)\fi}
\caption{Return Predictability of Exact-Match Orders: Robustness.}
\label{tab:matched_orders_predict_future_price_robustness}
\begin{minipage}{\textwidth}
\small
This table reports panel regressions of direction-adjusted forward mid-quote returns on reconstructed-quote matching indicators. 
The sample includes Shenzhen Stock Exchange common stocks from July 1, 2023 to October 31, 2025, excluding the smallest 30\% by prior-day market capitalization. 
We use an approximately 0.1\% systematically sampled subset of orders submitted during continuous trading hours. 
The dependent variable is measured in basis points: Column (1) uses $Ret120s$, and Columns (2)--(3) use $Ret10s$.
\textit{RQuoteMatchActive} and \textit{RQuoteMatchPassive} indicate orders submitted exactly at the best opposite-side and same-side quotes, respectively, reconstructed from millisecond-level order book data. 
Column (2) compares reconstructed-quote matching with three-second tick-quote matching: \textit{RQuoteTickMatchActive} and \textit{RQuoteTickMatchPassive} indicate orders matching both reconstructed and tick quotes, while \textit{TickQuoteMatchActive} and \textit{TickQuoteMatchPassive} indicate orders matching tick quotes. 
Column (3) controls for ultra-fast responses: \textit{RQuoteMatchActive100ms} and \textit{RQuoteMatchPassive100ms} indicate reconstructed-quote matched orders submitted less than 100 milliseconds after the latest prior best-quote price or quantity change, and \textit{ResponseIn100ms} indicates all orders submitted within this window.
Controls follow Table~\ref{tab:matched_orders_predict_future_price}. All models include stock and date fixed effects, with stock-clustered $t$-statistics in parentheses.
Each column uses its own estimation sample.  Columns (1) and (3) use precise-time returns; column (2) uses integer-second quote endpoints on the available snapshot sample.\end{minipage}\par\fontsize{9}{10}\selectfont
\paperfit{\textwidth}{
\begin{tabular}{lrrr}\toprule
 & (1) & (2) & (3) \\
 & Ret120s & Ret10s & Ret10s \\
\midrule
RQuoteMatchActive & 3.54050\sym{***} & 3.797\sym{***} & 0.42738\sym{***} \\
 & (60.32) & (56.28) & (16.58) \\
RQuoteMatchPassive & 0.32875\sym{***} & 1.420\sym{***} & -0.10676\sym{***} \\
 & (9.47) & (39.64) & (-5.47) \\
BuyOrder & 2.02270\sym{***} & 0.508\sym{***} & 0.50034\sym{***} \\
 & (25.83) & (41.03) & (43.92) \\
LaggedLogSize & 1.67622\sym{***} & 0.968\sym{***} & 1.08263\sym{***} \\
 & (8.78) & (10.79) & (8.76) \\
LaggedTurnoverRate & 3.25069\sym{***} & 10.42\sym{***} & 8.38827\sym{***} \\
 & (5.10) & (26.68) & (22.25) \\
LaggedInverseClosePrice & 13.18314\sym{***} & 7.201\sym{***} & 8.28837\sym{***} \\
 & (12.87) & (12.50) & (11.77) \\
OneCentLarger & 0.26006\sym{***} & -0.255\sym{***} & -0.65297\sym{***} \\
 & (3.36) & (-3.50) & (-9.03) \\
MultiCentLarger & -1.97742\sym{***} & -2.851\sym{***} & -2.93683\sym{***} \\
 & (-23.08) & (-26.62) & (-27.37) \\
NonRoundVol & 1.15560\sym{***} & 0.935\sym{***} & 0.57495\sym{***} \\
 & (46.42) & (68.50) & (49.63) \\
NonRoundPrice & 0.37988\sym{***} & 0.0761\sym{**} & 0.05107 \\
 & (4.49) & (2.47) & (1.04) \\
LargeOrder & 1.47428\sym{***} & 1.054\sym{***} & 1.15868\sym{***} \\
 & (16.29) & (30.74) & (33.57) \\
RQuoteTickMatchActive & & -1.028\sym{***} & \\
 & & (-22.20) & \\
RQuoteTickMatchPassive & & -2.125\sym{***} & \\
 & & (-49.95) & \\
TickQuoteMatchActive & & -1.248\sym{***} & \\
 & & (-26.46) & \\
TickQuoteMatchPassive & & -0.0560\sym{*} & \\
 & & (-1.78) & \\
RQuoteMatchActive100ms &  &  & 2.17432\sym{***} \\
 &  &  & (35.47) \\
RQuoteMatchPassive100ms &  &  & -0.79228\sym{***} \\
 &  &  & (-18.05) \\
ResponseIn100ms &  &  & 2.18203\sym{***} \\
 &  &  & (47.00) \\
\midrule
Stock FE & Yes & Yes & Yes \\
Date FE & Yes & Yes & Yes \\
Observations & 41,861,959 & 42,121,182 & 42,465,539 \\
\bottomrule\end{tabular}
}
\par\smallskip\begin{minipage}{\textwidth}\footnotesize $t$-statistics in parentheses. \sym{*} $p<0.10$, \sym{**} $p<0.05$, \sym{***} $p<0.01$.\end{minipage}
\end{table}

\clearpage

\begin{table}[p]\centering\singlespacing\footnotesize
\caption{Latency Crossings: Stock $\times$ Date Clustering.}\label{tab:did_stock_date}
\begin{minipage}{\textwidth}\small The sample contains 24 trading days and uses $G^F_t-10\le s<G^S_t$, with $G^S_t-G^F_t>1$, stock $\times$ date and date $\times$ calendar-second fixed effects. Standard errors are clustered by stock $\times$ date. Coefficients and observations match Tables~\ref{tab:did_latency_in_3s} and \ref{tab:did_market_expanded}. Pre-event tests omit $s-G^F_t=-1$ and jointly test nine leads, except ES, AS1s and RS1s: their forward-window restriction makes $s-G^F_t=-1$ unavailable, so they omit $s-G^F_t=-2$ and test eight leads. Price measures are in bp, shares in pp, and depth in log units. PE uses its valid filtered-transaction sample.\end{minipage}\par\medskip
\begin{tabular}{llrrrr}\toprule
Outcome & Unit & Coefficient & $t$ & $N$ & Pre-event $p$ \\\midrule
\multicolumn{6}{l}{\textbf{Panel A: Exact-Match Activity}} \\
Active count & orders & 0.44718$^{***}$ & 9.27 & 687,663 & 0.013 \\
Active share & pp & 0.50539$^{***}$ & 3.59 & 685,258 & 0.330 \\
Passive share & pp & 0.51496$^{***}$ & 3.57 & 685,258 & 0.094 \\
\midrule
\multicolumn{6}{l}{\textbf{Panel B: Liquidity and Price Variation}} \\
Spread & bp & -0.00330 & -0.05 & 685,371 & 0.117 \\
LogDepth & log & -0.01386$^{*}$ & -1.96 & 686,315 & 0.382 \\
Range & bp & 1.02884$^{***}$ & 17.37 & 685,371 & 0.101 \\
WideSpread & pp & 2.96447$^{***}$ & 10.78 & 685,554 & 0.518 \\
\midrule
\multicolumn{6}{l}{\textbf{Panel C: Transaction Costs and Pricing Error}} \\
ES$^{\dagger}$ & bp & -0.04906 & -0.53 & 482,510 & 0.197 \\
AS1s$^{\dagger}$ & bp & 0.64810$^{***}$ & 11.24 & 482,510 & 0.498 \\
RS1s$^{\dagger}$ & bp & -1.34527$^{***}$ & -10.08 & 482,510 & 0.750 \\
PEbps & bp & -0.04486$^{***}$ & -3.38 & 541,710 & 0.737 \\
\bottomrule\end{tabular}
\par\smallskip\begin{minipage}{\textwidth}\footnotesize $^{*}p<0.10$, $^{**}p<0.05$, $^{***}p<0.01$. $^{\dagger}$Forward-window-restricted outcomes.\end{minipage}
\end{table}

\clearpage

\begin{table}[p]\centering\singlespacing\footnotesize
\caption{Exclusion Tests: Market Orders and More Aggressive Limit Orders.}\label{tab:exclusion_tests}
\begin{minipage}{\textwidth}\small This table reports difference-in-differences estimates using the baseline three-second latency-crossing design. All specifications contain 685,258 stock-second observations across 24 trading days and use $G^F_t-10\leq s<G^S_t$, with $G^S_t-G^F_t>1$. Each measure divides the relevant order count by the number of limit orders with valid two-sided quotes in the same stock-second and multiplies the ratio by 100. OneCentLargerShare and MultiCentLargerShare refer to limit orders priced exactly one tick and more than one tick beyond the best opposite-side quote, respectively. MarketOrderShare measures opposite-side-best and same-side-best instructions relative to eligible limit orders, not total orders. All specifications include stock $\times$ date and date $\times$ calendar-second fixed effects, with two-way clustering by stock and date $\times$ calendar-second. Holm-adjusted $p$-values cover the three reported outcomes. Pre-event joint tests cover seconds $-10$ through $-2$, with second $-1$ omitted. Coefficients are reported in percentage points (pp), with $t$-statistics in parentheses.\end{minipage}\par\medskip
\begingroup\setlength{\tabcolsep}{4pt}
\begin{tabular}{lrrrrr}\toprule
Outcome & Coefficient & Raw $p$ & Holm $p$ & Pre-event $p$ & $N$ \\\midrule
OneCentLargerShare & 0.0615 & 0.4194 & 0.4194 & 0.3250 & 685,258 \\
 & (0.81) & & & & \\
MultiCentLargerShare & -0.3314$^{***}$ & 0.0005 & 0.0014 & 0.2494 & 685,258 \\
 & (-3.54) & & & & \\
MarketOrderShare & -0.0600$^{*}$ & 0.0956 & 0.1913 & 0.2489 & 685,258 \\
 & (-1.67) & & & & \\
\bottomrule\end{tabular}\endgroup
\par\smallskip\begin{minipage}{\textwidth}\footnotesize Stars indicate unadjusted significance: $^{*}p<0.10$, $^{**}p<0.05$, $^{***}p<0.01$.\end{minipage}
\end{table}

\clearpage

\begin{table}[p]\centering\singlespacing\footnotesize
\caption{Fixed-Grid Wide-Spread Share.}\label{tab:wide_share}
\begin{minipage}{\textwidth}\small The 24-day design uses $G_F-10\le s<G_S$, $G_S-G_F>1$, stock $\times$ date and date-by-calendar-second fixed effects. Each outcome uses its own valid sample. Shares are percentage points (pp). Pre-event tests use $s-G^F_t=-1$ as reference and jointly test $s-G^F_t=-10,\ldots,-2$. The outcome is the fraction of ten fixed 100ms grid points with spread greater than one tick. All ten quotes must be valid.\end{minipage}\par\medskip
\begin{tabular}{llrrrr}\toprule
Outcome & Unit & Coefficient & $t$ & $N$ & Pre-event $p$ \\\midrule
\multicolumn{6}{l}{\textit{Two-way clustering by stock and date $\times$ calendar-second}} \\
WideSpreadShare & pp & 0.51993$^{*}$ & 1.94 & 685,519 & 0.275 \\
\midrule
\multicolumn{6}{l}{\textit{Stock $\times$ date clustering}} \\
WideSpreadShare & pp & 0.51993$^{**}$ & 2.02 & 685,519 & 0.283 \\
\bottomrule\end{tabular}
\par\smallskip\begin{minipage}{\textwidth}\footnotesize $^{*}p<0.10$, $^{**}p<0.05$, $^{***}p<0.01$.\end{minipage}
\end{table}

\clearpage

\section*{Part III: Appendix Figures}

\clearpage

\clearpage

\clearpage

\clearpage

\end{document}